\documentclass[a4paper,fleqn,usenatbib]{mnras}
\usepackage[T1]{fontenc}
\usepackage{ae,aecompl}
\usepackage{amstext,natbib,bm}
\usepackage{url}
\bibpunct{(}{)}{;}{a}{}{,}
\usepackage{graphicx,epsfig,color,latexsym,nicefrac,textcomp}
\usepackage{amsmath}	
\usepackage{amssymb}	

\newcommand{\bea}{\begin{eqnarray}}
\newcommand{\eea}{\end{eqnarray}}
\newcommand{\be}{\begin{equation}}
\newcommand{\ee}{\end{equation}}
\newcommand{\rund}[1]{\left(#1\right)}

\newcommand{\eck}[1]{\left[ #1 \right]}

\renewcommand{\exp}{\mathrm{exp}}

\title[Dual plasma lenses]{Dual-Component Plasma Lens Models}
\author[Rogers and Er]
       {Adam Rogers$^1$ \thanks{E-mail: rogersa@brandonu.ca}
	 Xinzhong Er$^2$ \thanks{E-mail: xer@ynu.edu.cn}\\
$^1$ Department of Physics and Astronomy, Brandon University, Brandon, MB, R7A 6A9, Canada \\
$^2$ South-Western Institute For Astronomy Research, Yunnan University, Kunming, P.R.China\\
}

\date{Accepted XXX. Received YYY; in original form \today}
\pubyear{2018}

\begin{document}
\label{firstpage}
\maketitle

\begin{abstract}
In contrast to the converging, achromatic behaviour of axisymmetric gravitational lenses, diverging frequency-dependent lensing occurs from refraction due to a distribution of over-dense axisymmetric plasma along an observer's line of sight. Such plasma lenses are particularly interesting from the point of view of astronomical observations because they can both magnify and dim the appearance of background sources as a function of frequency. Plasma lensing is believed to be involved in a number of separate phenomena involving the scintillation of radio pulsars, extreme scattering events of background radio sources and may also play a role in the generation of fast radio bursts. These lensing phenomena are believed to occur in dense environments, in which there may be many density perturbations between an observer and background source. In this work we generalize individual plasma lens models to produce dual component lenses using families of plasma lens models previously studied in the literature, namely the exponential and softened power-law lenses. Similar to binary gravitational lens models, these dual component plasma lenses feature a rich and complex critical and caustic morphology, as well as generate more complicated light curves. We map the number of criticals formed for a given component separation and angular size, and highlight a relevant degeneracy between two particular models. This work provides an argument in favor of close monitoring of extreme scattering events in progress in order to break such model degeneracies.
\end{abstract}


\section{Introduction}
\label{sec:intro}

Axisymmetric gravitational lenses behave like astrophysical convex lenses, acting to converge the light from background sources. This process preserves the surface brightness of a source due to Liouville's theorem, but changes the apparent solid angle, potentially forming multiple images and magnifying the source \citep[e.g.][]{SEF,1999fsu..conf..360N,1986ApJ...310..568B}. In addition to gravitational lensing, plasma lensing occurs due to over-dense perturbations of the electron density along an observer's line of sight rather than mass. The optical effect of plasma over-densities produce this frequency-dependent lensing, in contrast to the achromatic behaviour of gravitational lenses and can act to magnify or dim a source based on the source-lens-observer geometry \citep[e.g.][]{cleggFey,1990ARA&A..28..561R,2004ARA&A..42..275S}. While the effects of both plasma and gravitation have been used to describe the environments surrounding massive compact objects \citep{BKT09,review,erMao,rogers15,rogers17a,rogers17b}, here we study purely plasma lens behaviour that occurs due to the inhomogeneity of the interstellar medium (ISM). We will follow previous work on the diverging effects of plasma lenses and consider two distinct families of lens density profiles, namely the softened power-law (SPL) and exponential lens models \citep{ErRogers18}.

Extreme scattering events (ESEs) are dynamical variations in the
brightness of background radio sources \citep{ESEIntro1,ESE0}. Rather than intrinsic fluctuations in the source it is believed that the cause of ESEs may be due to plasma lensing in the ISM, which refracts low-frequency radiation away from the observer \citep{ESE0, romani87, cleggFey}. If this scenario is correct, the required free-electron plasma density of symmetric lenses must be on the order of $\sim1000$cm$^{-3}$. Such lenses are therefore significantly overpressured compared to the typical diffuse ISM \citep[see e.g.][for more details]{ISMpressure,ESEreview}. Various alternatives have been suggested to address this overpressure problem, including sheets of plasma rather than isolated clumps \citep[e.g.][]{romani87,ESE2,ESE5}, or small localized inhomogeneities in the environs of hot stars rather than due to the ISM itself \citep{starESE}. For a thorough review of these issues, see \citet{ESEreview}. The exponential lens family that we consider includes the Gaussian lens density profile \citep{cleggFey} as a member, which is well-established and studied in the literature. The Gaussian lens has been applied to a large sample of ESEs \citep{ESE0, F94, cleggFey, ESE2D} and has been suggested for more exotic phenomena \citep[such as fast radio bursts;][]{FRBplasma1, FRBplasma2}. Thus, the exponential lens family is of general interest. The SPL lenses are a second family which produce similar magnifications so are also of astrophysical interest \citep{ErRogers18}.

In analogy with binary lens models in gravitational lensing, it is also possible to develop dual-component plasma lens models. These models describe lens components that are not bound to one another as in the gravitational lens case, but represent independent electron density enhancements that occupy positions near the observers line of sight and hence act in concert to affect the trajectories of passing light rays from distant objects. Such multi-component plasma lens models are physically motivated by an ESE recently observed in PSR J1740-3015 \citep{kerr2018}. This ESE shows two flux minima, with a central maximum peak. The observed flux was modeled using a dual-lobed electron column density profile that can be reproduced by simply adding two gaussian components of different amplitudes. This observation shows conclusively that plasma lens models with simple, highly symmetric density profiles are not necessarily sufficient to reproduce the entire morphology of observed ESE light curves, and as such a study of dual component analytical plasma lenses is particularly relevant. In this work we will focus on two families of plasma lenses that we have previously studied, the softened power-law and exponential lenses, including an investigation of the modifications to the caustic and critical curves that dual-component plasma lenses produce.

Plasma lenses are involved in a variety of interesting phenomena. Recent examples have included the discovery of frequency-dependent eclipses of a black widow pulsar by a brown dwarf companion that is emitting an ionized outflow due to Roche lobe overflow \citep{penNature}. The outflow from the planet produces a shroud of plasma which periodically eclipses the neutron star and results in frequency-dependent variations of brightness due to refraction of the pulsar emission. The basic mechanism of plasma lensing may also be related to fast radio bursts \citep[FRBs;][]{FRBplasma1}. These short pulses of high energy radio waves may be the result of magnification from plasma lensing due to compact structures in the dense environment of the FRBs host galaxy. Due to the time delay effects of refraction on pulsars, plasma lensing is also relevant to pulsar timing networks such as NANOGrav \citep{nanoGravPSR, nanoGrav11yr}.

We begin by reviewing the basic plasma lens theory and discuss the previously studied model families in Section\,\ref{sec:theory}. We then discuss the dual-component exponential lens and softened power-law plasma lenses in Section\,\ref{sec:multiLens}. We discuss our results in Section\,\ref{sec:discussion} and summarize our conclusions in Section\,\ref{sec:conclusions}. A gallery of criticals and caustics for a selection of dual component lens examples is discussed in \ref{appA}.

\section{Plasma Lens Theory and Model Families}
\label{sec:theory}

Plasma lens models follow the general gravitational lens formalism discussed in \citet{SEF} and \citet{narayan}. For astrophysically relevant situations, the deflection angle is taken to be small. Due to the great distances between the source and lens ($D_\text{ds}$) and the distances from lens and source to the observer ($D_\text{d}$ and $D_\text{s}$ respectively), the thin lens approximation can be adopted. Although the magnetic field can cause some observational effects, especially to the photon polarizations \citep{penNature, cordesBirefringence, turimovLens}, in this work we consider the plasma to be cold and neglect any magnetic field using the index of refraction
\be
n_\text{r}^2 = 1-\frac{\omega_\text{e}^2}{\omega^2}
\ee
where $\omega$ is the frequency of the ray \footnote{Note that $\omega$ is not a constant in general when both plasma and gravitational lensing effects are present due to the gravitational redshift effect. We do not consider gravitational effects in this work.}, and the plasma frequency
\be
\omega_\text{e}^2 = \frac{e^2 n_\text{e}(r)}{\epsilon_0 m_\text{e}}
\ee
depends on the electron number density $n_\text{e}(r)$ with the electron charge $e$, the mass of the electron $m_e$, and $\epsilon_0$ is the permittivity of free space.

The coordinates that the observer uses to describe the lensed image positions are $\theta=\sqrt{\theta_\text{x}^2+\theta_\text{y}^2}$. The effective lens potential in the lens plane is given by
\begin{equation}
\psi(\theta)=\frac{D_\text{ds}}{D_\text{s}D_\text{d}}\frac{1}{2 \pi} r_\text{e} \lambda^2 N_\text{e}(\theta)
\end{equation}
where $\lambda=2\pi c/\omega$ is the wavelength of the ray and the classical electron radius is
\begin{equation}
r_\text{e}=\frac{e^2}{4 \pi \epsilon_0 m_\text{e} c^2}.
\end{equation}
The lens potential also depends on the projected electron density over the lens plane,
\begin{equation}
N_\text{e}(\theta)=\int n_\text{e}dz.
\end{equation}
The deflection of rays occurs as they pass through the plasma density on the lens plane between the source and observer. The image coordinates are related to the coordinates that describe the position of the source $\beta=\sqrt{\beta_\text{x}^2+\beta_\text{y}^2}$ through the lens equation
\be
\mathbf{\beta} = \mathbf{\theta} - \frac{D_\text{ds}}{D_\text{s}}\mathbf{\hat{\alpha}}= \mathbf{\theta}-\nabla_\theta \psi(\theta)
\label{TLE}
\ee
where $\mathbf{\hat{\alpha}}$ is the deflection angle given in terms of the effective lens potential. Thus, for a given electron density distribution $n_\text{e}(r)$, we find the corresponding projected density $N_\text{e}$ and effective lens potential $\psi(\theta)$. The lens potential gives the corresponding deflection angle as in Eq.\,\ref{TLE} and all other relevant quantities in analogy with the gravitational lensing formalism.

Plasma lenses are particularly interesting because they can produce both magnification and demagnification depending on the relative positions of the source, lens and observer. Especially in the high density regions of the lens, low-frequency radiation will be fully deflected away from the line of sight. In such regions, there will be no images formed, which we refer to as the exclusion region of the diverging lens. The light curves produced by diverging lenses thus have a U or W-shape, depending on the details of the plasma distribution along the line of sight \citep[e.g.][]{coles2015,kerr2018}. The scale and properties of the exclusion region depend on the specific details of the electron density, such as the density gradient. We will see in the following sections that the dual lens models can increase the lensing efficiency and thus affect the exclusion region.

\subsection{Exponential Lenses}
\label{sec:Exp}
Following the analysis of the Gaussian plasma lens \citep{cleggFey}, we define the family of exponential lenses specifying the projected electron density on the lens plane $N_\text{e}$ directly,
\be
N_\text{e}(\theta) = N_\text{0}\exp\left( - \frac{\theta^h}{h \sigma^h} \right)
\ee
with $N_\text{0}$ the maximum electron column density and $\sigma$ the width of the lens for $h>0$. The normalization is chosen to match \citet{newESE2017} and \citet{ErRogers18} for the Gaussian lens which is realized for the exponent $h=2$.
The projected electron density gives the potential
\be
\psi(\theta)= \theta_0^2 \exp\left( -\frac{\theta^h}{h\sigma^h} \right)
\ee
and deflection angle
\be
\alpha_\text{exp}(\theta)=-\theta_\text{0}^2 \frac{\theta^{(h-1)}}{\sigma^h}\exp\left( -\frac{\theta^h}{h \sigma^h} \right)
\label{deflExp}
\ee
with the characteristic angular scale
\be
\theta_0 = \lambda\left( \frac{D_\text{ds}}{D_\text{s} D_\text{d}} \frac{1}{2\pi} r_\text{e} N_\text{0} \right)^\frac{1}{2},
\ee
where $\lambda$ is the observing wavelength, and $r_e$ is the classical electron radius.

The exponential lens can produce a variety of image configurations. For each $h$ value, we can define a critical limit below which sub-critical lenses produce no critical curves, and therefore only a single image $\theta_0<f(h)\sigma$. The super-critical lenses $\theta_0>f(h)\sigma$ produce multiple images. When $h=1$, super-critical lenses can form up to $2$ images. The $h=2$ and $h=3$ lenses can form three images. The critical values for $h=1,2,3$ are given in Table \ref{tab:exp}, which shows an example for $\sigma=1$. With $\sigma=1$, the condition for critical lenses with ($h\neq 1$) is
  \be
  f(h)=\eck{F^{\frac{h-2}{h}}\, \rund{F+1-h}\, e^{\frac{-F}{h}}}^{-1/2},
  \ee
  where the factor $F$ is
  \be
  F=\frac{1}{2}\eck{3(h-1)+\sqrt{(h-1)(5h-1)}}.
  \ee

\begin{table}
  \begin{tabular}{c|c|c|c}
    exponent &$h=1$  &$h=2$  &$h=3$\\
    \hline
    $f(h)$  &$1$   &$1.49$  &$1.006$\\
    \hline
  \end{tabular}
  \caption{The requirement on $\theta_0$ to produce critical curves for an exponential single lens with $\sigma=1$ \citep{ErRogers18}.}
  \label{tab:exp}
\end{table}

\subsection{Power-Law Lenses}
\label{sec:PL}
The family of power-law lenses is produced by a three-dimensional electron density given by
\be
n_\text{e}(r) = n_\text{0}\frac{R_0^h}{r^h}
\ee
with the power-law index $h$ and the characteristic radius $R_\text{0}$ at which $n_\text{e}(R_\text{0}) = n_\text{0}$. This electron density profile produces an effective lens potential \citep{ErRogers18}:
\begin{equation}
\psi(\theta) = \left\{
\begin{array}{ll}
 -\theta_{0}^2 {\rm \ln} \theta, & h=1\\
\\
 \dfrac{\theta_{0}^{h+1}}{(h-1)} \dfrac{1}{\theta^{h-1}}, & h \neq 1
\end{array}\right.
\label{softpl}
\end{equation}
and gives the deflection angle
\be
\alpha_\text{PL}(\theta) = - \frac{\theta_\text{0}^{h+1}}{\theta^h}
\label{deflPL}
\ee
which is written in terms of the characteristic angular scale
\be
\theta_0 = \left( \lambda^2 \frac{D_\text{ds}}{D_\text{s} D_\text{d}^h} \frac{r_\text{e} n_\text{0} R_\text{0}^h}{\sqrt{\pi}} \frac{\Gamma\left( \frac{h}{2} + \frac{1}{2} \right)}{\Gamma \left( \frac{h}{2} \right)} \right)^\frac{1}{h+1},
\ee
The deflection angle for this density distribution can be found in \citet{BKT09}, who included an additional contribution from gravitational lensing to study lensing by a compact object embedded in a non-uniform plasma.

A further modification of the power-law lens deflection angle comes from the inclusion of a finite core $\theta_\text{c}$, which acts to soften the
singularity at the origin. The softened power-law (SPL) lens is also
used in gravitational lens models, often called the Plummer lens. It
is trivial to soften a power-law lens using a finite core by simply
making the transformation $\theta \rightarrow \sqrt{\theta^2 +
  \theta_\text{c}^2}$, giving the deflection angle \footnote{ Note that the softened deflection angle given by Eq.\,\ref{deflPLFinite} is not equivalent to the deflection angle
  derived from a three-dimensional softened electron density, i.e. $n_0R_0^h/(r+r_c)^h$.}
\be
\alpha_{SPL}(\theta) = - \theta_\text{0}^{h+1}\frac{\theta}{\left( \theta^2+\theta_\text{c}^2 \right)^{h+1}}.
\label{deflPLFinite}
\ee
When $h=0$, the profile yields a deflection angle analogous to the Plummer lens \citep{1911MNRAS..71..460P}.

In a similar manner to the exponential lens, we can also define a critical value of the core size required to produce a critical curve. This limit is given in terms of the characteristic angular scale, such that
\be
\theta_\text{crit}(h) = \theta_0 \left[ 2 \left(\frac{3}{h} +1 \right)^{-\frac{h+3}{2}} \right]^\frac{1}{h+1}.
\ee
For $\theta_c<\theta_\text{crit}$, the SPL lens forms two critical curves. For core size equal to $\theta_\text{crit}$, one critical curve is formed, and when the core size is in excess of $\theta_\text{crit}$ a single SPL lens does not produce any criticals. We show the results for $h=1$, $2$ and $3$ with $\theta_0=1$ in Table \ref{tab:SPL}.
\begin{table}
  \begin{tabular}{c|c|c|c}
    exponent &$h=1$  &$h=2$  &$h=3$\\
    \hline
    $\theta_\text{crit}(h)$  &$0.354$   &$0.587$  &$0.707$\\
    \hline
  \end{tabular}
  \caption{The requirement $\theta_\text{c}<\theta_\text{crit}$ to produce critical curves for an SPL single lens with $\theta_0=1$ as a function of $h$ \citep{ErRogers18}.}
  \label{tab:SPL}
\end{table}

\section{Dual-Component Lens Models}
\label{sec:multiLens}

For a dual-component lens, we locate both lens components on the $\theta_x$ axis, with their centers at $(\theta_\text{xj}, 0)$ and $j=1,2$. Radial distances in the image coordinates from the center of each lens are then $\Theta_j = \sqrt{ (\theta_\text{x} - \theta_\text{xj})^2 + \theta_\text{y}^2}$.
We will generally arrange the lenses on the $\theta_\text{x}$ axis such that $\theta_\text{yj}=0$. Moreover, for simplicity the lenses are equally spaced from the origin, i.e.,
$\theta_\text{x1}=-\theta_\text{x2}$. Let us call the distance between the lenses $s$. We will denote the individual lens components characteristic scale radii as $\theta_{01}$ and $\theta_{02}$ respectively. The Cartesian two-dimensional thin lens equation is written component-wise,
\bea
\beta_\text{x} &=& \theta_\text{x} - \alpha_\text{x1}(\Theta_1) - \alpha_\text{x2}(\Theta_2)\\
\beta_\text{y} &=& \theta_\text{y} - \alpha_\text{y1}(\Theta_1) - \alpha_\text{y2}(\Theta_2).
\eea
The Cartesian components of the deflection angle are given by
\begin{equation}
\alpha_\text{xj}(\Theta_j, \theta_\text{xj} )=\alpha(\Theta_j) \frac{(\theta_\text{x} - \theta_\text{xj})}{\Theta_\text{j}}
\end{equation}
\begin{equation}
\alpha_\text{yj}(\Theta_j, \theta_\text{xj})=\alpha(\Theta_j) \frac{\theta_\text{y}}{\Theta_\text{j}}.
\end{equation}
We have explicitly included the unit vectors in these expressions, which point toward the center of each individual lens at $(\theta_{\text{x}j}, 0)$. We use this general scheme for both the exponential and the power-law lens.

The magnification produced by the thin lens equation is the inverse of the Jacobian determinant. In general for a given lens model, the Jacobian is
\begin{equation}
\mathbf{A}= \| \dfrac{\partial \mathbf{\beta}}{\partial \mathbf{\theta}} \| = \left(
\begin{array}
[c]{cc}
1-\dfrac{\partial \alpha_\text{x1} }{\partial \theta_\text{x}} -\dfrac{\partial \alpha_\text{x2} }{\partial \theta_\text{x}}& -\dfrac{\partial \alpha_\text{x1} }{\partial \theta_\text{y}} -\dfrac{\partial \alpha_\text{x2} }{\partial \theta_\text{y}} \\
\\
-\dfrac{\partial \alpha_\text{y1}}{\partial \theta_\text{x}} -\dfrac{\partial \alpha_\text{y2} }{\partial \theta_\text{x}}& 1-\dfrac{\partial \alpha_\text{y1}}{\partial \theta_\text{y}} -\dfrac{\partial \alpha_\text{y2}}{\partial \theta_\text{y}}
\end{array}
\right)
\label{magTensor}
\end{equation}
with the magnification of a particular image given as
\begin{equation}
  \mu_k=\frac{1}{{\rm det} \mathbf{A}}
  \label{eq:detamag}
\end{equation}
and the total magnification $\mu_\text{T} = \sum |\mu_k|$ as the sum over the $k$ images produced by the lens.

In the case of a single lens component, both the exponential and softened power-law lenses can be solved exactly for conditions when the lenses will transition from having no critical curves to producing one or more \citep{ErRogers18}. For dual component lenses from both model families, the analytical expression for the Jacobian determinant is formidable and does not provide such a tidy analytical result, except in the simplest cases when the lens is widely separated and the two components can be treated independently from one another. For general considerations of the criticals of the dual-component lenses, we employ numerical methods to study their properties. Due to the richness of the mathematics even for the gravitational binary point lens \citep[e.g.][]{2012RAA....12..947M}, we present numerical results in this work.

\subsection{The Dual-Component Exponential Lens}
\label{sec:dualExp}

The exponential lens has deflection angle given in Eq.\,\ref{deflExp}, which gives\begin{equation}
  \dfrac{\partial \alpha_\text{xj}}{\partial \theta_\text{x}} =
  E_j\left[ (2-h+ \dfrac{\Theta_j^h}{\sigma^h})\dfrac{(\theta_\text{x}-\theta_\text{xj})^2}{\Theta_{j}^2}   - 1 \right],
\end{equation}
with $E_j=\theta_{0j}^2 \dfrac{\Theta_j^{h-2}}{\sigma^h}e^{-\dfrac{\Theta_j^h}{h\sigma^h}} $,
and similarly for $\partial \alpha_\text{yj} / \partial \theta_\text{y}$. The mixed terms have $\partial \alpha_\text{xj} / \partial \theta_\text{y} = \partial \alpha_\text{yj} / \partial \theta_\text{x}$ and
\begin{equation}
\dfrac{\partial \alpha_\text{xj}}{\partial \theta_\text{y}} = E_j \dfrac{(\theta_\text{x} - \theta_\text{xj}) \theta_\text{y} }{\Theta_j^2} \left[ 2 - h +\dfrac{\Theta_j^h}{\sigma^h} \right].
\end{equation}
To simplify and compare the action of the dual-component lenses, we will use $\sigma=1$ (we omit the unit of arcsec if not mentioned) for both families of dual models unless otherwise stated.

\begin{center}
\begin{figure}
  \includegraphics[width=8cm]{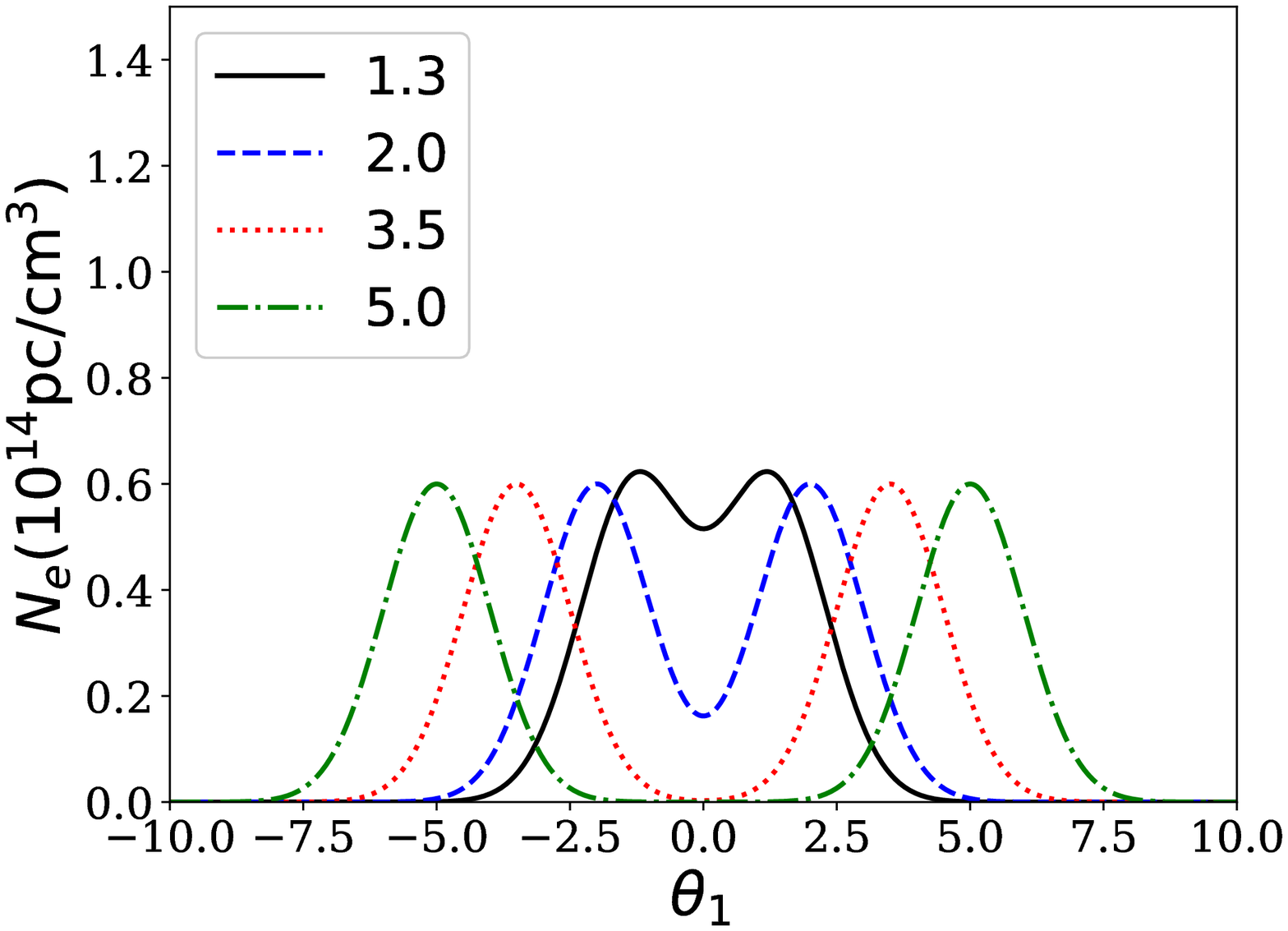}
  \includegraphics[width=8cm]{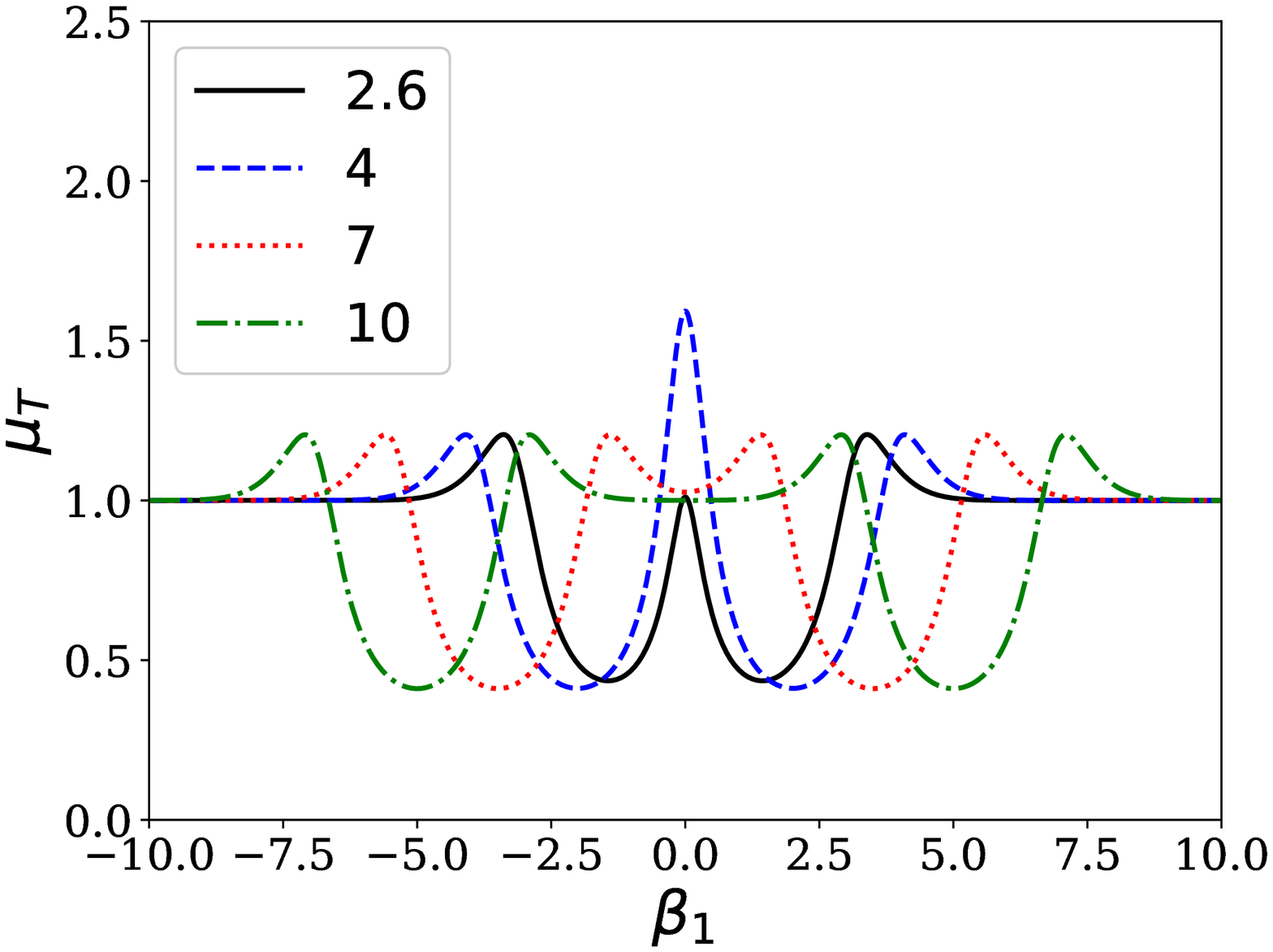}
  \caption{Top panel: The two Gaussian ($h=2$) electron density components $N_\text{e}(\theta)$ are shown here in units of pc/cm$^3$. The Gaussian components are placed equidistant from the origin with $\theta_\text{x2}=-\theta_\text{x1}=s/2$ (as labeled on the legend) and $\theta_\text{y1}=\theta_\text{y2}=0$. From solid, dashed, dotted and dash-dotted line, the components are separated by $s=2.6$, $4$, $7$ and $10$ respectively. Bottom panel: The resulting magnification for the two components. The lens is chosen to be sub-critical such that $\theta_\text{0j}=0.5 f(2)\sigma$ with $\sigma=1$ for this example. From this $\theta_\text{0j}$, we find the maximum plasma density $N_0$ assuming $D_\text{s}=1$ kpc, with the lens equidistant between observer and source. In this example, we chose an arbitrary observation frequency of $800$ MHz.}
\label{figGaussian}
\end{figure}
\end{center}

In Fig.\,\ref{figGaussian} we demonstrate the lightcurve produced by
dual Gaussian lenses ($h=2$), with characteristic radius
$\theta_\text{0}=0.5 f(2)\sigma$. As expected from the model
fit to the ESE in PSR J1740-3015 \citep{kerr2018}, when both of the
Gaussian components are placed closeby one another ($s=2.6$), the
lightcurve of a source that passes directly behind the lenses produces
a symmetrical lightcurve, showing a local maximum bracketed by two
minima. As the Gaussian components are separated from one another to
$s=4$, the light curve changes, with the central peak becoming the
global maximum. When the lens components are further separated from one
another ($s=7$), the light curve shows three dips. Finally, with the
components sufficiently separated at $s=10$, the lenses behave individually. Both components affect the light curve, which shows two essentially independent Gaussian lenses.

\begin{center}
\begin{figure}
  \includegraphics[width=4.1cm]{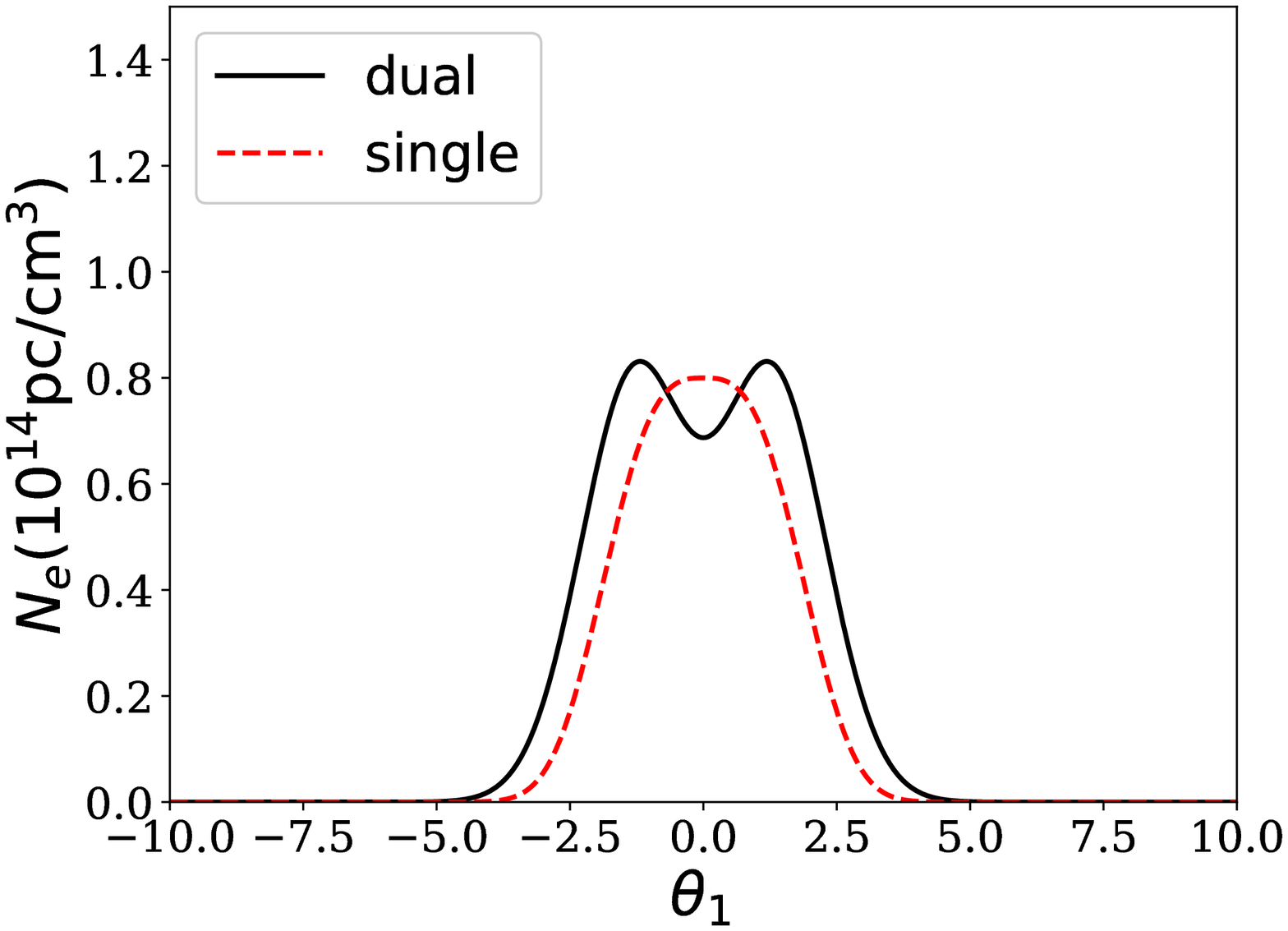}
  \includegraphics[width=4.1cm]{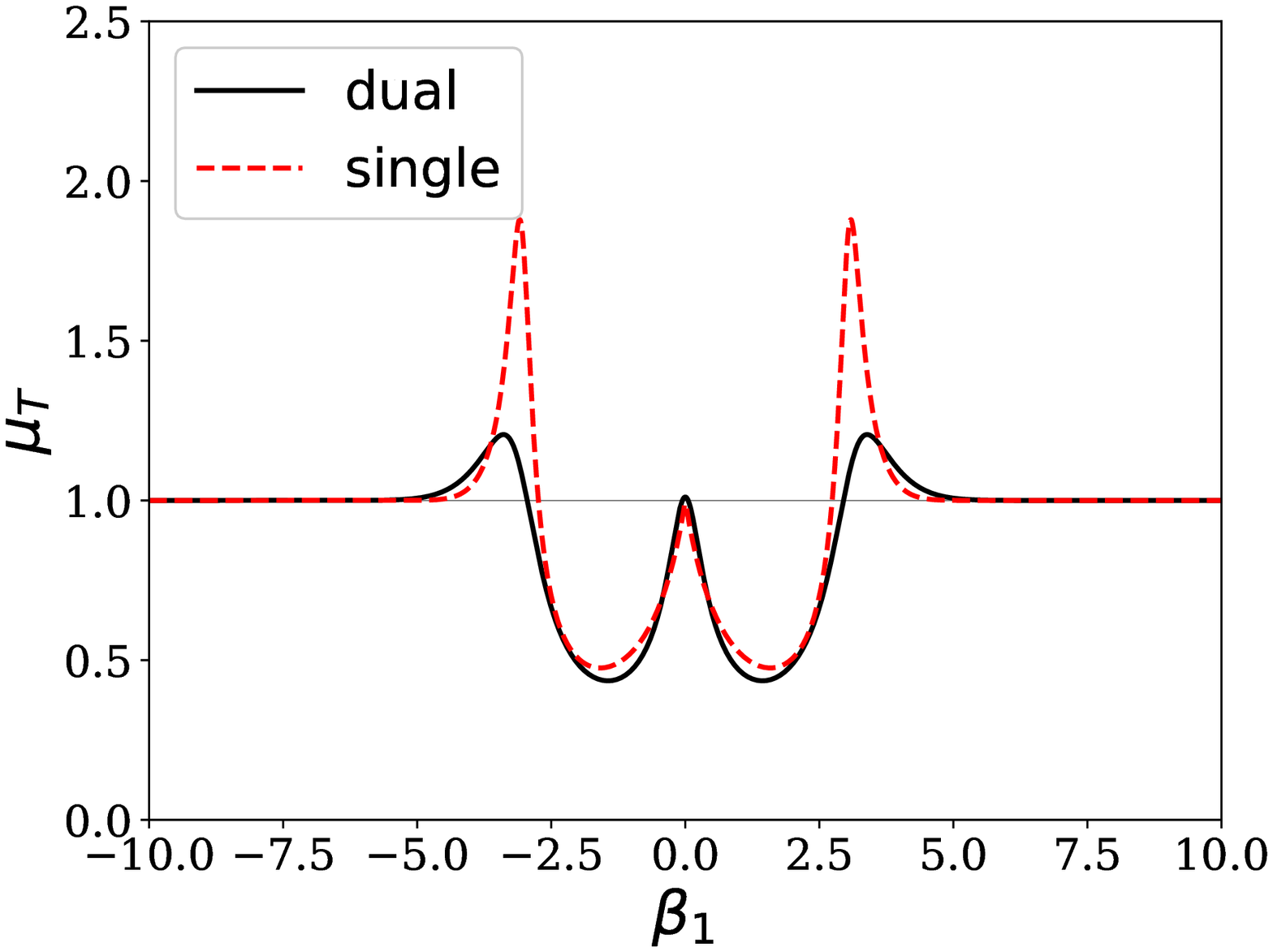}
  \caption{In this figure we compare the behaviour of the dual-component Gaussian lens (black) and a single $h=3$ exponential lens (red). Left column: The Gaussian components are placed equidistant from the origin with $s=2.6$ and $\sigma=1$. The $h=3$ component is centered on the origin with $\sigma=1.5$. Right column: The resulting magnification for the two components. The Gaussian lens is chosen to be sub-critical such that $\theta_0=0.5 f(h)\sigma$ and the $h=3$ lens has $0.73 f(3)\sigma$, chosen arbitrarily to emphasize the similarity between the models. Physical details of the density scale are given in Figure \ref{figGaussian}.}
\label{figCompare}
\end{figure}
\end{center}

An interesting degeneracy arises in the case of the closely separated ($s=2.6$), sub-critical dual Gaussian lens shown by the solid black lines in Fig.\,\ref{figGaussian}. This symmetric configuration produces a light curve that is nearly identical to a sub-critical spherically symmetric $h=3$ exponential lens with $\sigma=1.5$ and $\theta_\text{0}=0.73 f(3)\sigma$, chosen arbitrarily to match the dual Gaussian components. We demonstrate the degeneracy between these models in Fig.\,\ref{figCompare}, with the electron density shown in the left column and the light curve for a source passing directly behind the lenses on the right hand side. In the figure, the black curve represents the dual component Gaussian, and the red curve is the $h=3$ exponential. The major difference in these models comes from the magnification at the edge of the exclusion region of the $h=3$ exponential lens. The dual Gaussian does not contain a sharp boundary and the magnification is smooth in this area. However, particularly at the lens center, the degeneracy between these models is striking and surprising given the difference between the density distributions. This example emphasizes the importance of close and careful monitoring of ESEs in progress across multiple frequencies to distinguish between such models. The super-critical lens ($\theta_\text{0j}>f(h)\sigma$) will generate a more complicated arrangement of critical curves on the edge of the exclusion region, producing multiple magnification peaks, and can be easily distinguished from the single lens case with the $h=3$
exponential provided the light curve is well-sampled in this region. The similarity between Gaussian lenses and a single $h=3$ exponential lens is broken when the dual lens components are asymmetric ($\theta_\text{01} \neq \theta_\text{02}$), which produces an asymmetric light curve as seen in the case of the ESE modelled in the pulsar PSR J1740-3015 \citep{kerr2018}. However, this example of model degeneracy provides a relevant argument for the ongoing close monitoring of ESEs in progress, especially at multiple wavelengths and during caustic crossing events, which can distinguish between these lens models. In addition, other similar degeneracies may exist for plasma lens models not yet developed or explored in the literature.

\begin{figure}
  \includegraphics[width=8cm]{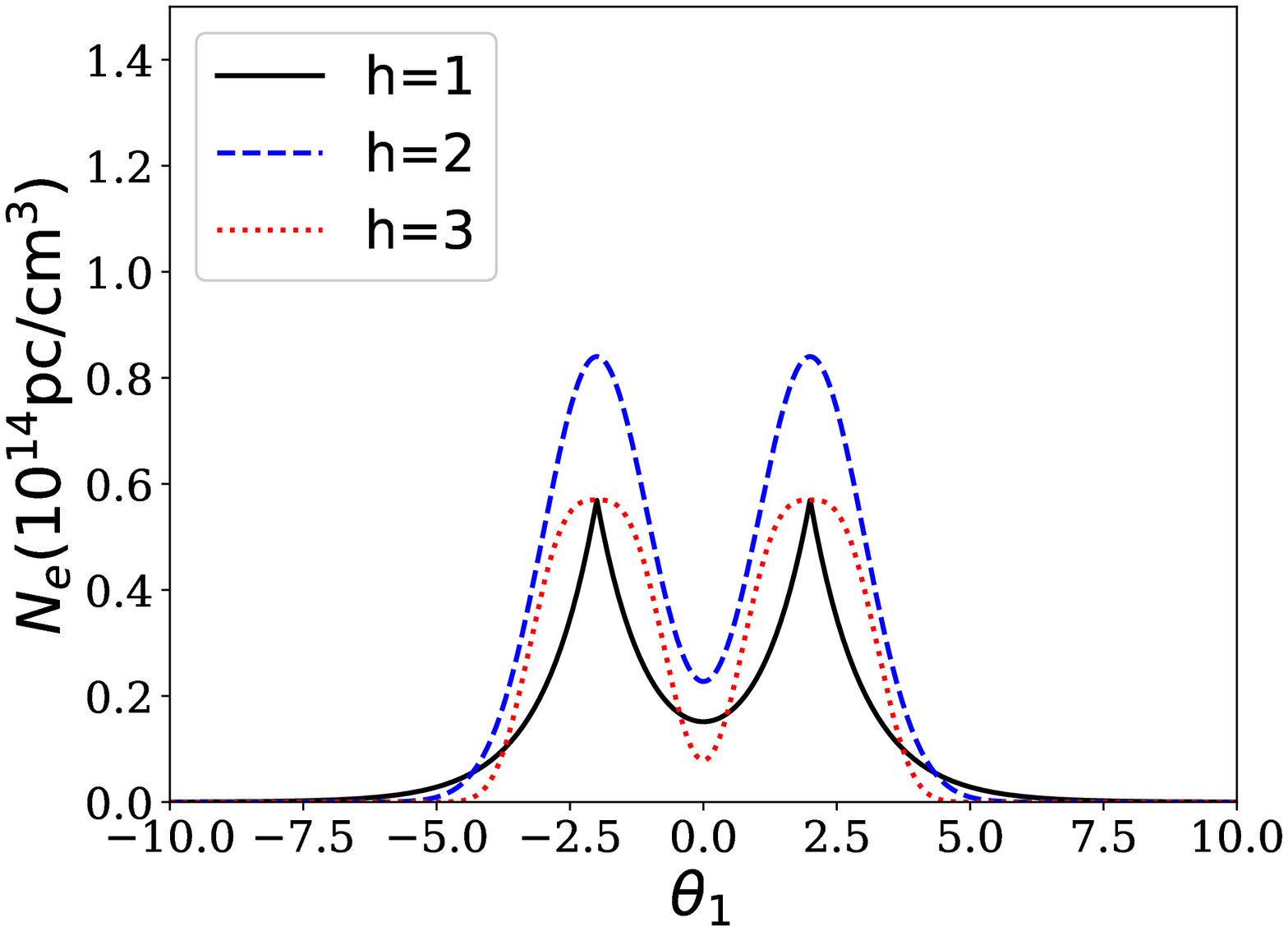}
  \includegraphics[width=8cm]{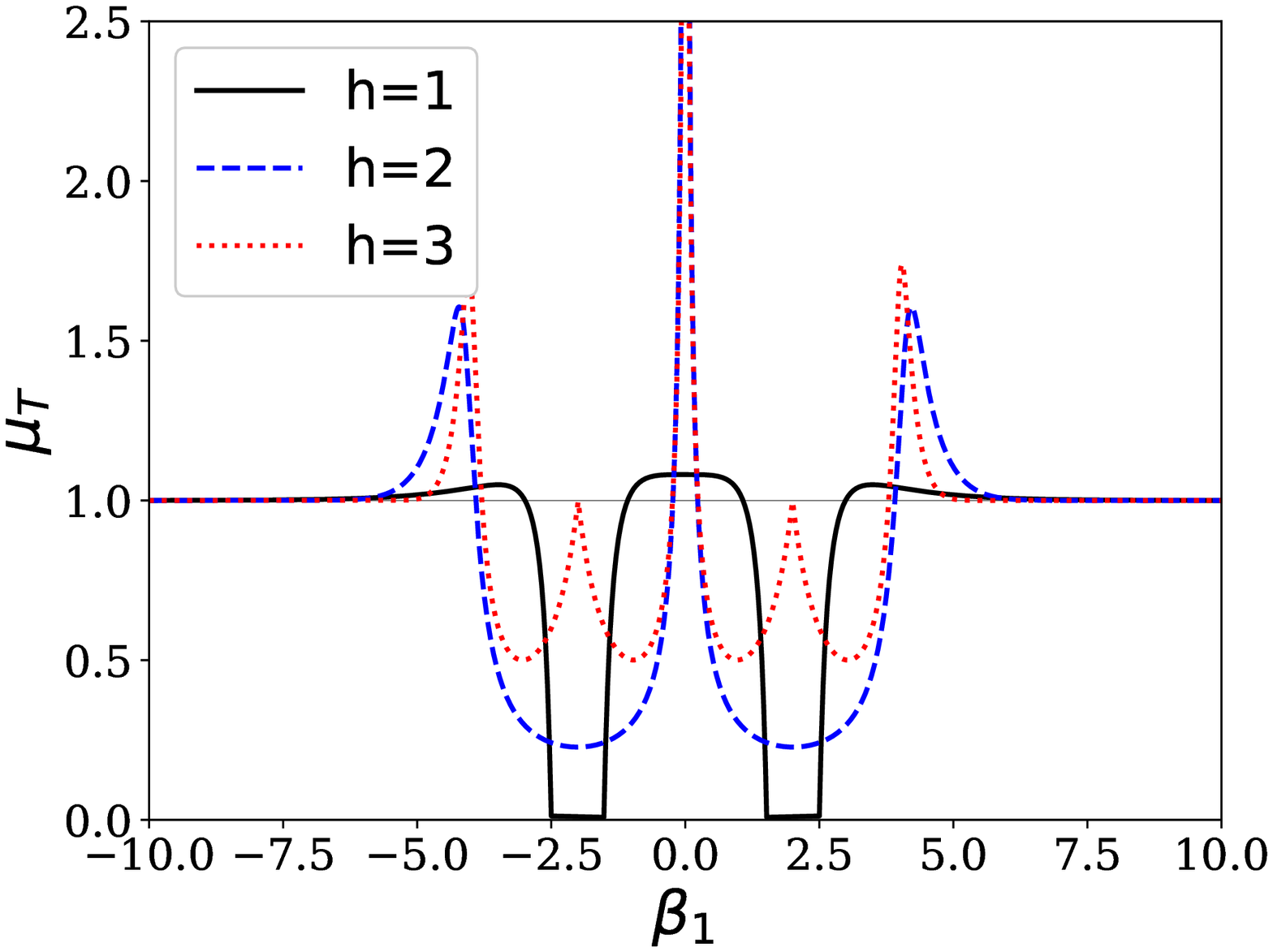}
  \caption{In this figure we compare the behaviour of the
    dual-components lens with different exponent $h=1,2,3$. Top panel:
    the surface electron density. Bottom panel: The resulting
    magnification for the two components. All the lenses are chosen to
    have same width ($\sigma=1$), separation $s=4$ and sub-critical
    with $\theta_0=0.7 f(h)\sigma$ (See table \ref{tab:exp} for the
    value of $f(h)$). Physical details of the density scale are given in Figure
    \ref{figGaussian}.}
\label{fig:comph123}
\end{figure}

\begin{figure*}
  \centerline{\includegraphics[width=4.7cm]{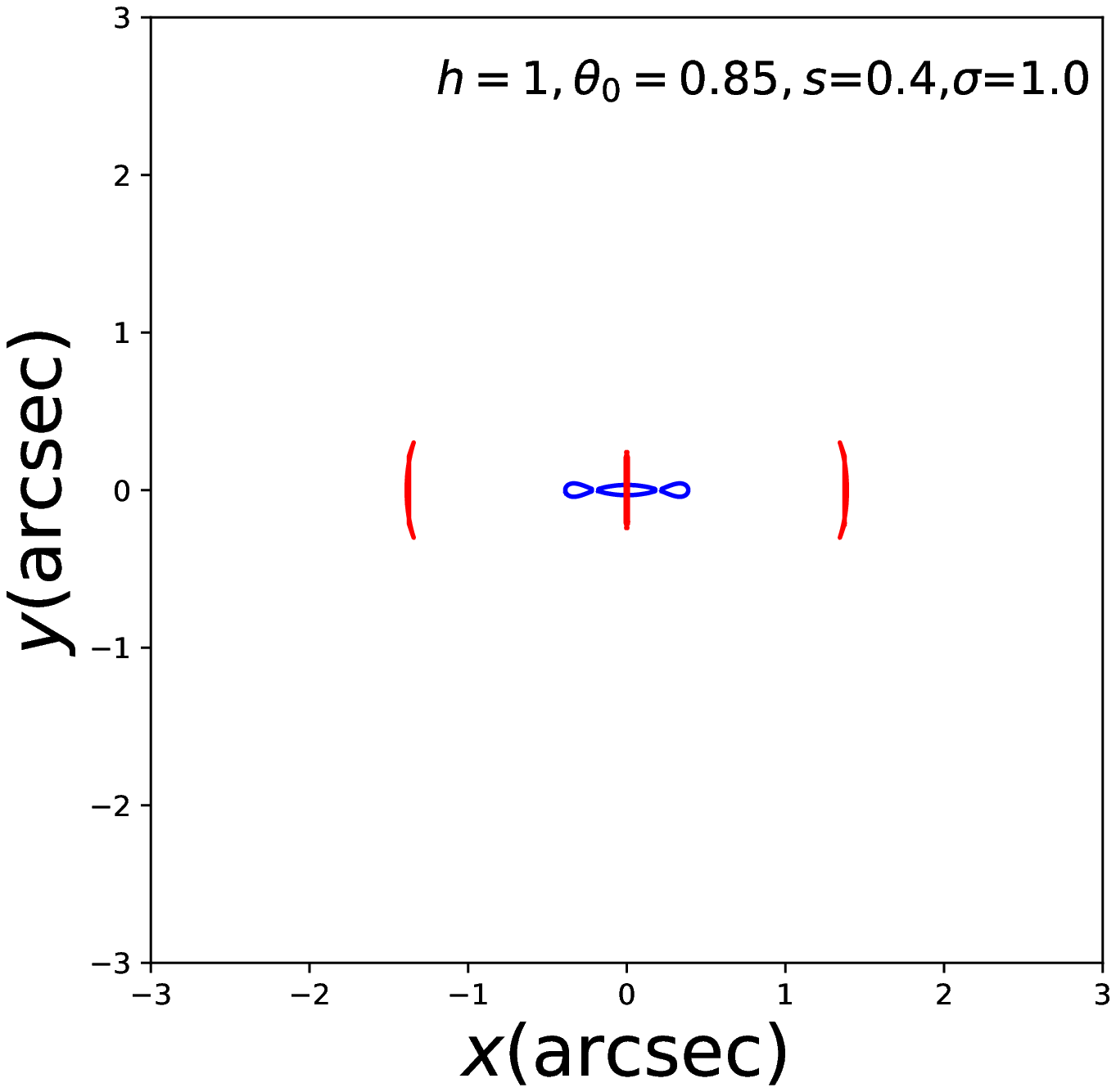}
  \includegraphics[width=4.7cm]{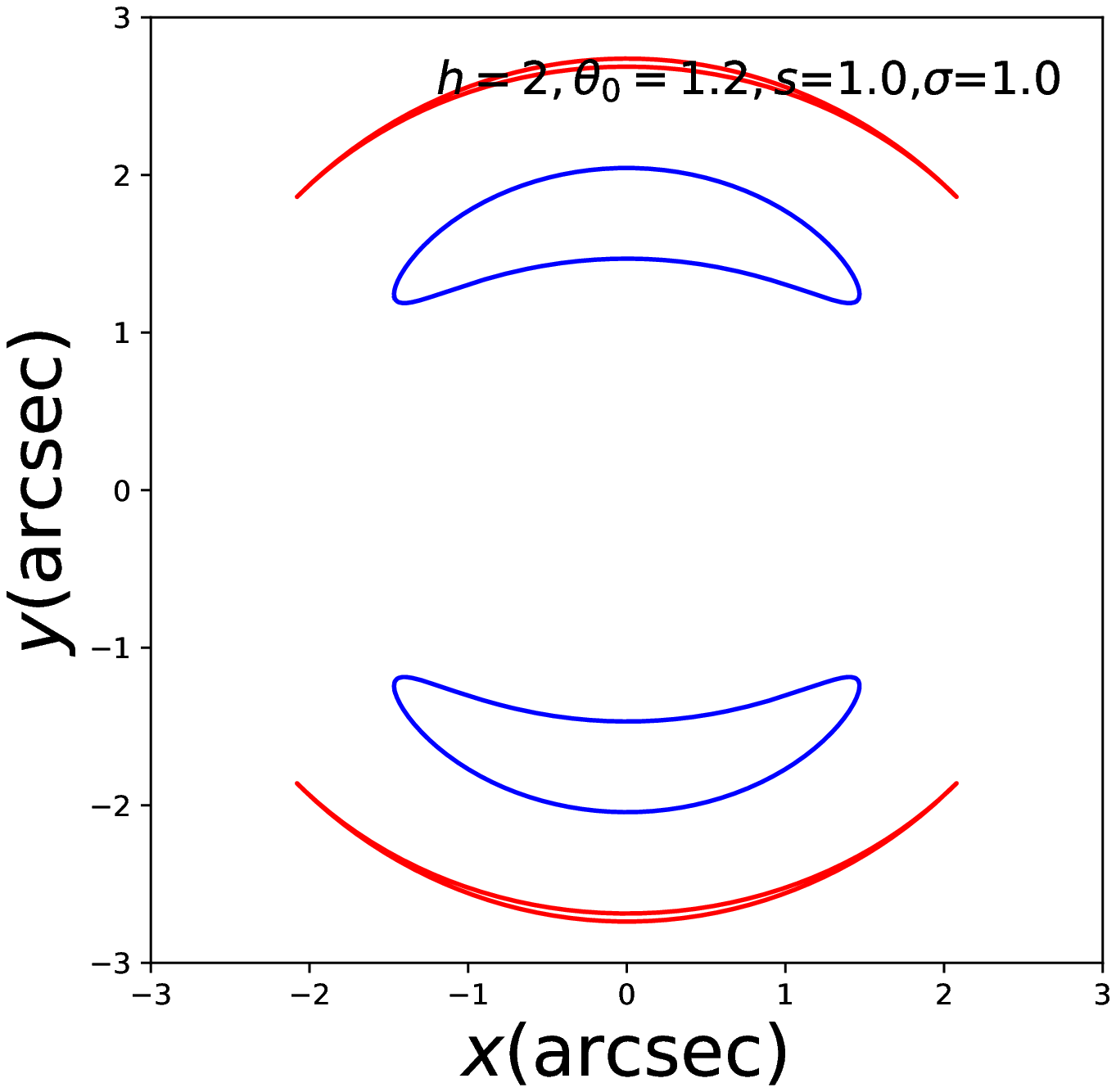}
  \includegraphics[width=4.7cm]{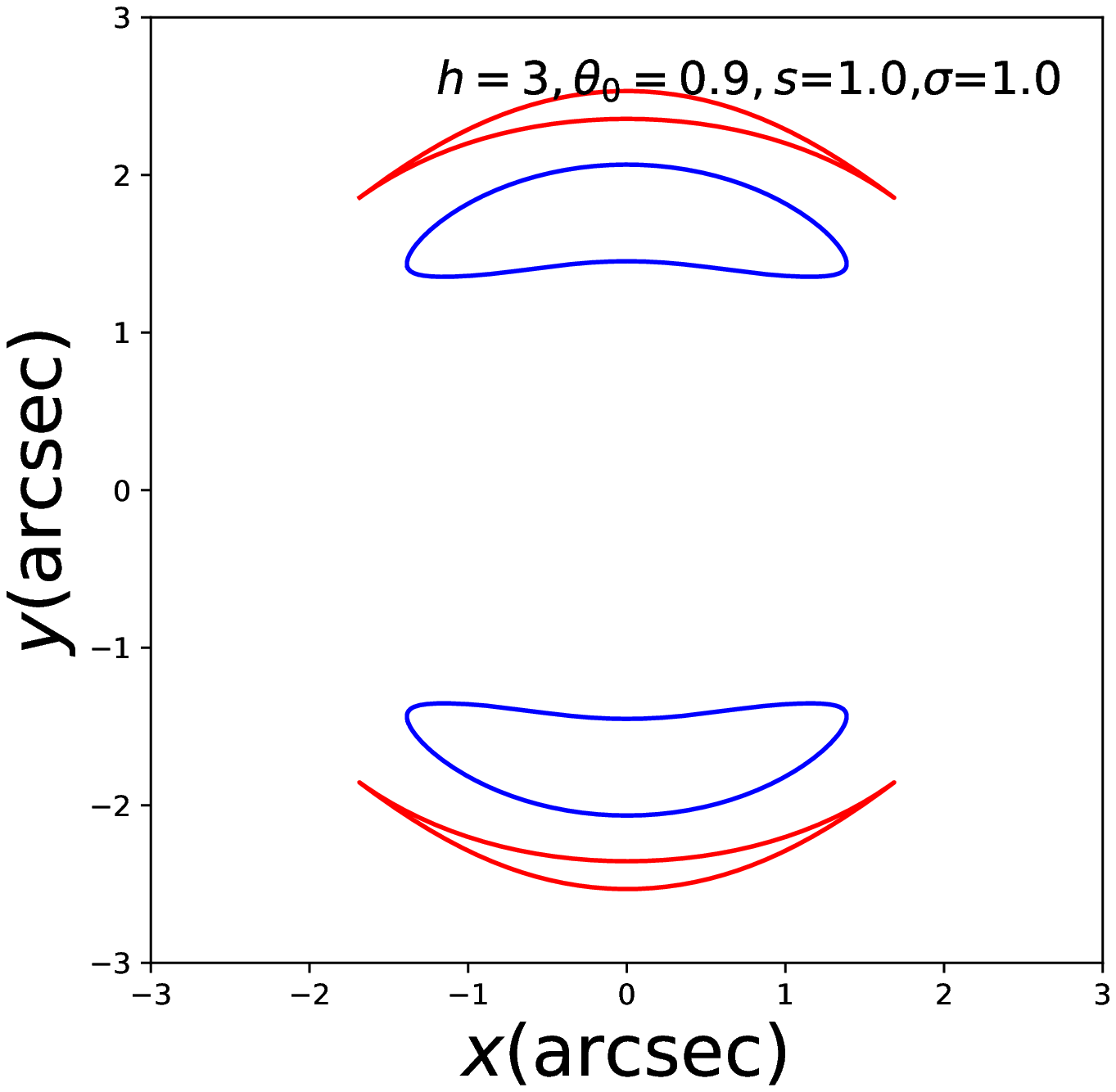}}
  \centerline{\includegraphics[width=4.7cm]{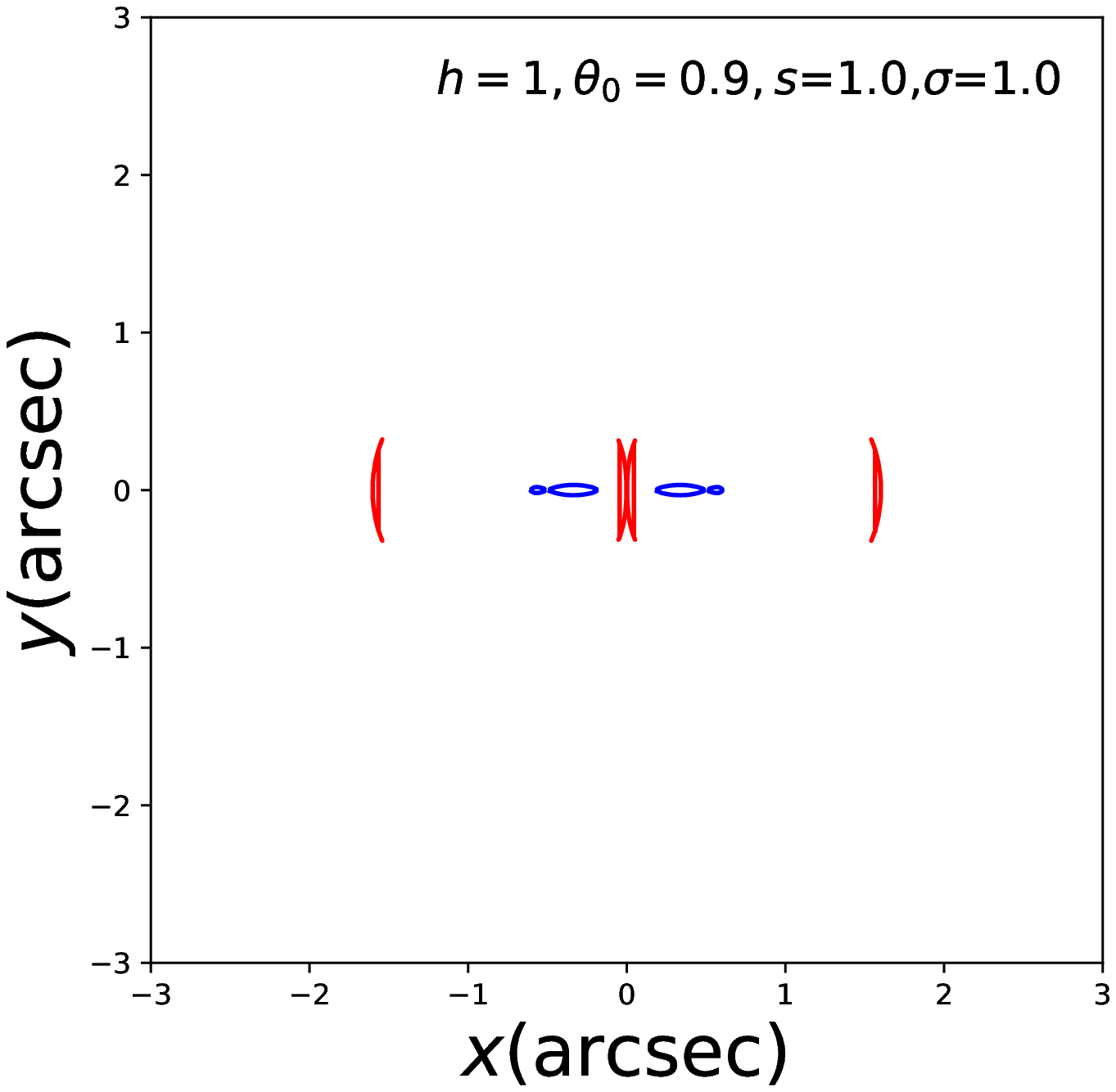}
  \includegraphics[width=4.7cm]{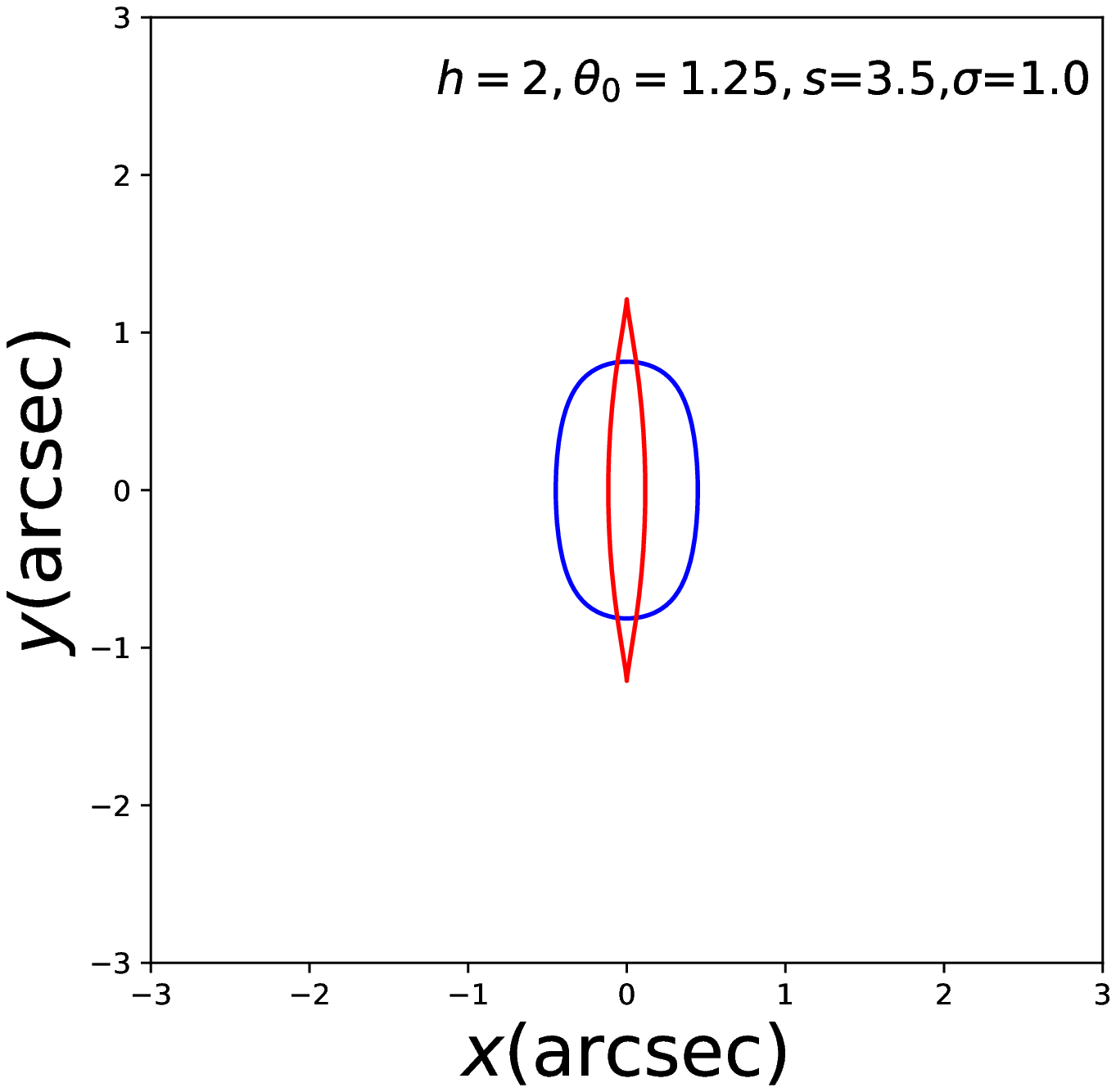}
  \includegraphics[width=4.7cm]{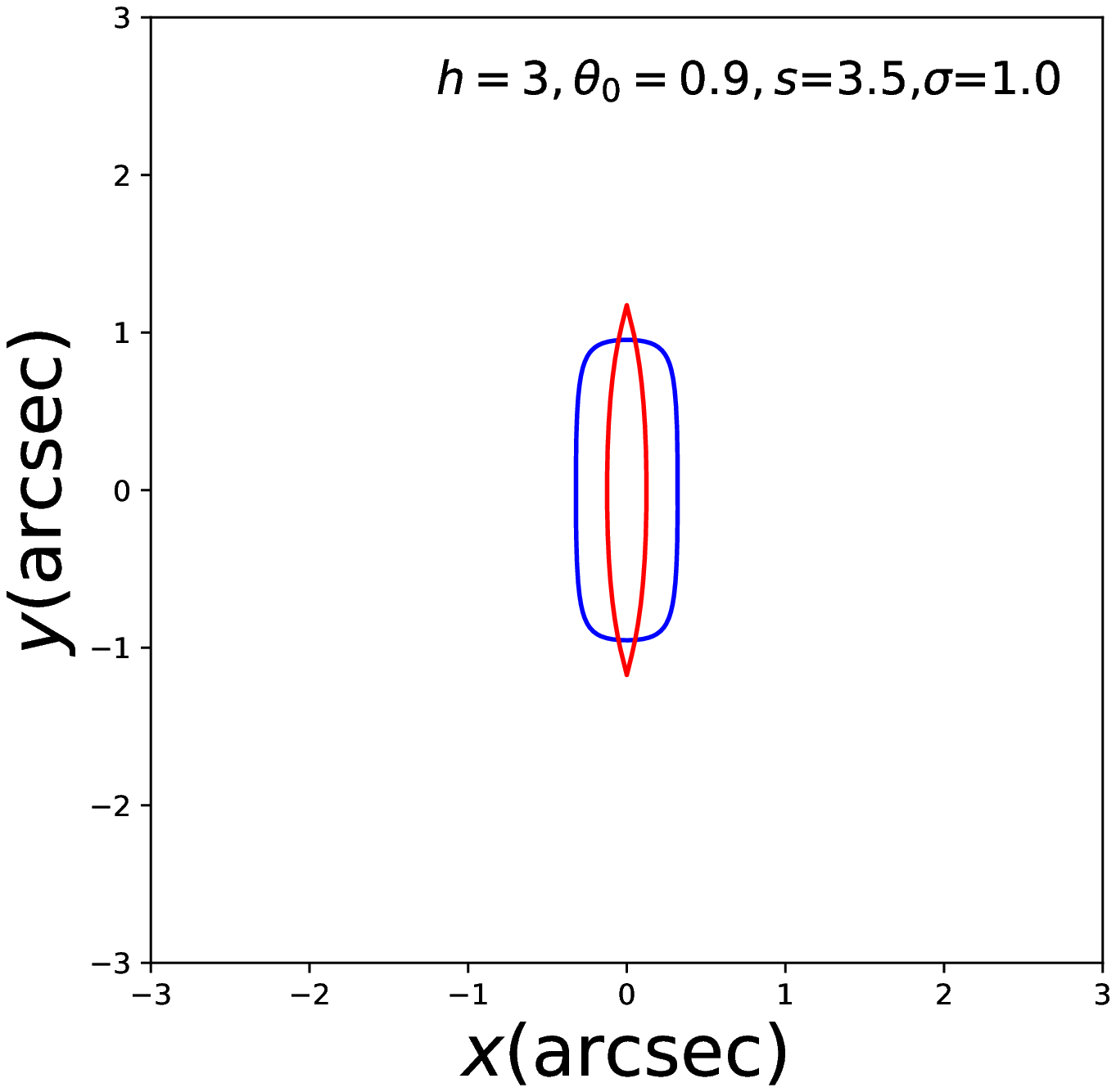}}
  \caption{A selection of sub-critical exponential lenses. The
    left-hand column has $h=1$, middle column has $h=2$ and right-hand
    column has $h=3$. The top row has lens components at low
    separation, and the bottom row has lens components at large
    separation. The $h=2$ and $h=3$ cases are similar.}
  \label{fig:subCriticals}
\end{figure*}

Along with more complicated light curves, the dual lenses also produce
more complicated sets of critical curves in the lens plane. In
Fig.\,\ref{fig:subCriticals}, we show examples of the critical curves
(blue) and caustics (red) of the dual exponential lens with $h=1$
(left-hand column), $h=2$ (middle column) and $h=3$ (right-hand
column). The characteristic radii for a single lens is given in Table
\ref{tab:exp}. While the distinction between sub and super-critical
lenses is clear for single lenses, no well-defined division exists for
the case of dual component lenses. In contrast to the individual case,
when two lens components are placed sufficiently near one another,
sub-critical lens components can also generate critical
curves. Examples are shown in Fig.\,\ref{fig:subCriticals}, in which
individually sub-critical lenses produce some surprising critical
morphologies. The $h=1$ cases with $\theta_0=0.85$ (top left) and
$\theta_0=0.90$ (bottom) produce three and two critical curves
respectively. The critical curves appear along the axis of symmetry
and form narrow regions on the source plane where multiple images
occur. As the characteristic angle is increased and the lens
approaches the critical limit for single lenses ($\theta_0=0.90$),
these regions become larger. The bottom left
panel shows the components moving apart from one another. As the lens
components are increased to farther separations they vanish entirely,
reproducing the result for two independent sub-critical $h=1$
lenses. Holding the separation constant and increasing the
characteristic angle causes the criticals to merge together.

In contrast with the $h=1$ case, for $h=2,3$ (shown in the middle and right-hand columns of Fig.\,\ref{fig:subCriticals} respectively), the sub-critical lens produces two kidney-shaped regions counter-intuitively located on the y-axis. These critical curves begin as nested circular criticals, where the sub-critical components are not resolvable independently. As the separation between the lens components increases the inner critical grows more elliptical, until the inner and outer pair merge forming the kidney-like arcs seen in the figure. The lower panels show the effect of further increasing the separation of the lens. The sub-critical lenses do not produce any criticals when sufficiently separated, as is the case for the $h=1$ sub-critical lens. In this case the criticals merge together and shrink down to the origin between the components of the lens system. The critical curves assume an elliptical morphology as they shrink before vanishing entirely.

\begin{figure*}
\centerline{\includegraphics[scale=1]{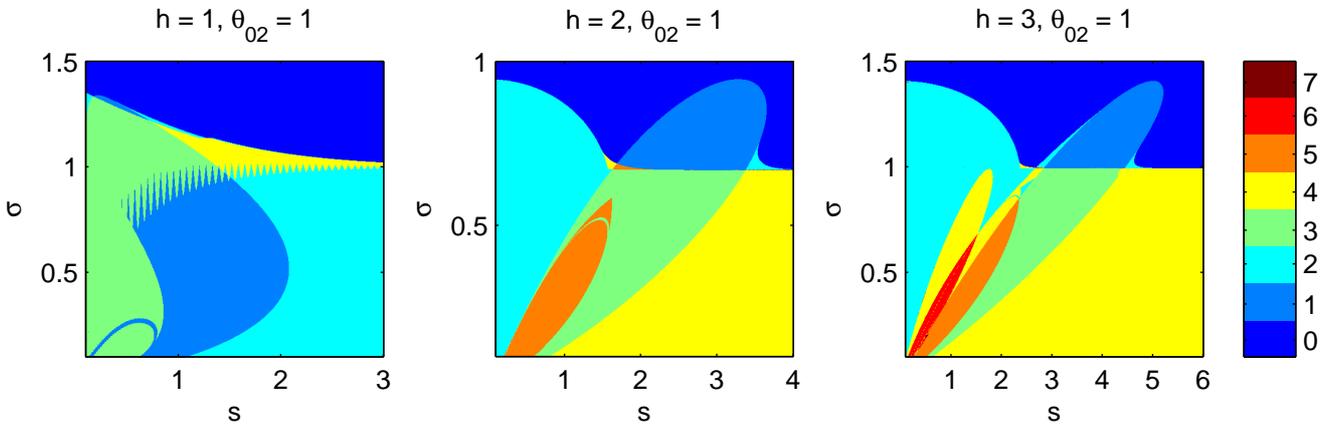}}
\caption{This figure maps the number of critical curves for the dual component exponential lens with equal strength components $\theta_{01}=\theta_{02}$ as a function of angular width $\sigma$ and component separation $s$. The left panel shows the $h=1$ case, middle panel $h=2$ and right panel $h=3$. The maps are color-coded such that the number of critical curves per lens configuration is shown following the colorbar on the right of the figure.}
\label{fig:exponential_map}
\end{figure*}

To characterize the behavior of the dual exponential lens, we have mapped the parameter space of the lens model as a function of $\sigma$ and $s$. Setting the lens components equal to one another $\theta_{01}=\theta_{02}$, we have calculated the number of critical curves for each $s$, $\sigma$ pair by contouring the Jacobian where it vanishes. Quantifying regions in this parameter space where the number of criticals change is difficult, so we have used a numerical approach. For each lens configuration, we calculate the Jacobian on a finite coordinate grid ($\theta_{x}$, $\theta_{y}$) and contour the Jacobian to reveal the curves over which it vanishes. This set of curves are the lens criticals. We then count the number of contours that a given ($s$, $\sigma$) pair provides. We save the number of criticals in an array corresponding to each of the $s$, $\sigma$ pairs.

\begin{figure*}
\centerline{\includegraphics[scale=1]{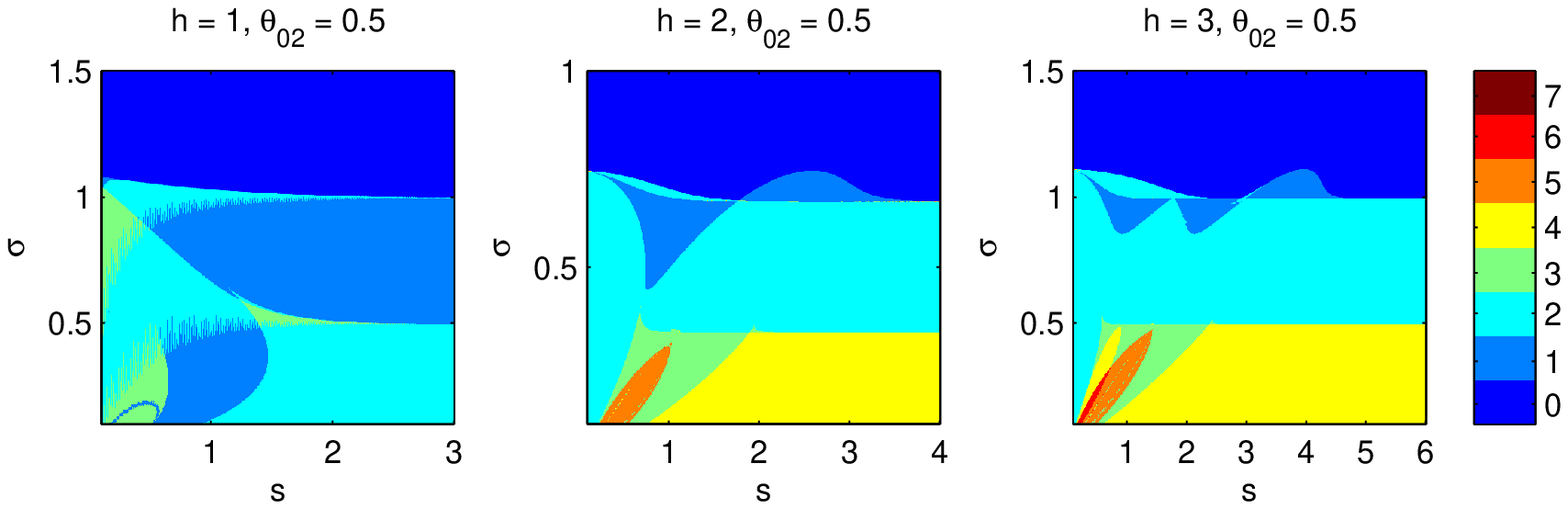}}
\caption{This figure maps the number of critical curves for the dual component exponential lens with unequal strength components $\theta_{01}=1$, $\theta_{02}=0.5$ as a function of angular width $\sigma$ and component separation $s$. The left panel shows the $h=1$ case, middle panel $h=2$ and right panel $h=3$. The number of critical curves is color-coded according to the colorbar on the right of the figure.}
\label{fig:exponential_map_asymm}
\end{figure*}

We show the results of our ($s$, $\sigma$) parameter space map in Fig.\,\ref{fig:exponential_map}. This figure shows the number of critical curves for equal strength dual exponential components $\theta_{01}=\theta_{02}$ as a function of the $\sigma$, $s$ parameter space for the $h=1$ (left panel), $h=2$ (middle panel) and $h=3$ (right panel) exponential lenses. The $h=1$ case produces between $0$, $1$, $2$, $3$ and $4$ criticals. This plot shows a ``fringing'' effect above $\approx 0.5 \sigma$. This fringing is a numerical artifact in our calculation since the critical curves are found by contouring the Jacobian determinant on a finite coordinate grid. As the critical curves shrink, at some point they become smaller than the size of a grid cell and may not be detected depending on the details of their exact position. This occurs when the number of critical curves change, and small critical curves merge, appear or vanish. In such cases it may be difficult to resolve the critical curves reliably, since the calculation becomes sensitive to the positions of critical curves that enclose a small area with respect to the coordinate grid. This produces the fringing effect since the calculation is sensitive to the exact position of the grid cells. The accuracy of the calculation scales with the resolution of the coordinate grid at the expense of computational time. To avoid these numerical difficulties as much as possible, we have done the calculations on a coordinate grid of $500 \times 500$ pixels for each ($s$, $\sigma$) pair. Portions of the map show this numerical difficulty at resolutions of up to $5000 \times 5000$ pixels. Even this fine resolution is unable to avoid the problem for the $h=1$ exponential. The $h=2$ case can produce configurations with $5$ criticals, while the $h=3$ lens can produce $6$ and up to $7$ criticals. In Fig.\,\ref{fig:exponential_map_asymm} shows the case for an asymmetrical arrangement of lenses with $\theta_{01}=1$, $\theta_{02}=0.5$.

\begin{figure*}
\centerline{\includegraphics[scale=1]{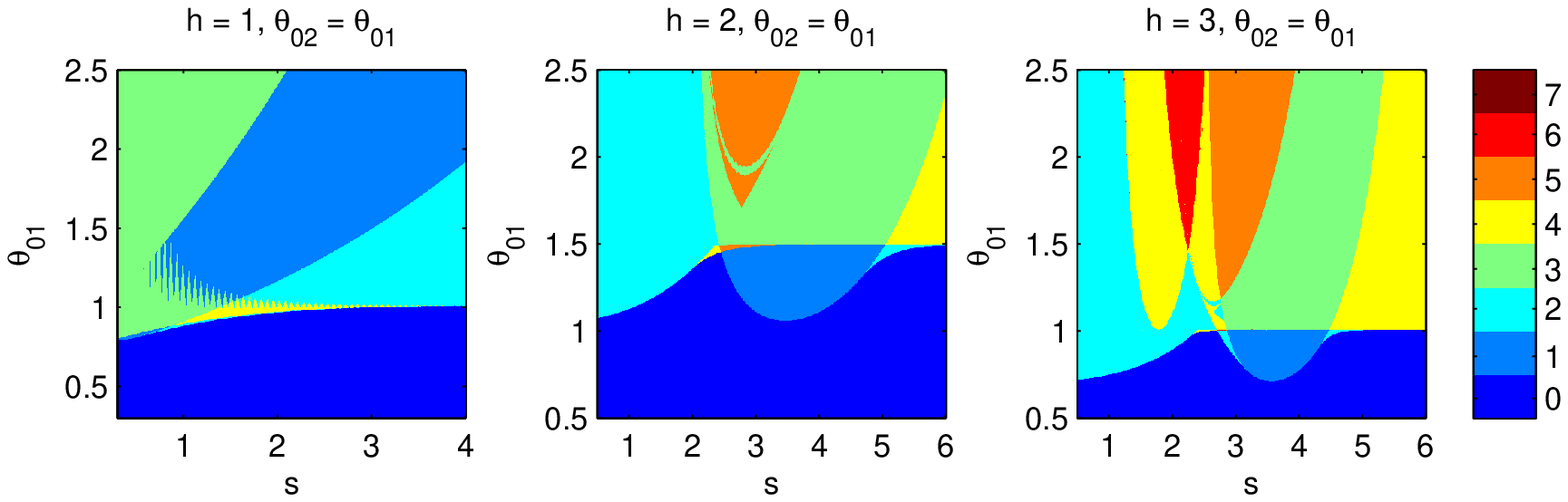}}
\caption{This figure maps the number of critical curves for the dual component exponential lens with equal strength components $\theta_{01}=\theta_{02}$ as a function of characteristic angular scale $\theta_{01}$ and component separation $s$ with $\sigma=1$. The left panel shows the $h=1$ case, middle panel $h=2$ and right panel $h=3$. The number of critical curves is color-coded according to the colorbar on the right of the figure.}
\label{fig:exponential_map_2_1}
\end{figure*}

\begin{figure*}
\centerline{\includegraphics[scale=1]{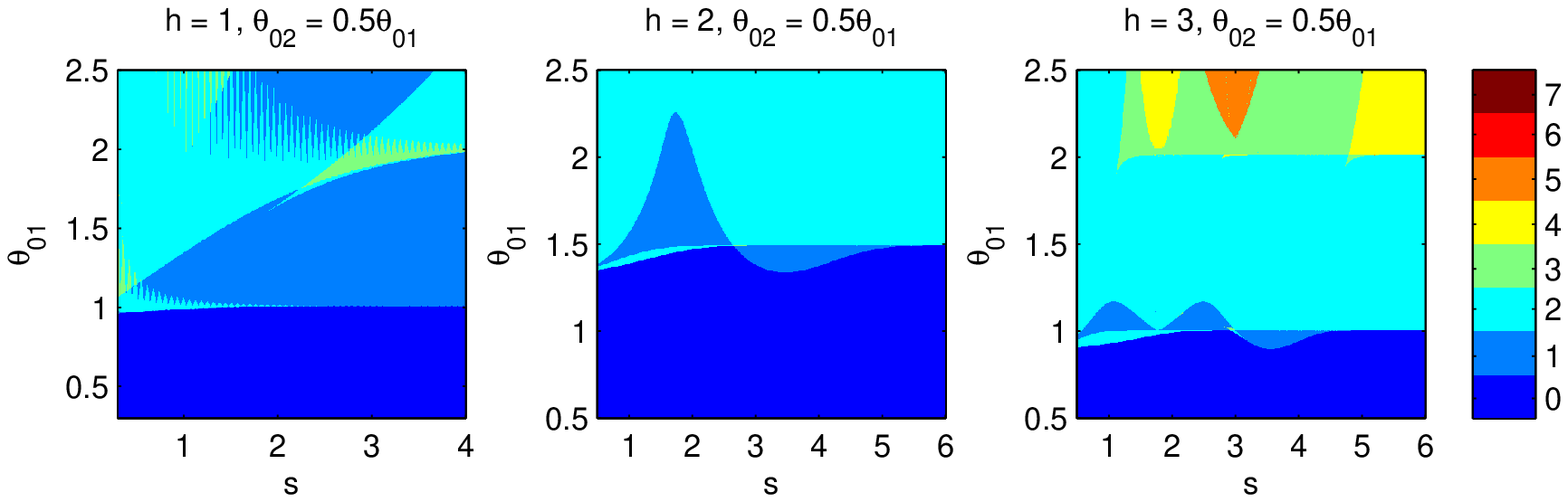}}
\caption{This figure maps the number of critical curves for the dual component exponential lens with unequal lens strength $\theta_{02}=\frac{1}{2}\theta_{01}$ as a function of characteristic angular scale $\theta_{01}$ and component separation $s$ with $\sigma=1$. The left panel shows the $h=1$ case, middle panel $h=2$ and right panel $h=3$. The number of critical curves is color-coded according to the colorbar on the right of the figure.}
\label{fig:exponential_map_2_2}
\end{figure*}

In Fig.\,\ref{fig:exponential_map_2_1} we show an alternative parameterization ($\theta_{01}$, $s$) for the critical curve map, but now assume $\sigma=1$, which is a particularly appropriate choice since $\sigma$ is a physical scale that is independent of the wavelength. We show the number of critical curves for equal strength dual exponential components $\theta_{01}=\theta_{02}$. The parameter space for the $h=1$ (left panel), $h=2$ (middle panel) and $h=3$ (right panel) exponential lenses are color-coded as in Fig.\ref{fig:exponential_map}. Fig.\,\ref{fig:exponential_map_2_2} shows the critical map for an asymmetrical arrangement of lenses with $\theta_{02}=0.5 \theta_{01}$ and $\sigma=1$.

\subsection{The Dual-Component Softened Power-Law Lens}
\label{sec:dualPL}
The dual-component softened power-law lens (SPL) is related to the binary gravitational lens in the case where the finite core $\theta_\text{c}$ vanishes. Let us first develop the SPL lens with a finite core $\theta_\text{c}$, such that we set $\Theta_j \rightarrow \sqrt{ \Theta_j^2 + \theta_\text{c}^2 }$. After we have established the lens formulae, we specialize to the singular case when $\theta_\text{c} \rightarrow 0$. The singular model is substantially simpler than the case when the core size is finite.

In general, the thin lens equation for the power-law lens is
\be
\beta_x = \theta_x + \theta_{01}^{h+1}\frac{\theta_x - \theta_{x1}}{\Theta_1^{h+1}} + \theta_{02}^{h+1}\frac{\theta_x - \theta_{x2}}{\Theta_2^{h+1}}
\ee
and similarly for the y-component with $\theta_x-\theta_{xj}$ replaced by $\theta_{yj}$. This leads to the components of the inverse magnification
\be
\frac{\partial \beta_x}{\partial \theta_x} = 1+ \sum_{j=1,2} \left( \frac{\theta_{0j}^{h+1}}{\Theta_j^{h+1}} - (h+1) \frac{\theta_{0j}^{h+1}}{\Theta_{j}^{h+3}} \left(\theta_x - \theta_{xj} \right)^2 \right),
\ee
with analogous expressions for the y-component. The off-diagonal elements are
\be
\frac{\partial \beta_x}{\partial \theta_y} = - \sum_{j=1,2} (h+1) \left(\frac{\theta_{0j}^{h+1}}{\Theta_j^{h+3}}\left(\theta_x - \theta_{xj} \right) \theta_y \right).
\ee
The magnification of the lens can be calculated from Eq.\,\ref{eq:detamag} using the Jacobian determinant.

\begin{figure*}
\centerline{\includegraphics[scale=1]{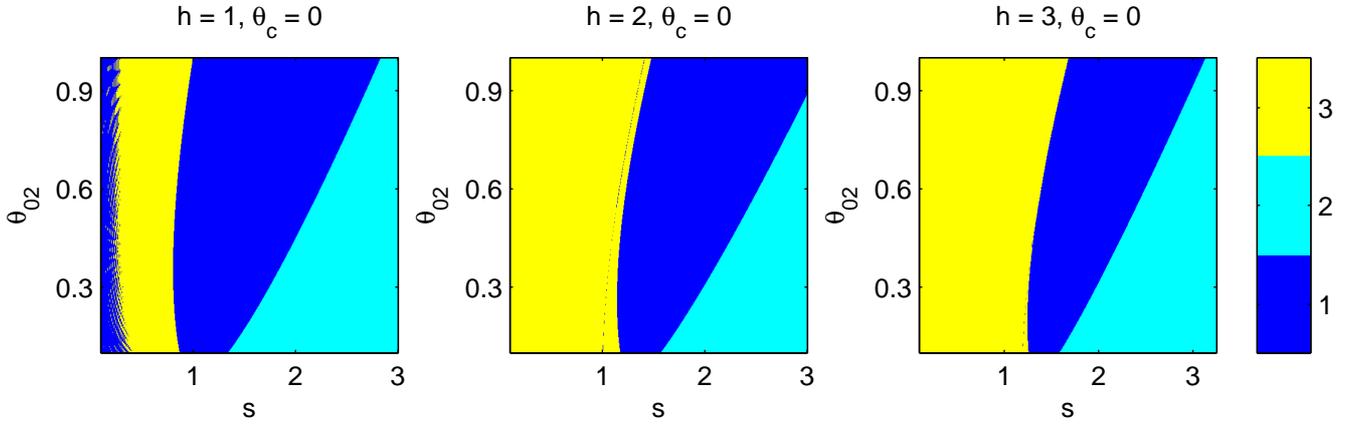}}
\caption{A map of the number of critical curves for the dual component singular $\theta_\text{c}=0$ SPL lens with primary component $\theta_{01}=1$ as a function of the angular scale of the secondary $\theta_{02}$ and component separation $s$. The left panel shows the $h=1$ case, middle panel $h=2$ and right panel $h=3$. The maps are color-coded such that the number of critical curves per lens configuration is shown for $1$ critical (dark blue), $2$ criticals (green) and $3$ criticals (red).}
\label{fig:SPL_map_singular}
\end{figure*}

For the singular case of the SPL lens $\theta_\text{c} \rightarrow 0$, we can fully characterize the imaging properties of the lens. In analogy with the exponential lens, we map the number of critical curves that occur for a given pair of lens parameters $s$ and the characteristic angular scale of the secondary lens, $\theta_{02}$, keeping the primary fixed at $\theta_{01}=1$. Since we are free to scale angular distances, any pair of characteristic angles can be used to transform a lens system to an analogous form in which the primary has $\theta_{01}=1$. The parameter space map for the singular SPL dual component lens is shown in Fig. \ref{fig:SPL_map_singular} for $h=1$ (left panel), $h=2$ (middle panel) and $h=3$ (right panel). As we increase the separation of the lens components, we move from left to right on these diagrams. At small separations ($h=1$), there is only a single critical curve. As the separation increases two small criticals form on the x-axis, within the larger exterior curve, and move apart in the y direction. For equal strength lenses, these small interior curves are symmetric along the y-axis (red areas). Further separating the lenses causes these criticals to merge with the exterior critical and a single dumb-bell or peanut shaped critical curve results (central blue region). As the lens components are further separated, they begin to behave as two independent lenses. The dumb-bell shape splits at the center, resulting in two critical curves that become increasingly circular as the lens components move apart from one another. Regardless of the power-law index, each singular power-law lens behaves this way, although the initial region with a single critical at small separation is not easily seen on the figures for the $h=2$ and $h=3$ cases.

\begin{figure*}
\centerline{\includegraphics[scale=1]{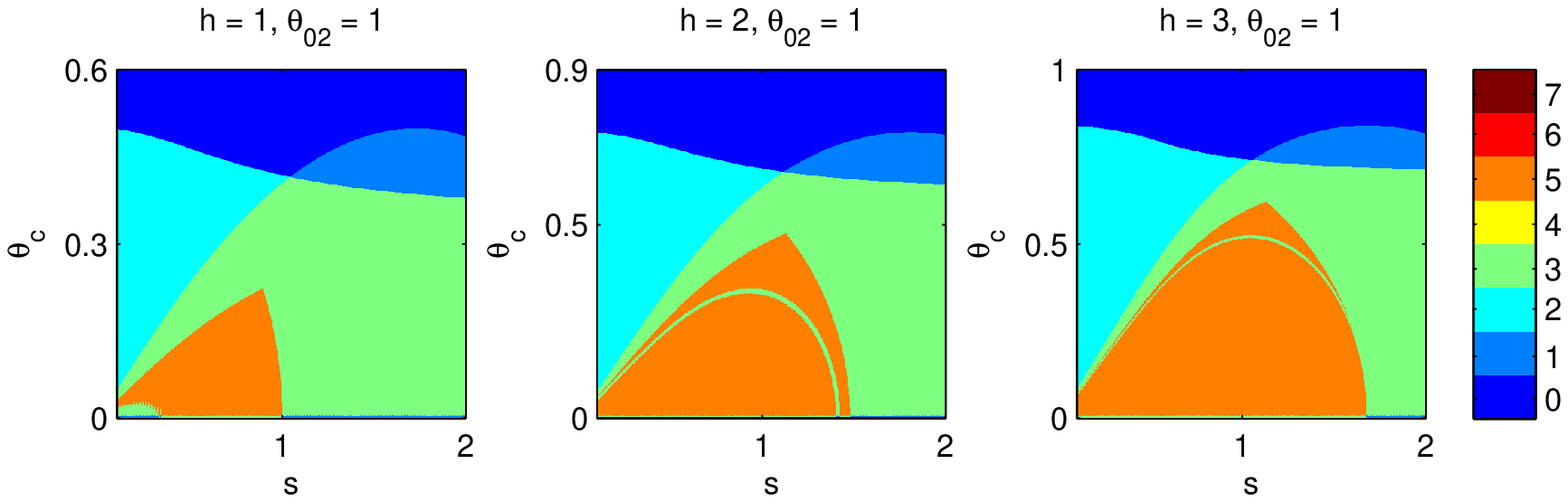}}
\caption{This figure maps the number of critical curves for the dual component softened power-law lens with equal strength components ($\theta_\text{01} = \theta_\text{02}$) as a function of angular core radius $\theta_c$ and component separation $s$. The left panel shows the $h=1$ case, middle panel $h=2$ and right panel $h=3$. The maps follow the coloring scheme in Fig. \ref{fig:exponential_map}. The $4$-critical case is a transitionary state between configurations and is not directly visible on the plots.}
\label{fig:SPL_map}
\end{figure*}

\begin{figure*}
\centerline{\includegraphics[scale=1]{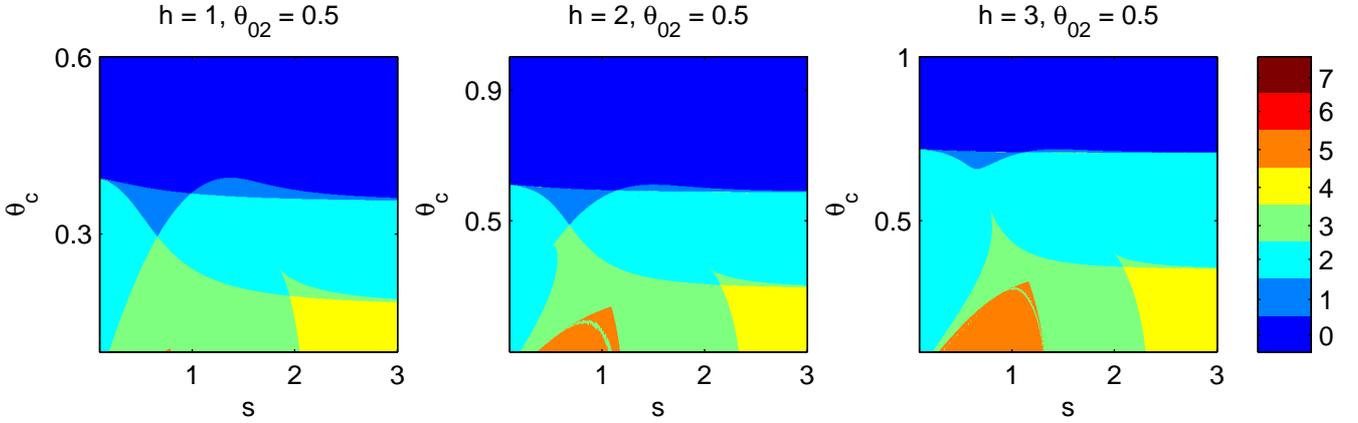}}
\caption{This figure maps the number of critical curves for the dual component softened power-law lens with unequal strength components ($\theta_\text{01} =1$, $\theta_\text{02}=0.5$) as a function of angular core radius $\theta_c$ and component separation $s$. The left panel shows the $h=1$ case, middle panel $h=2$ and right panel $h=3$. The maps follow the coloring scheme in Fig. \ref{fig:exponential_map}.}
\label{fig:SPL_map_asymm}
\end{figure*}

\begin{figure*}
\centerline{\includegraphics[scale=1]{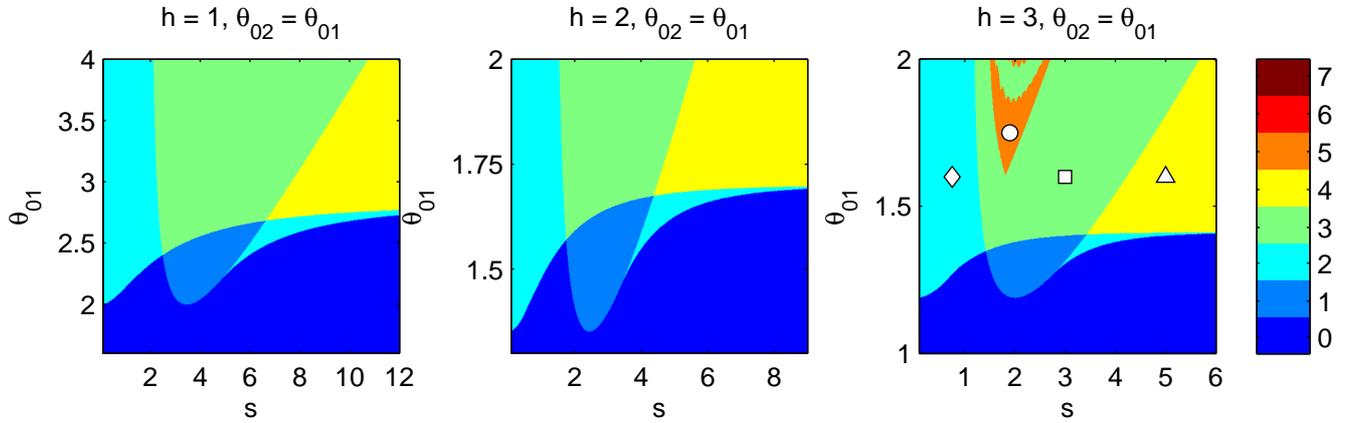}}
\caption{This figure maps the number of critical curves for the dual component softened power-law lens with equal strength components ($\theta_\text{02} = \theta_\text{01}$) as a function of angular scale $\theta_\text{01}$ and component separation $s$. The core radius is fixed at a constant value $\theta_c=1$. The left panel shows the $h=1$ case, middle panel $h=2$ and right panel $h=3$. The points that are marked in the $h=3$ case are shown in figure \ref{fig:traces}. The maps follow the coloring scheme in Fig. \ref{fig:exponential_map}.}
\label{fig:SPL_theta_11_map}
\end{figure*}

\begin{figure*}
\centerline{\includegraphics[scale=1]{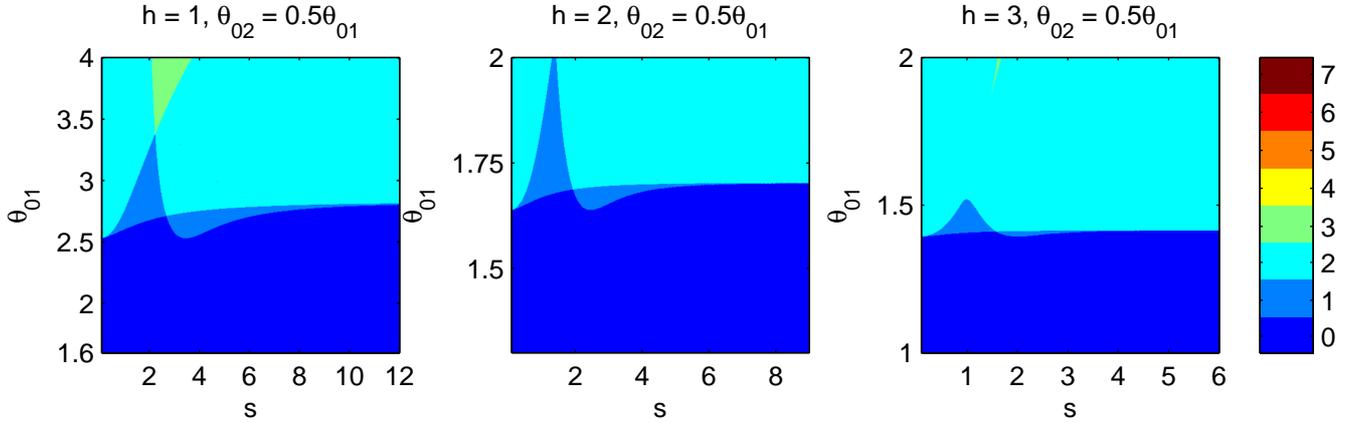}}
\caption{Map of the number of critical curves for the dual component softened power-law lens with asymmetric components ($\theta_\text{02} = \frac{1}{2}\theta_\text{01}$) as a function of angular scale $\theta_\text{01}$ and component separation $s$. The core radius is fixed at a constant value $\theta_c=1$. The left panel shows the $h=1$ case, middle panel $h=2$ and right panel $h=3$. The maps follow the coloring scheme in Fig. \ref{fig:exponential_map}. }
\label{fig:SPL_theta_105_map}
\end{figure*}

\begin{figure*}
\centerline{\includegraphics[bb= 44 307 555 480, clip, scale=0.8]{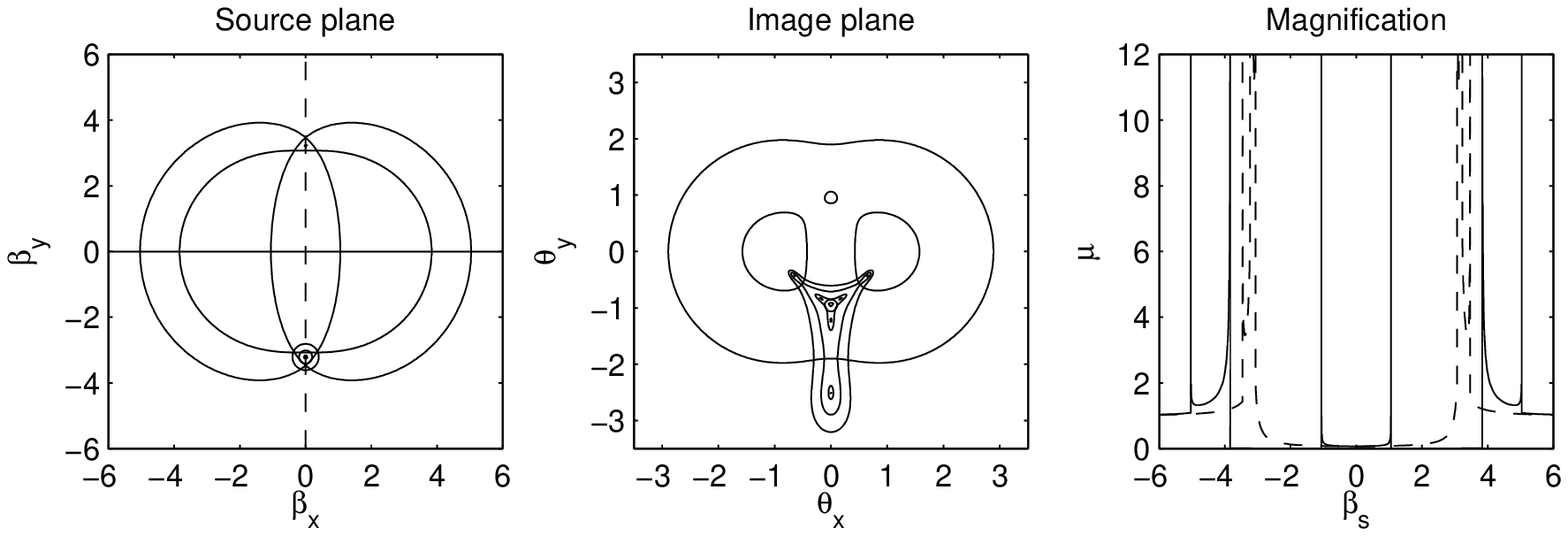}}
\centerline{\includegraphics[bb= 44 307 559 480, clip, scale=0.8]{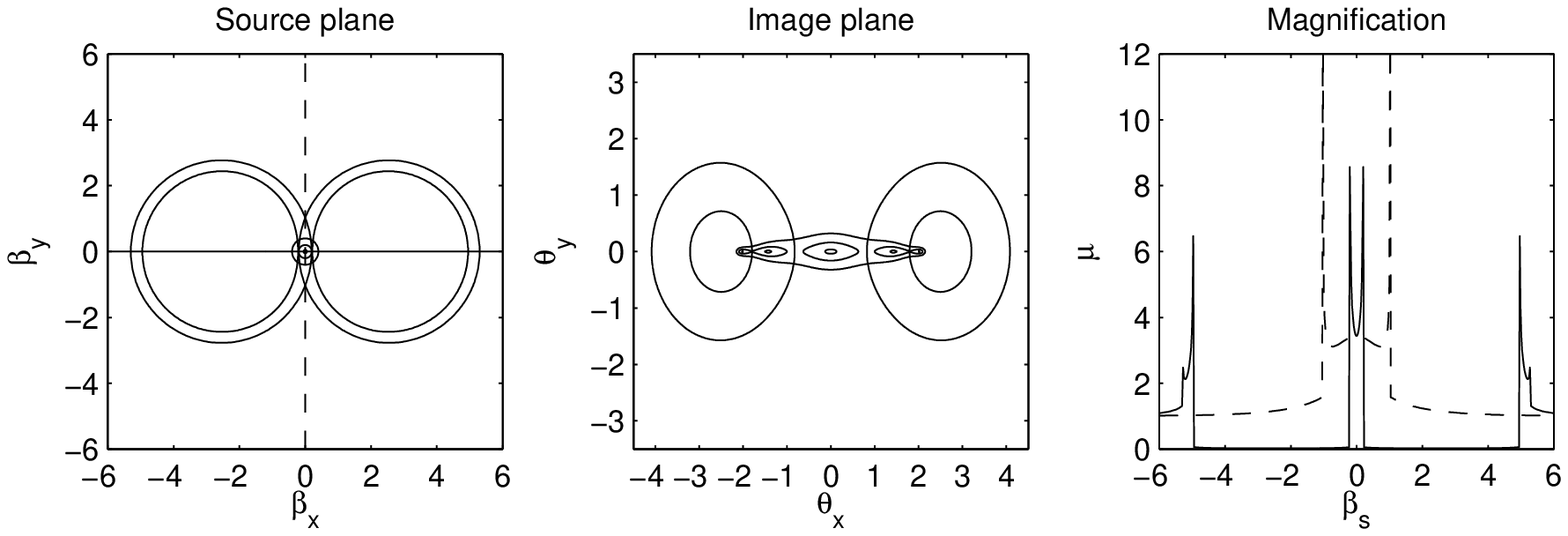}}
\centerline{\includegraphics[bb= 44 307 555 480, clip, scale=0.8]{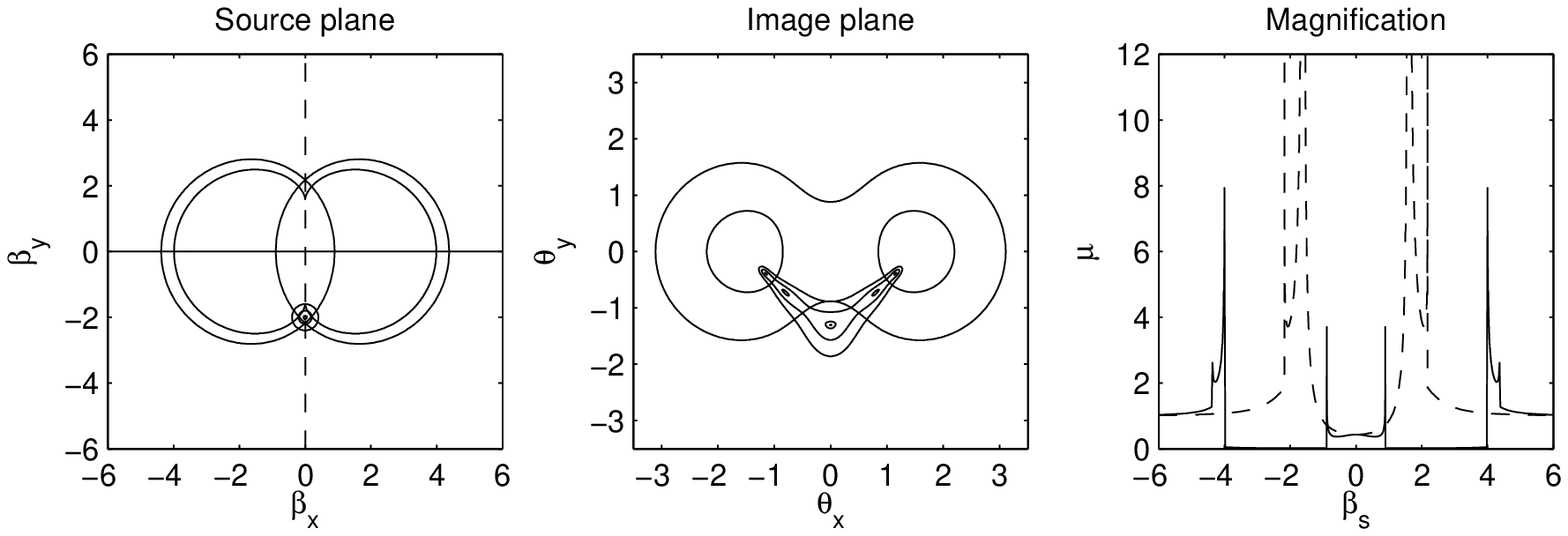}}
\centerline{\includegraphics[bb= 44 307 555 480, clip, scale=0.8]{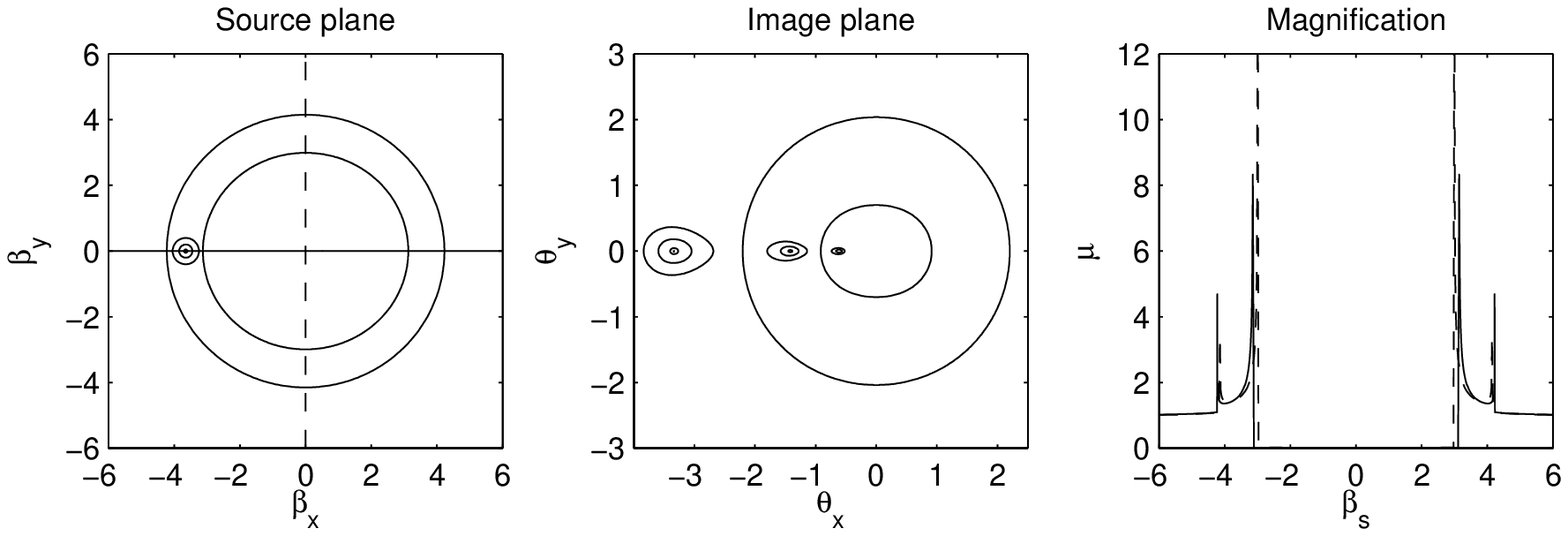}}
\caption{From top to bottom, solutions marked in figure \ref{fig:SPL_theta_11_map} with circle ($\theta_{01}=1.75$, $s=1.90$), triangle ($\theta_{01}=5.00$, $s=1.60$), square ($\theta_{01}=3.00$, $s=1.60$), diamond ($\theta_{01}=0.75$, $s=1.60$). The left-hand column shows the source plane with the corresponding caustics, and the positon of a circular source. Two paths on these figures are shown, one along the $\beta_x$ direction (solid), and the second (dashed) path along $\beta_y$. The corresponding lensed image contours are plotted in the center column with the critical curves. The right-hand column shows the magnification along the paths on the source plane. The dashed path along $\beta_y$ is the dashed magnification curve and the solid magnification curve is the magnification along the $\beta_x$ path. Note that the maximum number of images in these examples is an odd integer less than or equal to $N_\text{crit}+2$.}
\label{fig:traces}
\end{figure*}

We map the number of criticals for a softened power-law lens with a finite core $\theta_\text{c} \geq 0$ in Fig.\,\ref{fig:SPL_map}. We consider the number of criticals as a function of core size $\theta_c$ and lens separation $s$ for the $h=1$ case (left panel), $h=2$ (middle panel) and $h=3$ (right panel). In all panels, the coloring follows the exponential lens critical maps, given by the color bar on the right hand side of the figure. Regions in the parameter space can produce between $0$ and $5$ criticals. The $4$-critical models are a transitionary state between configurations. We do not find any substantial regions of the parameter space that produce four criticals at this resolution ($500 \times 500$ pixels). The major change between the each of the power-law exponents is the region in parameter space in which $5$ critical curves occur. In the $h=2$ and $h=3$ cases, the $5$ critical curve regions are comprised of two disconnected areas. Between these areas $3$ critical curves are produced. In Fig. \ref{fig:SPL_map_asymm}, we plot the asymmetric case for $\theta_{01}=1$, $\theta_{02}=0.5$. As an alternative, in Figures \ref{fig:SPL_theta_11_map} and \ref{fig:SPL_theta_105_map}, we fix the angular core size as constant, $\theta_\text{c}=1$, and map the number of critical curves as a function of the characteristic angular scale of the primary, $\theta_{01}$. In Figure \ref{fig:SPL_theta_11_map} we set the primary and secondary lens scales equal, $\theta_\text{01}=\theta_\text{02}$. In addition we display a variety of critical curve morphologies for the $h=3$ case. We expand on these solutions in Figure \ref{fig:traces}. In Figure \ref{fig:SPL_theta_105_map}, we set $\theta_\text{02}=\frac{1}{2}\theta_\text{01}$. These figures are directly comparable to the exponential cases shown in Figures \ref{fig:exponential_map_2_1} and \ref{fig:exponential_map_2_2}, and show that when the analogous width of both lenses are held constant (ie, $\sigma$ and $\theta_c$), the binary exponential lens produces a richer variety of critical configurations than the binary SPL lens.

We expand on the solutions shown in Figure \ref{fig:SPL_theta_11_map} in the $h=3$ case with $\theta_\text{01}=\theta_\text{02}$ in Figure \ref{fig:traces}. These solutions show examples of lens configurations with $N_\text{crit}=5$, $4$, $3$, $2$ critical curves from the top to bottom row, respectively. The left column plots the caustics on the source plane, as well as two source paths, along the $\beta_x$ (solid line) and $\beta_y$ (dashed line) axes. In addition, we have arbitrarily chosen the position of a circular source to be located at a point where the density of the caustics on the source plane is the greatest. In the middle column we show the image plane with critical curves and the corresponding image of the circular source. On the right-hand column we show the magnification of a point source along each path aligned with the $\beta_x$ (solid curve) and $\beta_y$ (dashed curve) axes. The magnifications due to the binary lens are more complex than for a single SPL lens, as expected. These examples show that the $h=3$ SPL lens with $\theta_{01}=\theta_{02}$ and $\theta_\text{c}=1$ produces an odd number of images $N_\text{images}$. We note that the maximum number of images seems to be bounded by $\max(N_\text{images})=N_\text{crit}+2$. We have tested this apparent relation for a large number of SPL lens configurations and found that it seems to hold for all SPL lenses regardless of $h$ and the relationship between the angular scales of the binary components. However, there is no guarantee that this maximum number of images will be reached for a particular source configuration, nor do we claim that this bound is universal for all dual component SPL lenses. However, these examples demonstrate the usefulness of the critical curve maps presented here. In general, the number of critical curves is a proxy for the maximum number of images a lens can produce. The regions of the parameter space that produce the largest number of critical curves also tend to produce a large number and more complicated configuration of images. Thus, the critical maps provide an overview of the parameter space capable of producing the most complex image configurations for a given set of lens assumptions (here for $\theta_{01}=\theta_{02}$, $\theta_\text{c}=1$ and $h=3$).

\subsection{Complex Formulation for SPL lenses}
Let us consider using complex notation to describe the criticals and caustics of the dual component SPL lens, similar to some more complicated gravitational lens situations that have been studied by \citet{1990A&A...236..311W}. Our intention here is to compare with the analogous treatment of exotic gravitational lens binaries discussed in \citet{bozzaBinary}. In this work, the authors study exotic objects with gravitational influence that decrease as a power-law $1/r^h$. We will show that the dual component SPL lens is closely related to this gravitational lens family.

In the case of the exotic gravitational lens binary, the total mass of the lens is denoted as $M=M_1+M_2$, where $M_1$ and $M_2$ are the mass of each individual binary partner. The resulting Einstein radius, $\theta_\text{E}$, is then used to express the strength of each individual lens. Thus, following \citet{bozzaBinary} we introduce the ratios $\epsilon_j$ such that the Einstein radius (characteristic angular scale) of each lens component is $\theta_\text{Ej}=\epsilon_j^{\frac{1}{1+h}} \theta_\text{E}$, and thus the $\epsilon_j$ appear to linear power in the thin lens equation. Additionally, the mass ratio of the lenses is given by $q=\epsilon_1/\epsilon_2$ and $\epsilon_1+\epsilon_2=1$, such that two equal mass lenses have $q=1$. The difference in this notation in moving from gravitational to plasma lensing is that we must consider the plasma density of each lens component in place of the mass, however the scale ratios between the lens components $\epsilon_j$ are interpreted analogously. To distinguish between the Einstein radius $\theta_\text{E}$ and characteristic scale of lensing in the plasma case we use the notation $\theta_0$ for the total plasma lens, and $\theta_\text{01}$, $\theta_\text{02}$ for each component, such that $\theta_\text{0j}=\epsilon_j^\frac{1}{1+h}\theta_0$.

Let us now introduce complex coordinates on the source plane,
\be
\xi=\frac{\beta_\text{x}+i \beta_\text{y}}{\theta_0}
\ee
and the image plane,
\be
z=\frac{\theta_\text{x}+i \theta_\text{y}}{\theta_0}.
\ee
With the lenses arranged symmetrically about the origin on the $\theta_x$-axis ($\theta_\text{x1}/\theta_\text{0}=-\theta_\text{x2}/\theta_\text{0}$), we once again call the separation between the lens centers $s$, such that $\theta_\text{x1}/\theta_\text{0}=-s/2$ and $\theta_\text{x2}/\theta_\text{0}=s/2$. Using the normalized core size $r_\text{c} = \theta_\text{c} / \theta_\text{0}$, we can express the thin lens equation in complex form as
\be
\xi = z + \sum_{j=1}^2 \epsilon_{j}\frac{(z+(-1)^{j+1}s/2)}{\left([z+(-1)^{j+1}s/2][\bar{z}+(-1)^{j+1}s/2]+r_c^2 \right)^\frac{h+1}{2}}
\ee
The  Jacobian in complex notation is then
\be
J(z,\bar{z}) = \left( \frac{\partial \xi}{\partial z} \right)^2 - \left| \frac{\partial \xi}{\partial \bar{z}} \right|^2.
\ee
Let $Z_1 = \sqrt{(z+s/2)(\bar{z}+s/2)+r_c^2}$ and $Z_2 = \sqrt{(z-s/2)(\bar{z}-s/2)+r_c^2}$. We can then write
\begin{equation}
\begin{array}{ll}
\frac{\partial \xi}{\partial z} = & 1 + \sum_{j=1}^2 \frac{\epsilon_j}{Z_j^{h+1}}  \\
 &  - \frac{h+1}{2} \sum_{j=1}^2  \epsilon_j\frac{(z+(-1)^{j+1}s/2)(\bar{z}+(-1)^{j+1}s/2)}{Z_j^{h+3}} \\
\label{J1}
\end{array}
\end{equation}
and
\be
\frac{\partial \xi}{\partial \bar{z}} = - \sum_{j=1}^2 \epsilon_j\frac{(z+(-1)^{j+1}s/2)^{2}}{Z_j^{h+3}}.
\label{J2}
\ee
This expression matches the corresponding equation in \citet{bozzaBinary} within sign differences. The exotic gravitational lens binary case is recovered by swapping the signs of the terms in eqs. \ref{J1} and \ref{J2}. This extends the results of \citet{bozzaBinary} to diverging plasma lenses.

\section{Discussion}
\label{sec:discussion}

While we have studied several different configurations of the dual component exponential and SPL lenses, there are a number of ways our work can be extended. First, we have assumed that two distributions of plasma along the line of sight can be represented by different characteristic angular sizes $\theta_{01}$ and $\theta_{02}$. However, a more general approach would be to extend this study into multiple lens planes. This would allow for a more general treatment of lenses with multiple components at different distances from the observer.

We have included only one value of the core size that applies to both lens components, but the most general case would have two components with unique core sizes $\theta_\text{c}$ and lens widths $\sigma$. In addition, we have also treated the lens components as having equivalent power-indices $h$. However, this is also an arbitrary choice, and in the most general case each lens can have an independent electron density profile. We have not considered these options, but they will greatly extend the morphology of possible critical curves, caustics, and light curves.

\section{Conclusions}
\label{sec:conclusions}
Motivated by the dual-lobed structure of the ESE recently observed in PSR J1740-3015 \citep{kerr2018}, we have developed models of dual component lenses using the SPL and exponential lens models which we studied in previous work \citep{ErRogers18}. These dual component lens models are interesting in a number of ways. First, we showed that a binary Gaussian ($h=2$ exponential lens) is degenerate with an $h=3$ exponential provided that the parameters are carefully chosen. The only place the light curves produced by these lenses substantially deviate from one another is on the edge of the exclusion region, which causes the most significant dimming and is itself frequency dependent. Thus, these binary lenses can be distinguished from one another by carefully monitoring the light curve over a range of frequencies when the source passes by the exclusion region. Additionally, we have numerically studied the exponential and SPL lens families for a variety of lens configurations. By mapping the parameter space of these lenses, we have explored the number of criticals $N_\text{crit}$ produced by the lenses, and we use these maps to explore particular parameter choices in the appendix. For the cases we have studied we find that the maximum number of visible images a non-singular SPL binary lens produces is an odd number related to the number of critical curves $N_{crit}$ that the lenses produce $\max(N_\text{image})=N_\text{crit}+2$ images. The singular lenses produce two images at the center of each lens component, with vanishing magnification. These images become observable as the lens is softened. Thus, the critical maps point toward regions of the binary lens parameter combinations capable of producing complicated criticals, caustics and image configurations. Just as binary gravitational lenses produce more complex and interesting image configurations and light curves than individual axisymmetric mass distributions, our work shows that dual component plasma lenses also continue this trend.

\appendix

\section{A Gallery of Critical Curves and Caustics}
\label{appA}
In the appendix, we present the critical curves and caustics for the
dual-component plasma lenses that have been discussed in this work. We have tried to display critical and caustics from each region shown in Figs.\,\ref{fig:exponential_map}, \ref{fig:SPL_map}. Some of the caustics extend beyond the scope of the panel. We omit those large caustics in order to present better resolution of the inner region.

\subsection{Dual-Component Exponential Lens}
\begin{figure*}
  \centerline{\includegraphics[width=4.7cm]{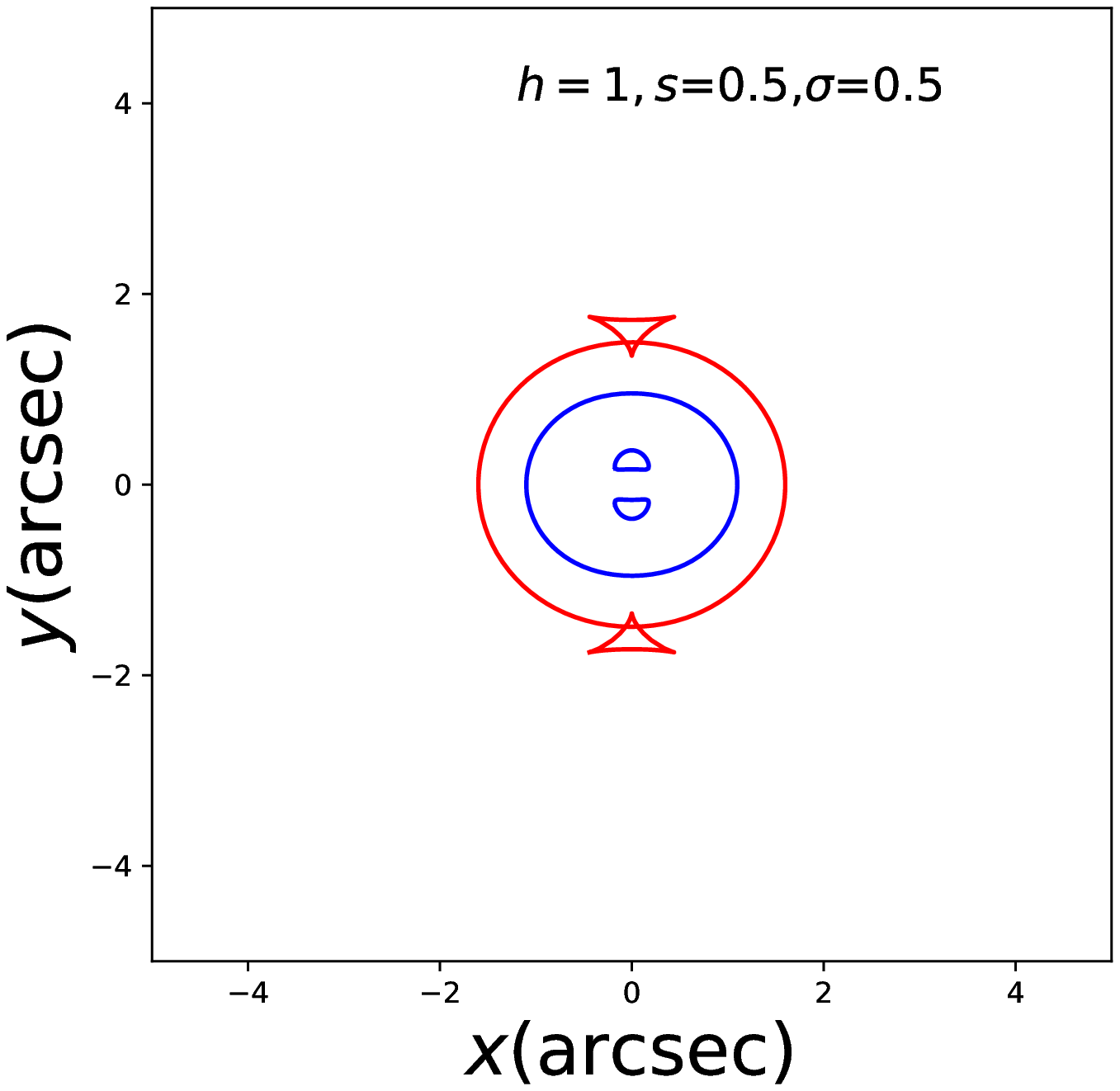}
  \includegraphics[width=4.7cm]{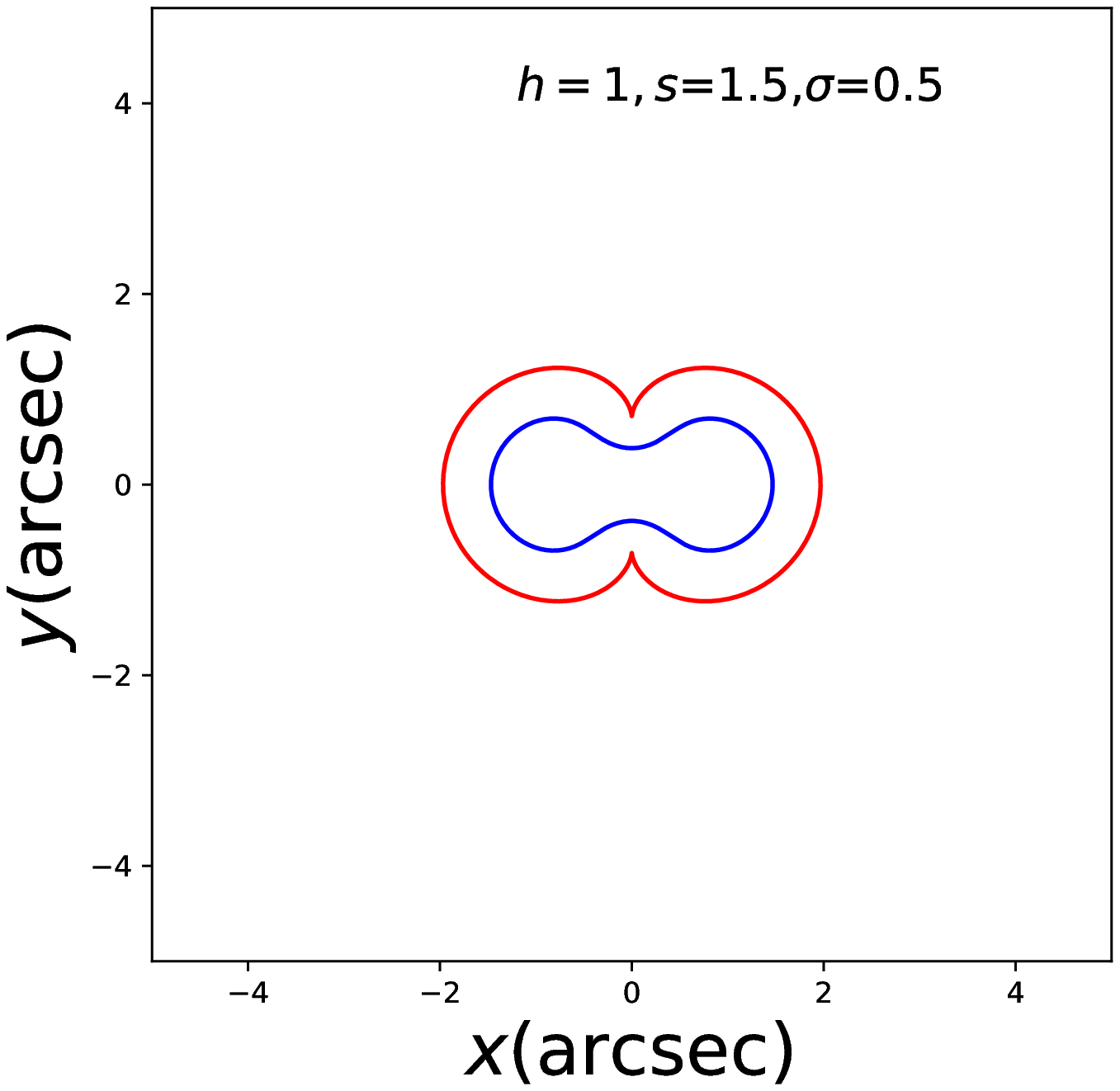}
  \includegraphics[width=4.7cm]{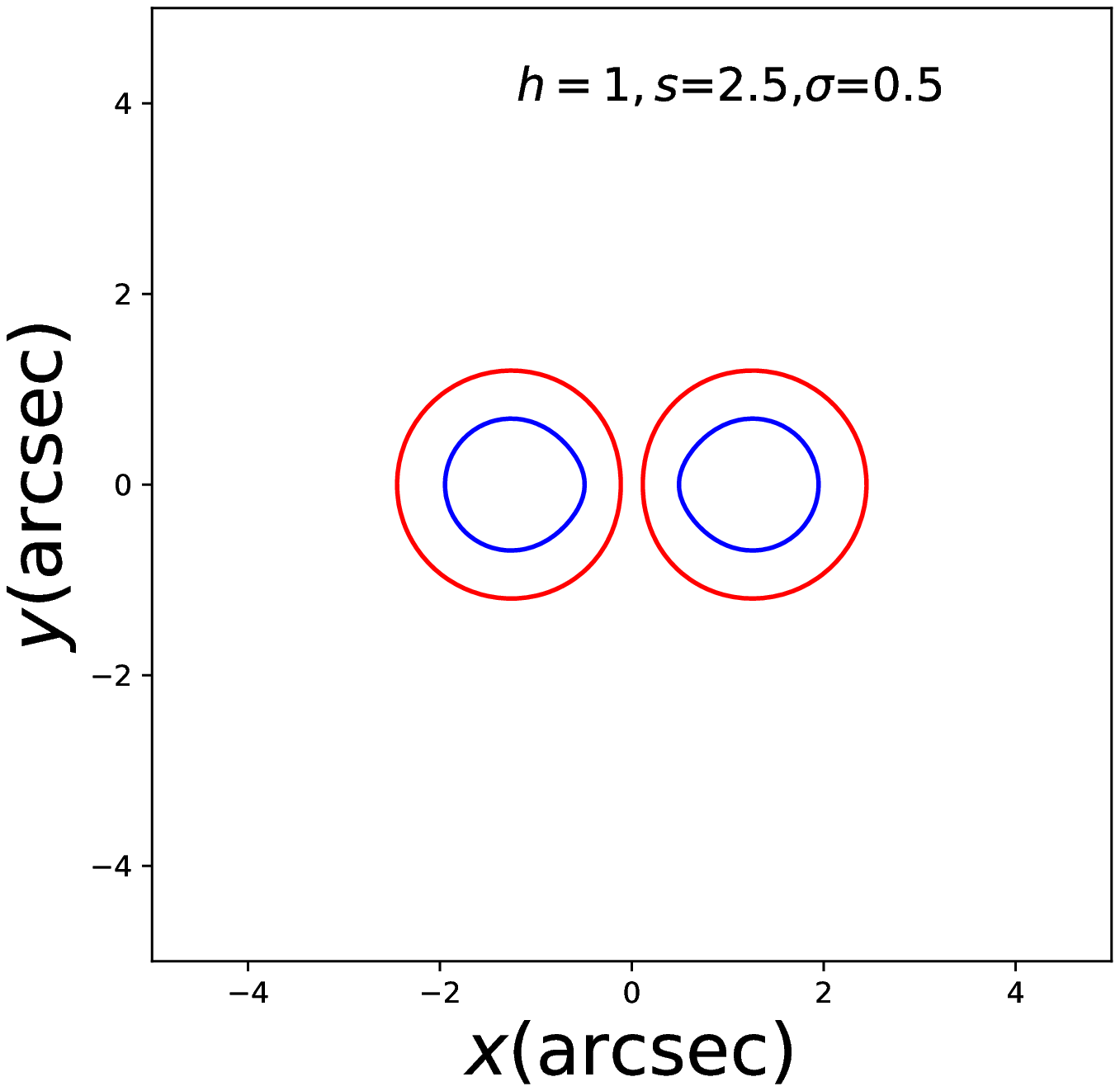}
  \includegraphics[width=4.7cm]{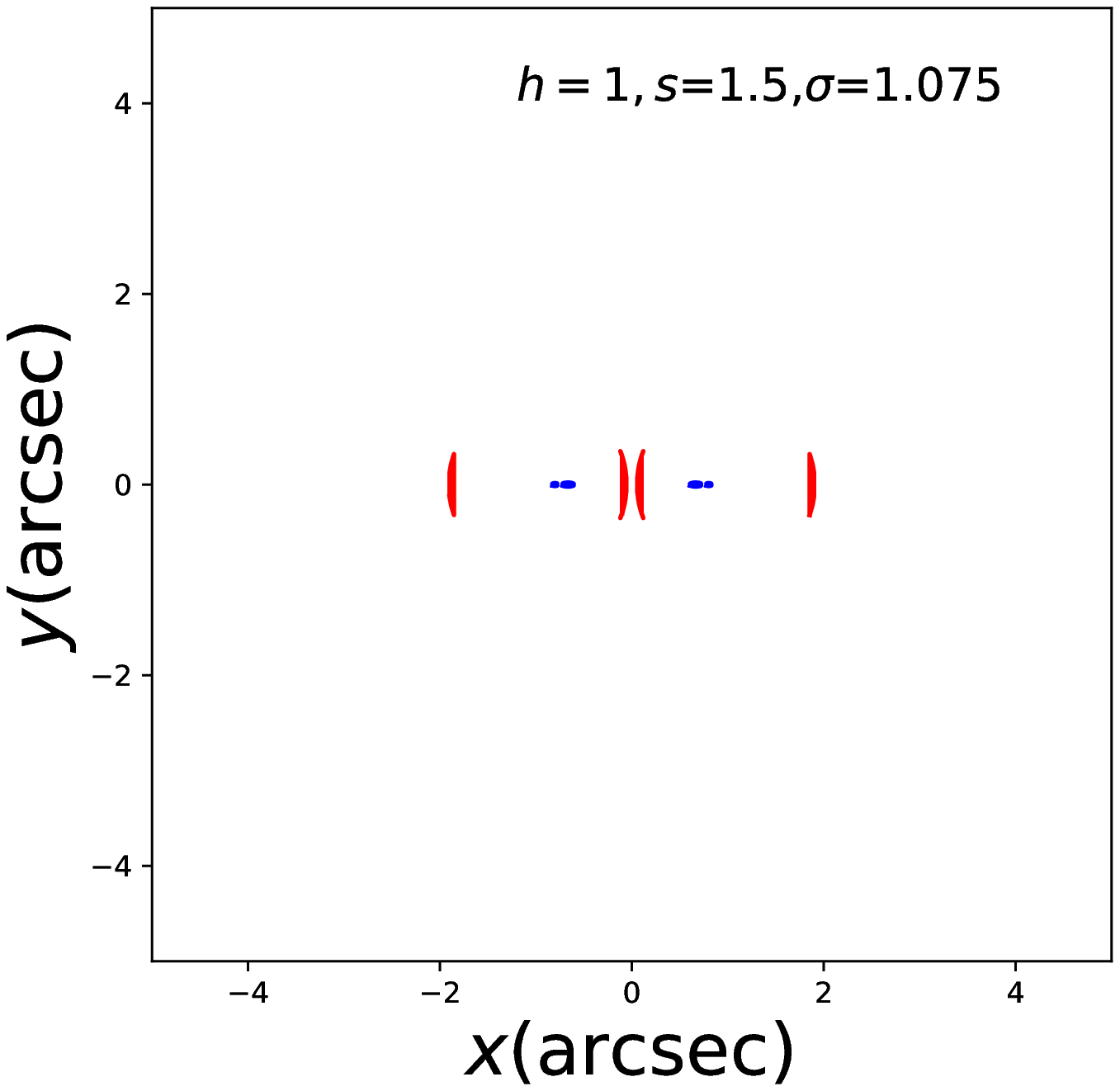}}
  \centerline{\includegraphics[width=4.7cm]{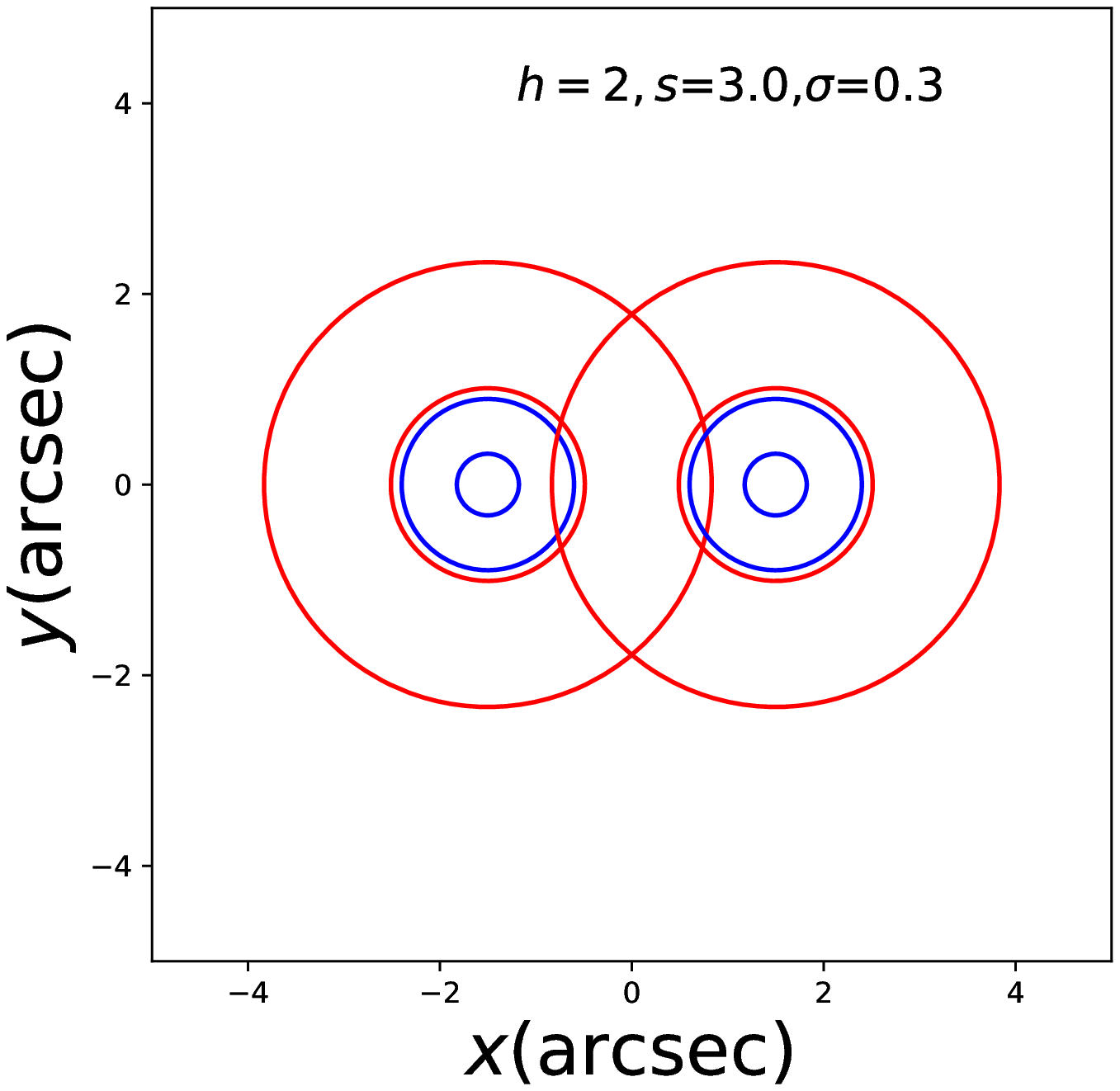}
  \includegraphics[width=4.7cm]{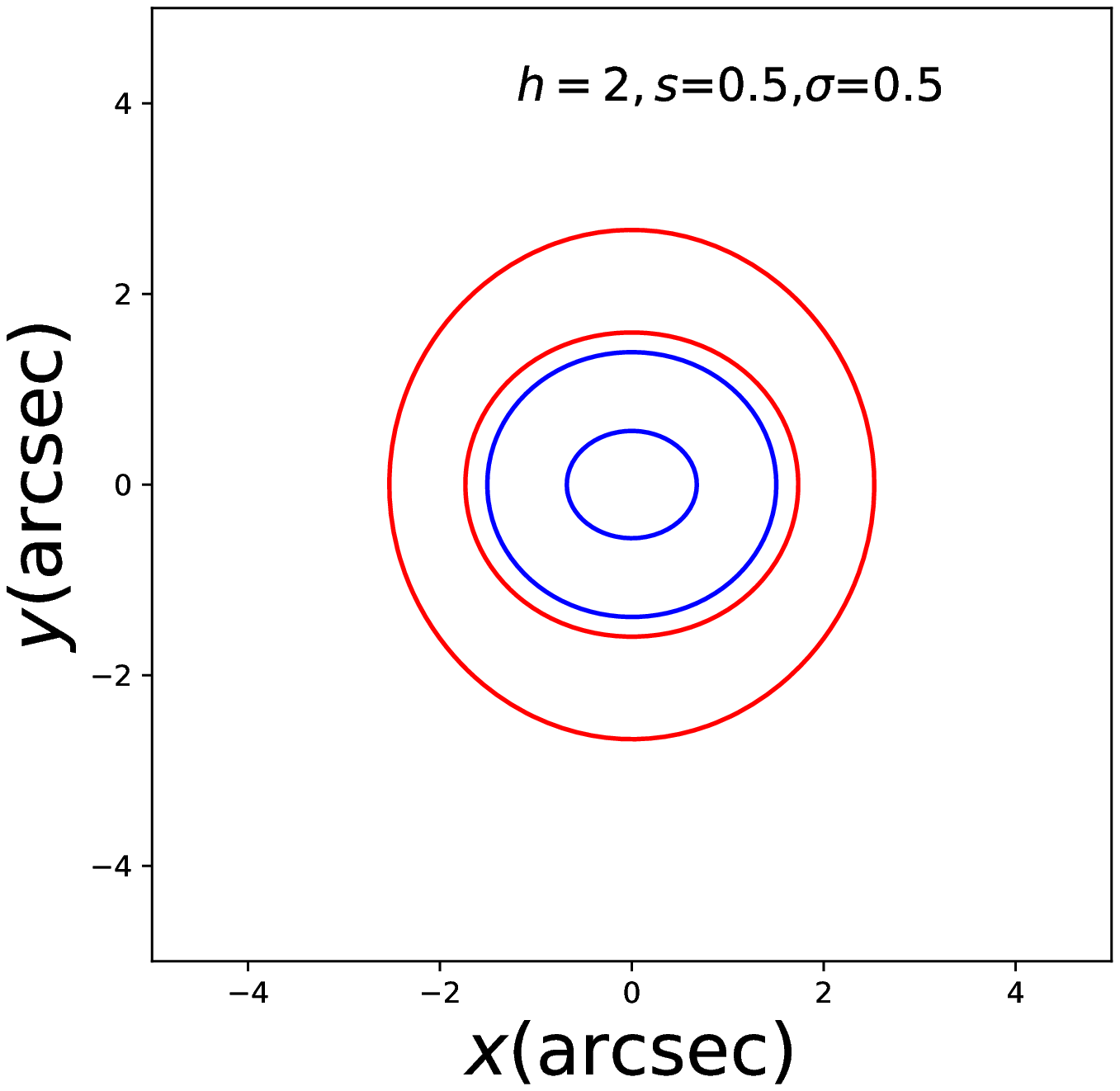}
  \includegraphics[width=4.7cm]{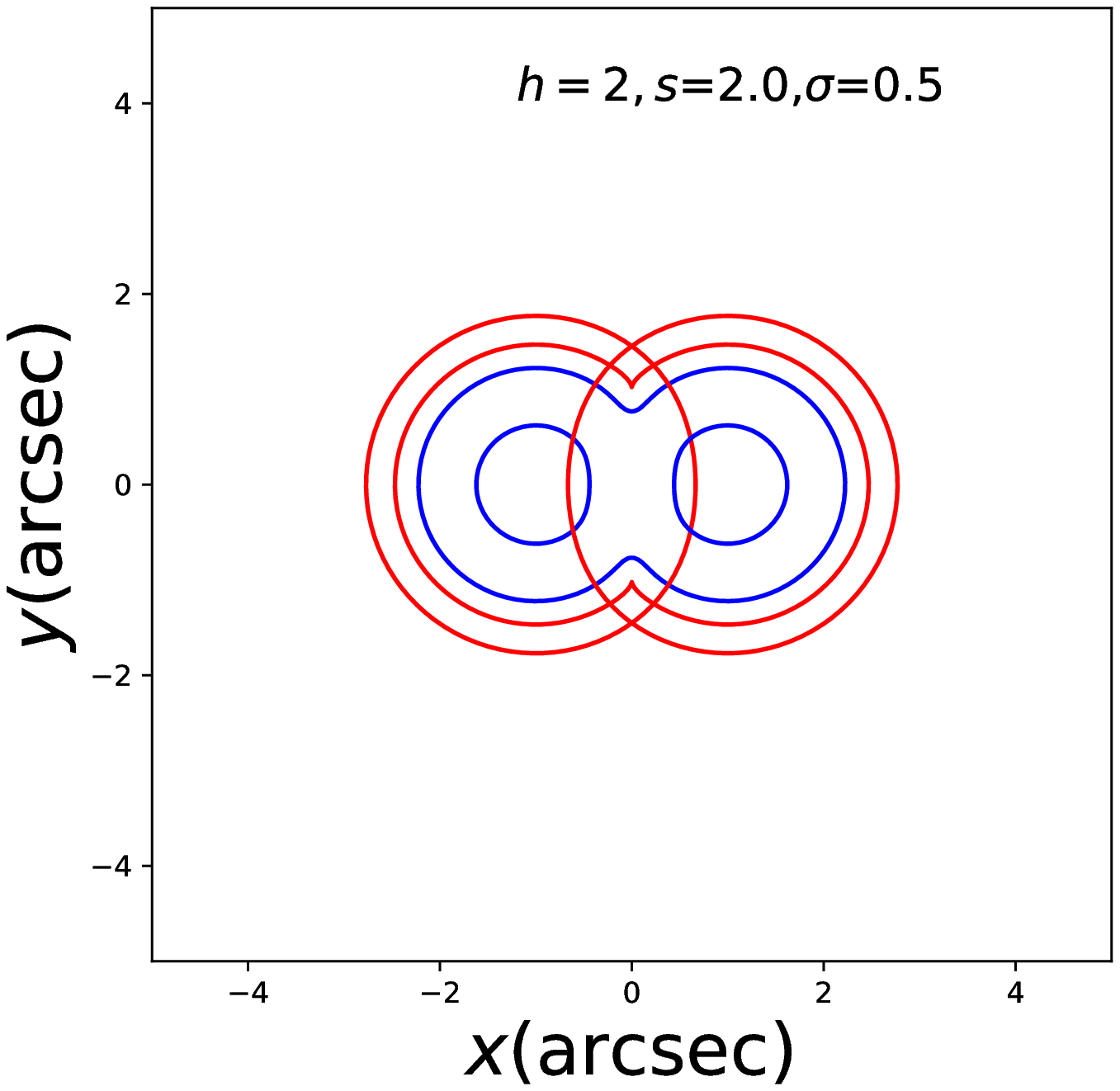}
  \includegraphics[width=4.7cm]{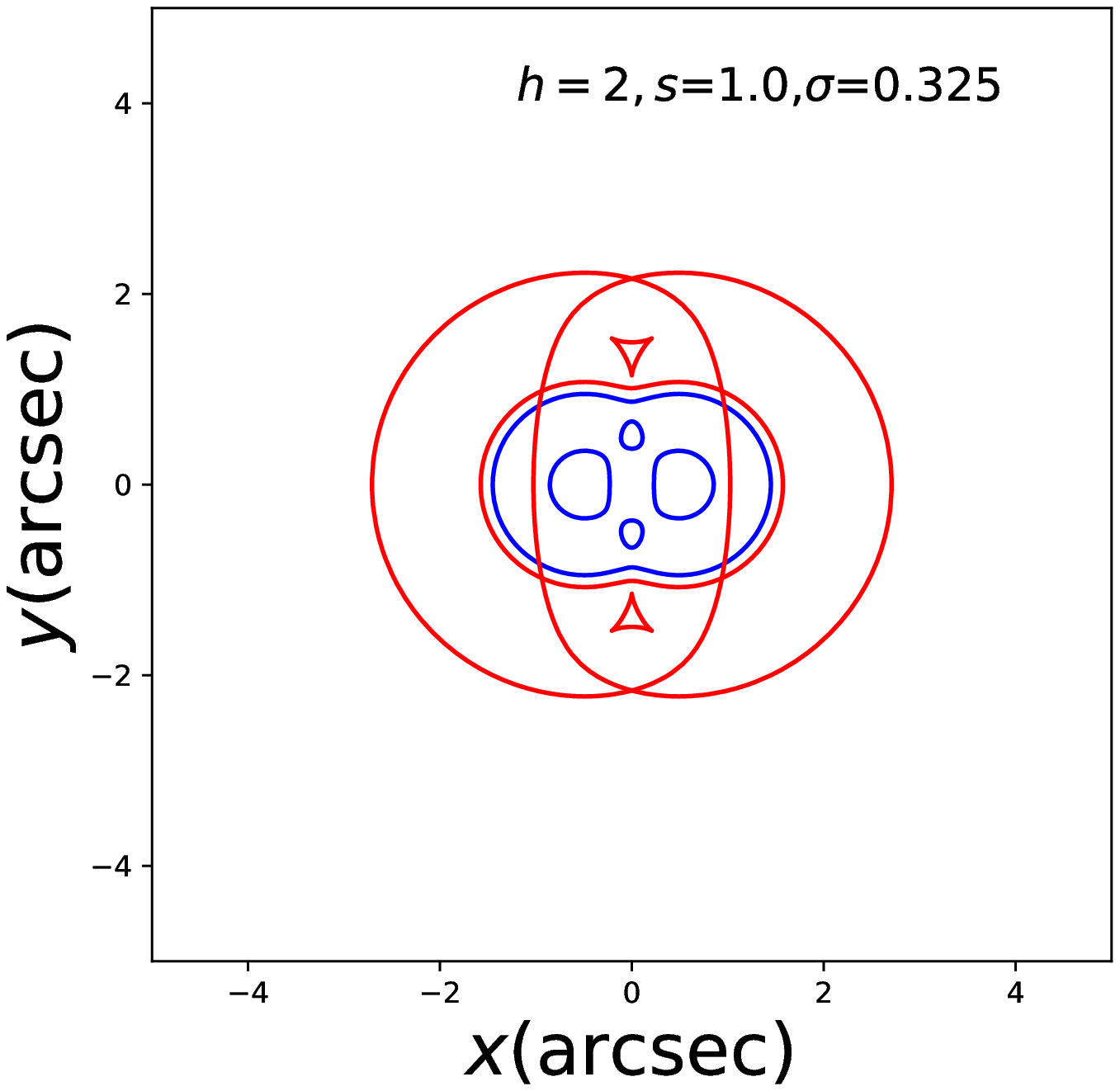}}
  \centerline{\includegraphics[width=4.7cm]{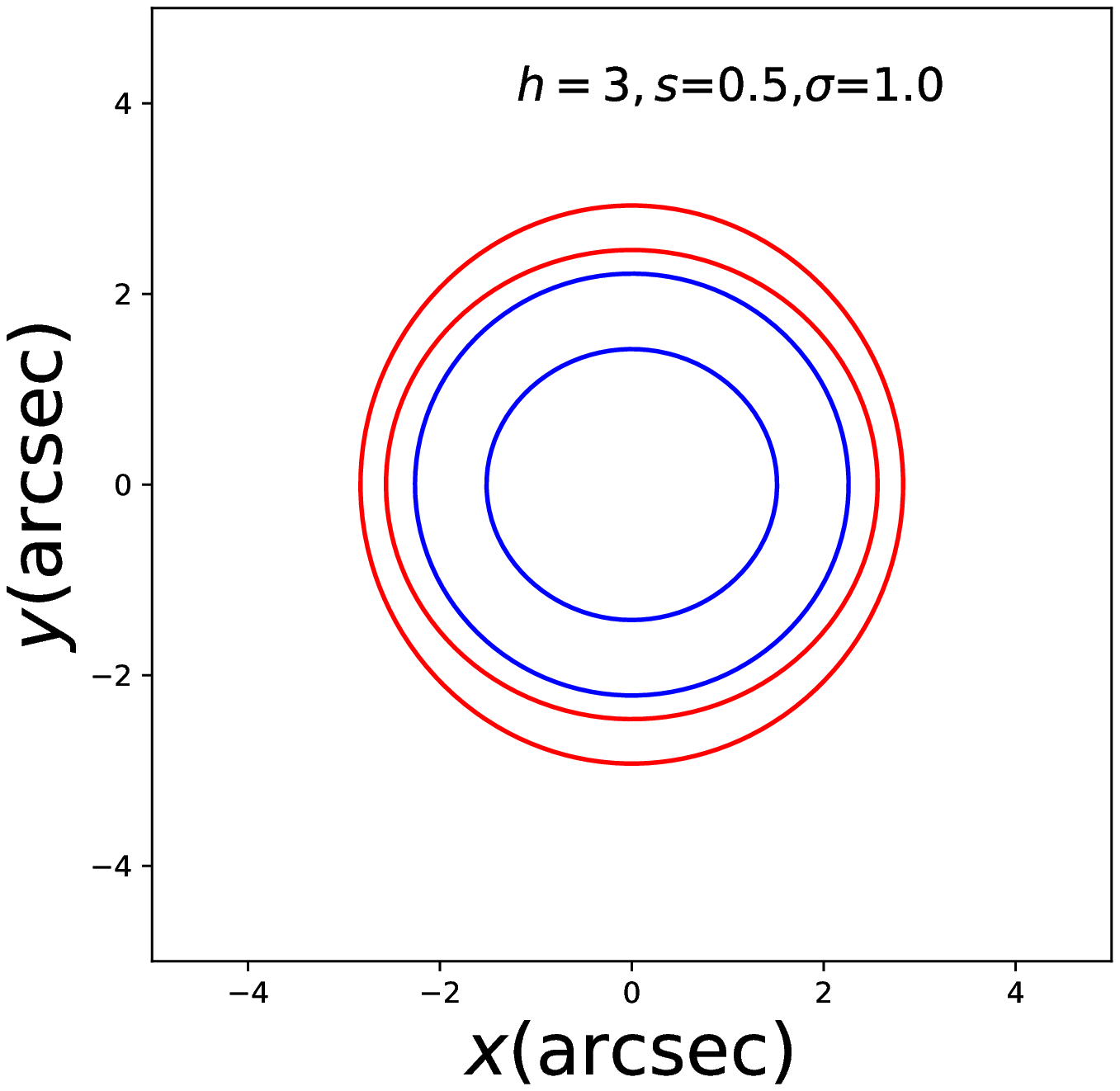}
  \includegraphics[width=4.7cm]{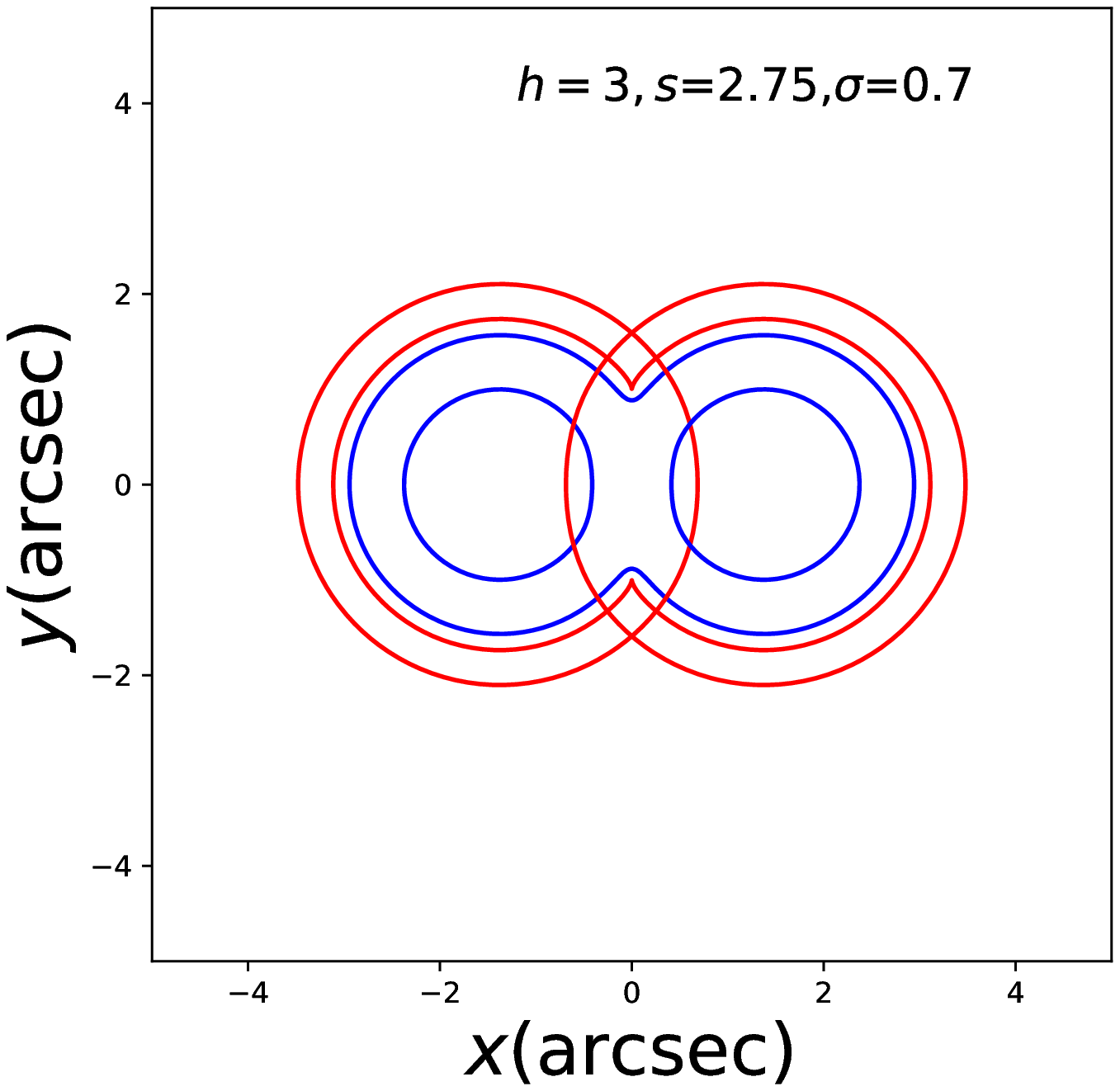}
  \includegraphics[width=4.7cm]{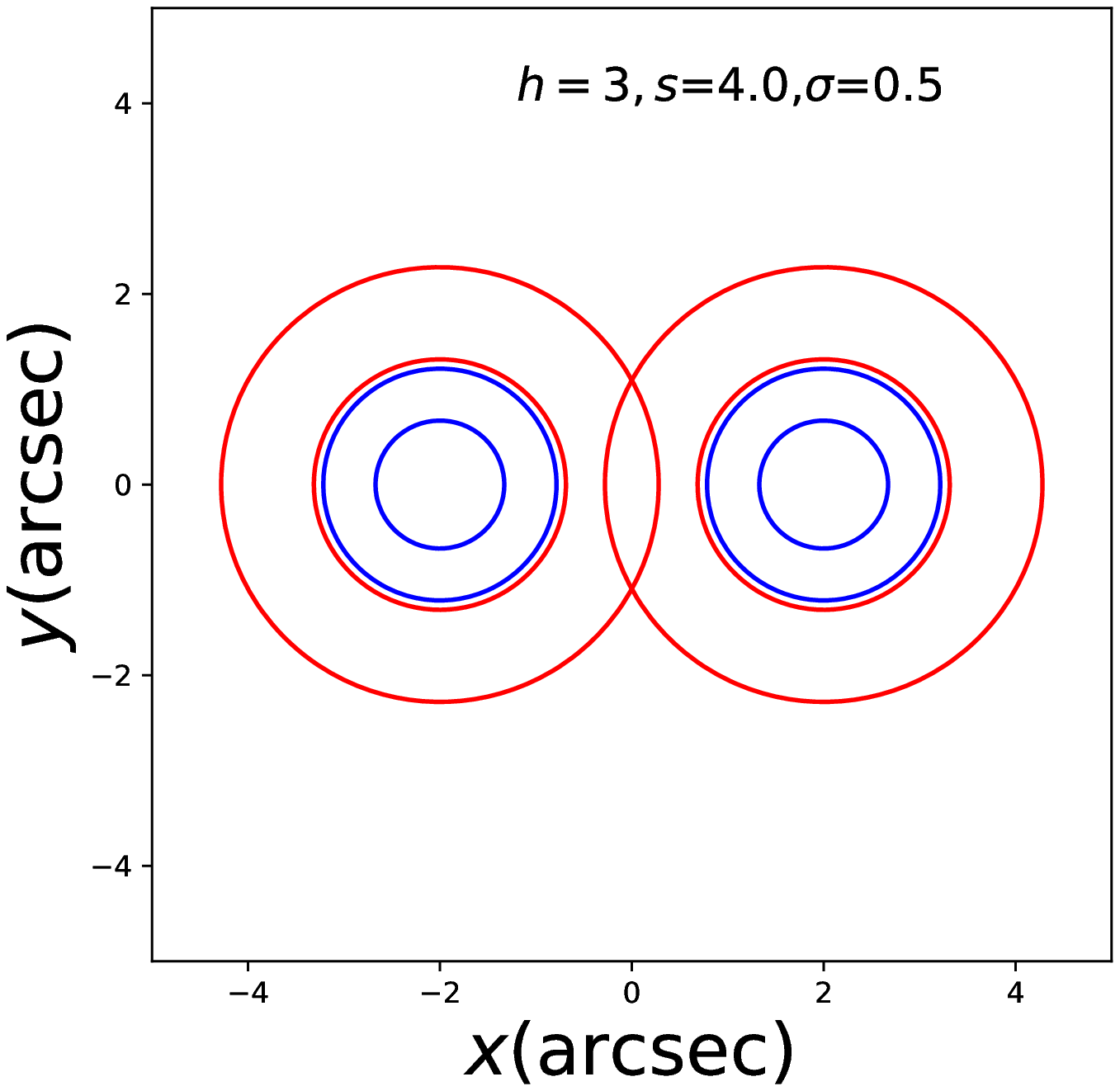}
  \includegraphics[width=4.7cm]{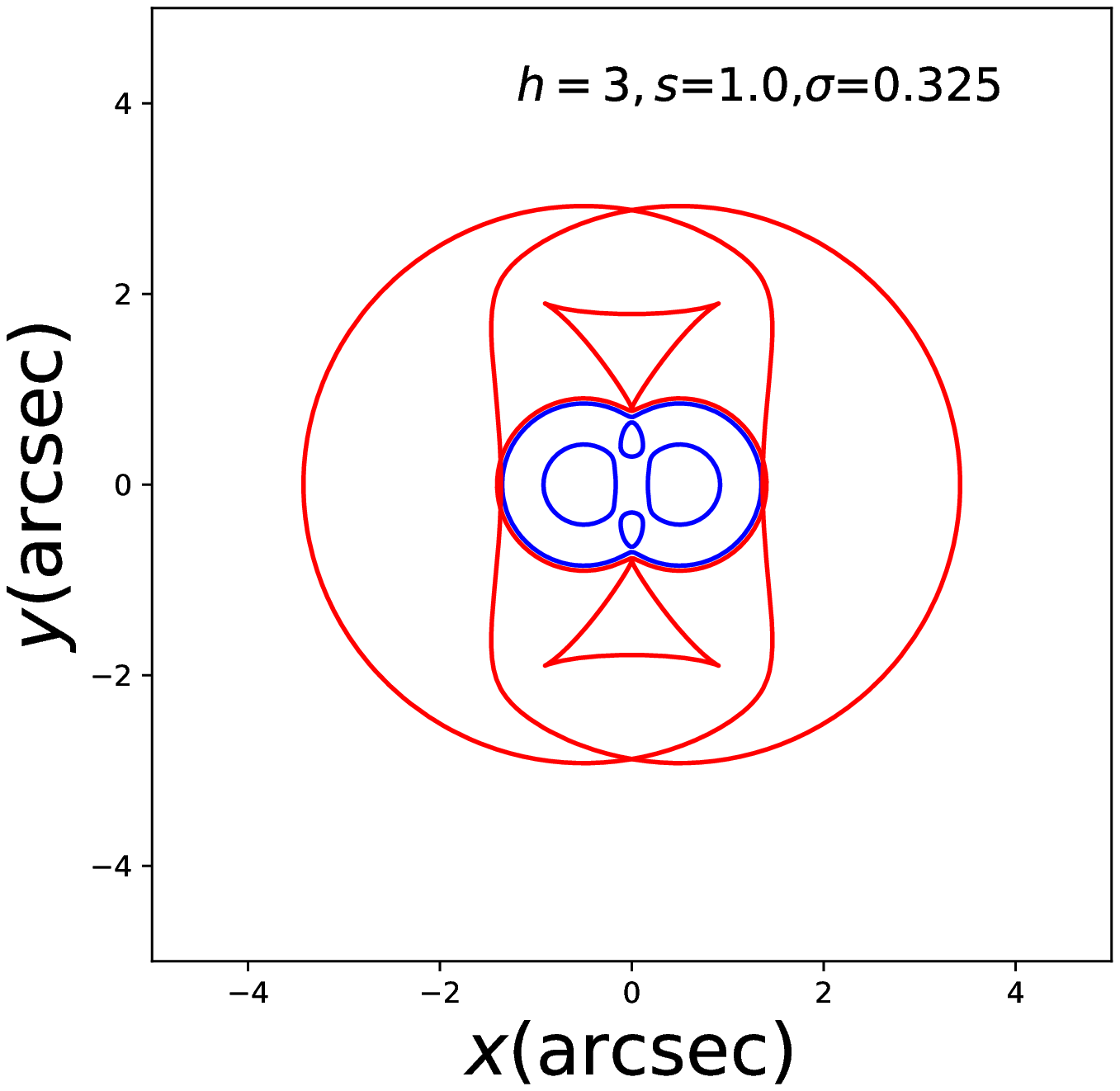}}
  \caption{The critical curve (blue) and caustics (red) of
    dual-component exponential lens. From top to the bottom row we
    show the lenses of exponent $h=1,2,3$ respectively. In this
    figure, the dual-components have the same $\theta_0=1$. The
    separation and the width are given at the top right corner in each
    panel.}
  \label{fig:expte11}
\end{figure*}

\begin{figure*}
  \centerline{\includegraphics[width=4.7cm]{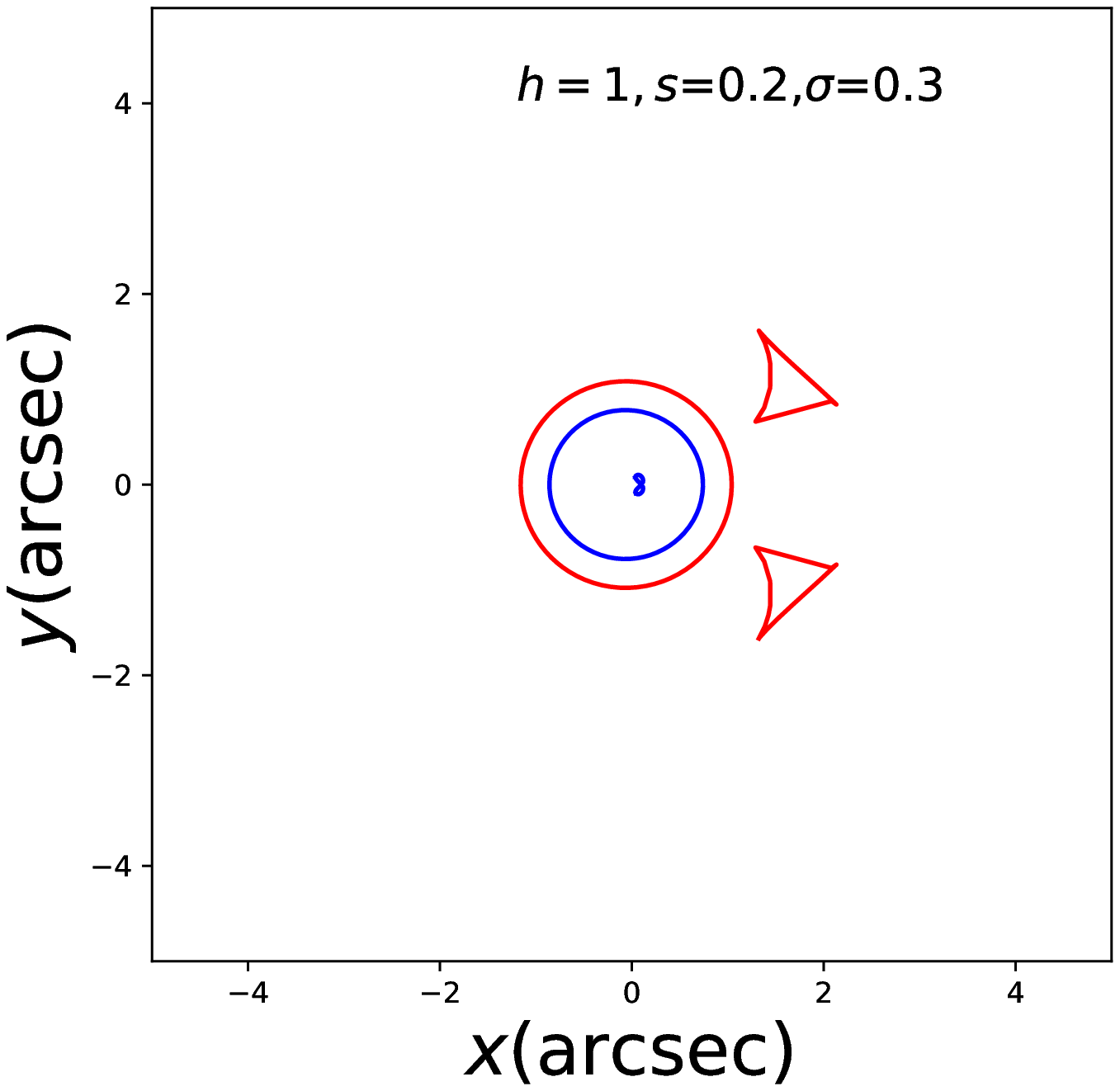}
  \includegraphics[width=4.7cm]{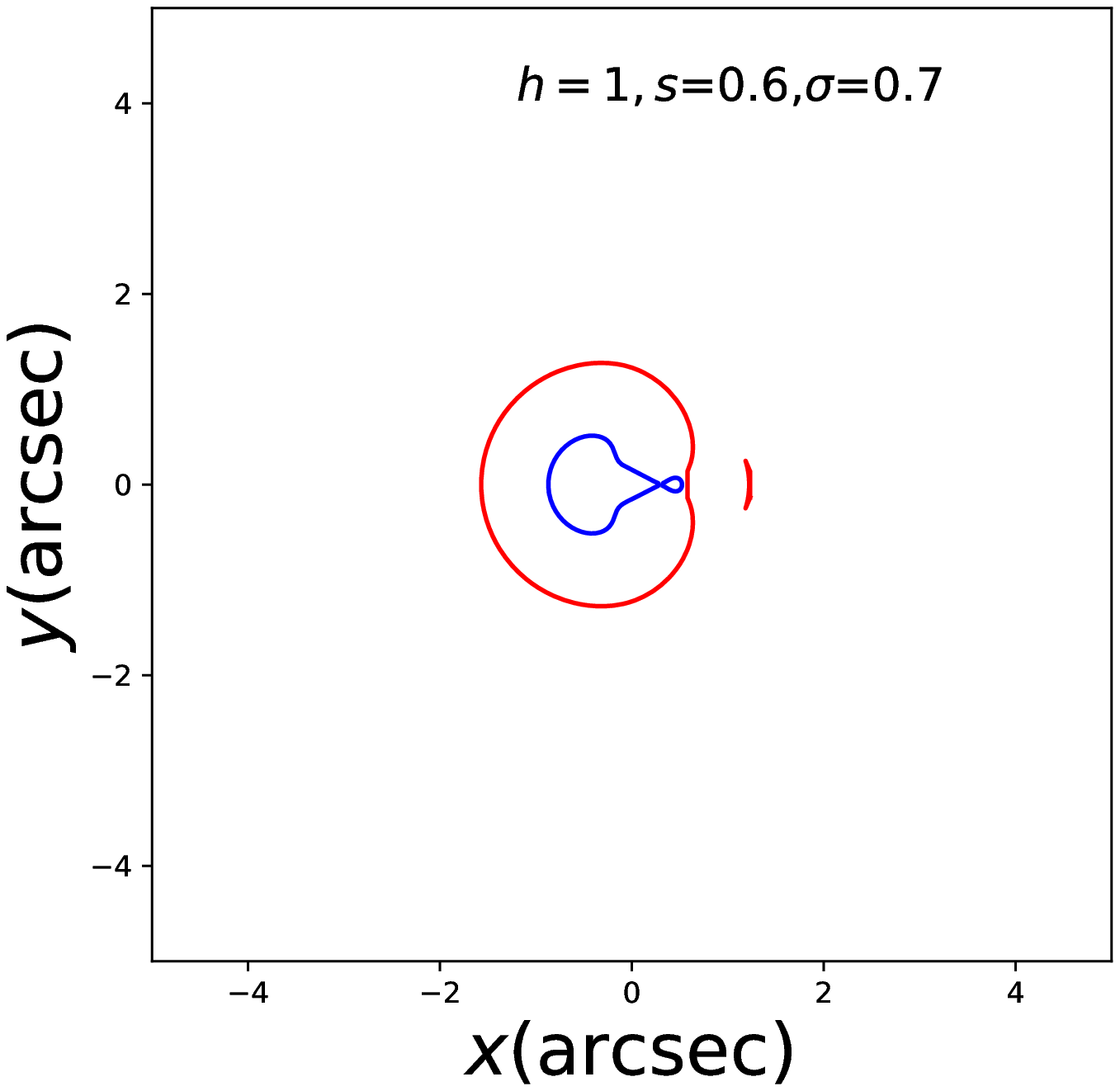}
  \includegraphics[width=4.7cm]{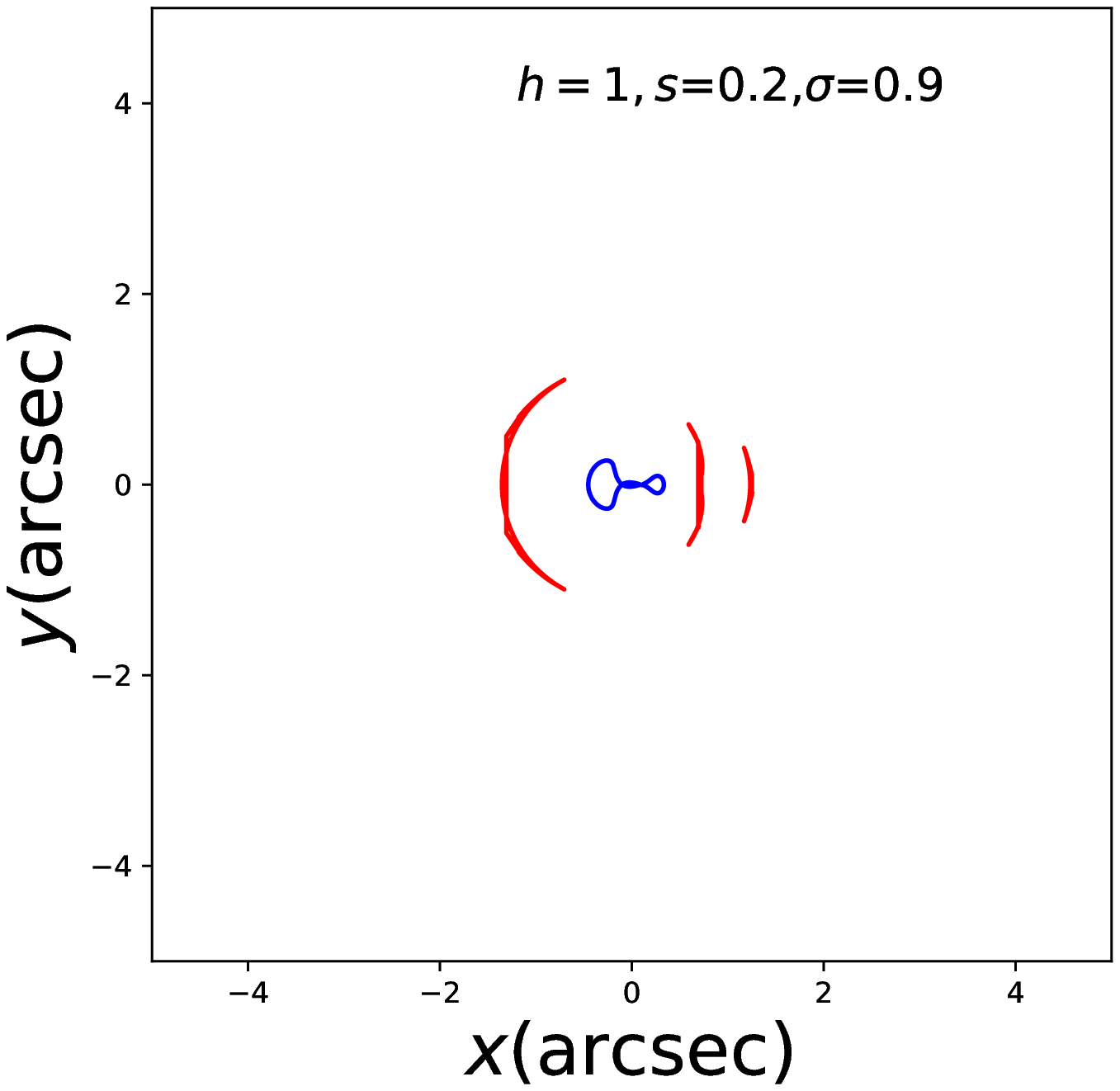}
  \includegraphics[width=4.7cm]{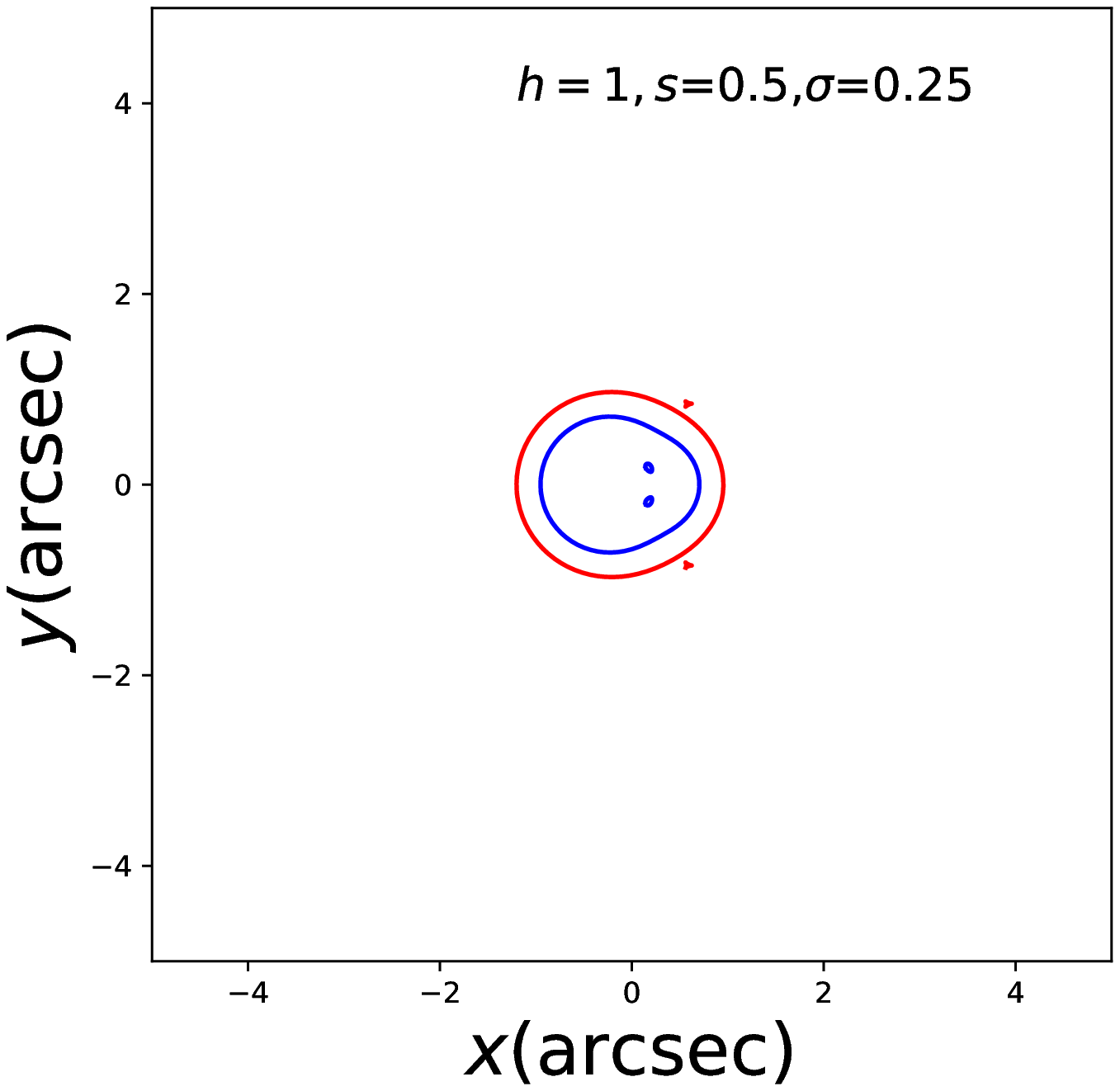}}
  \centerline{\includegraphics[width=4.7cm]{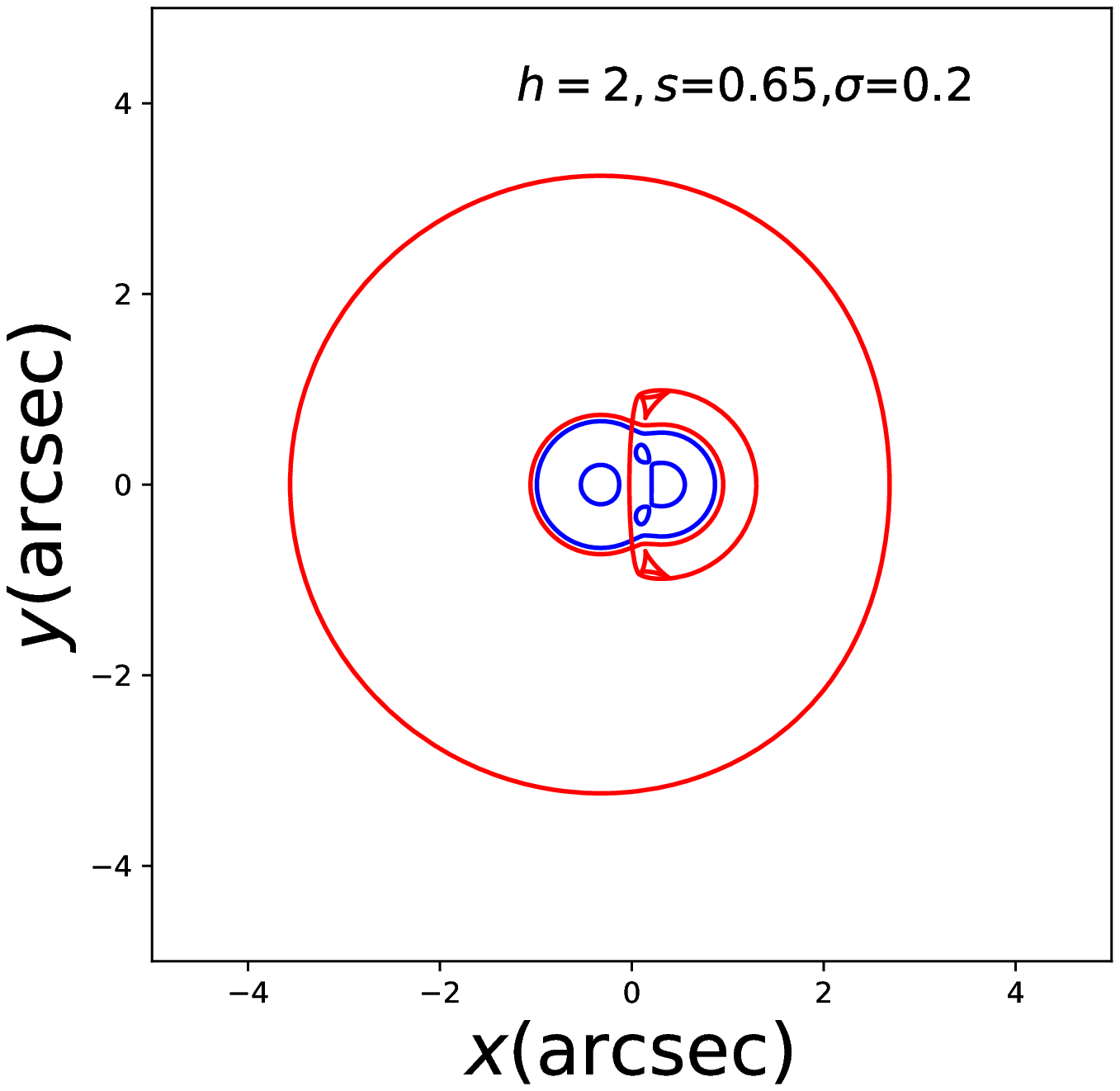}
  \includegraphics[width=4.7cm]{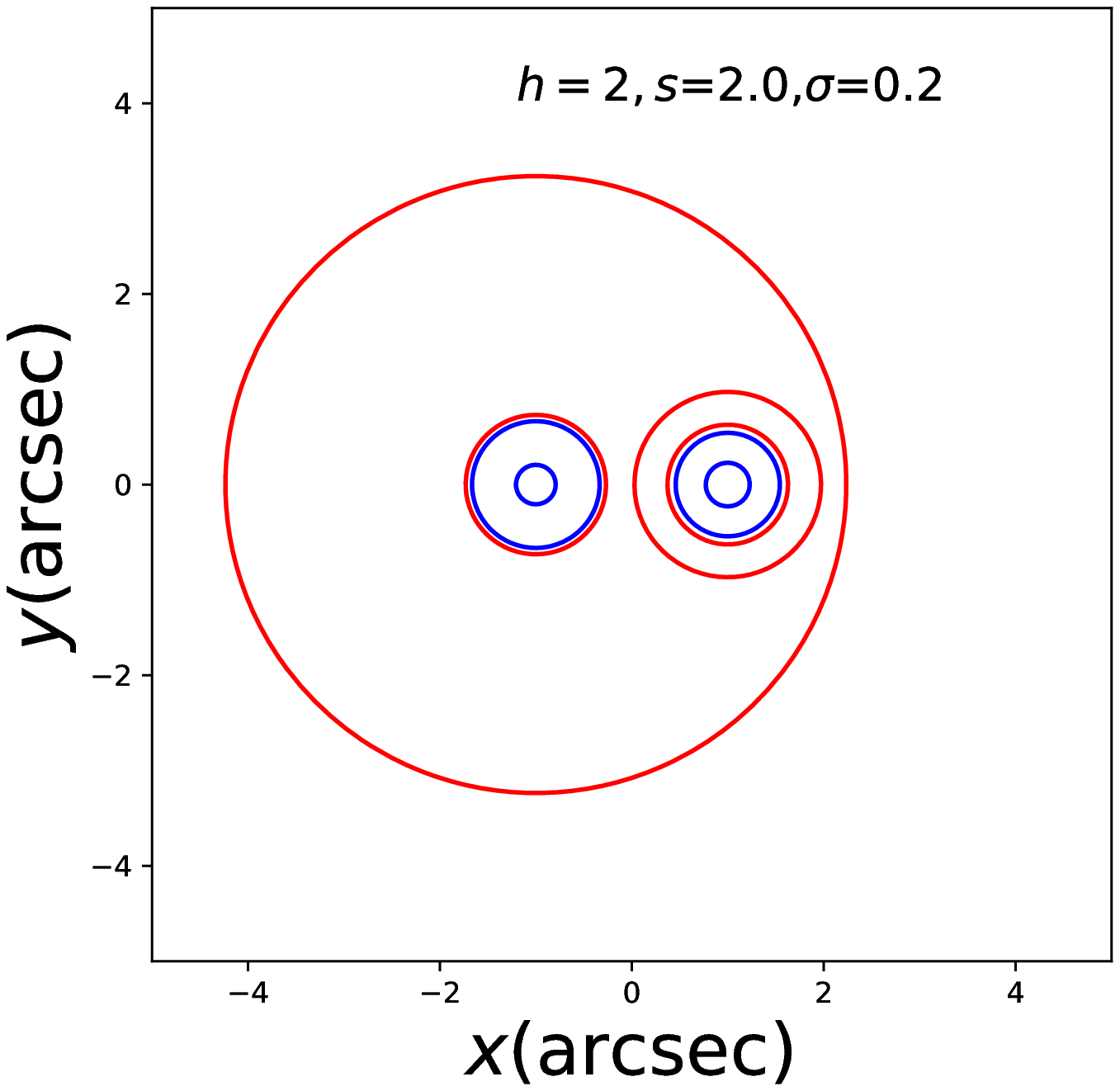}
  \includegraphics[width=4.7cm]{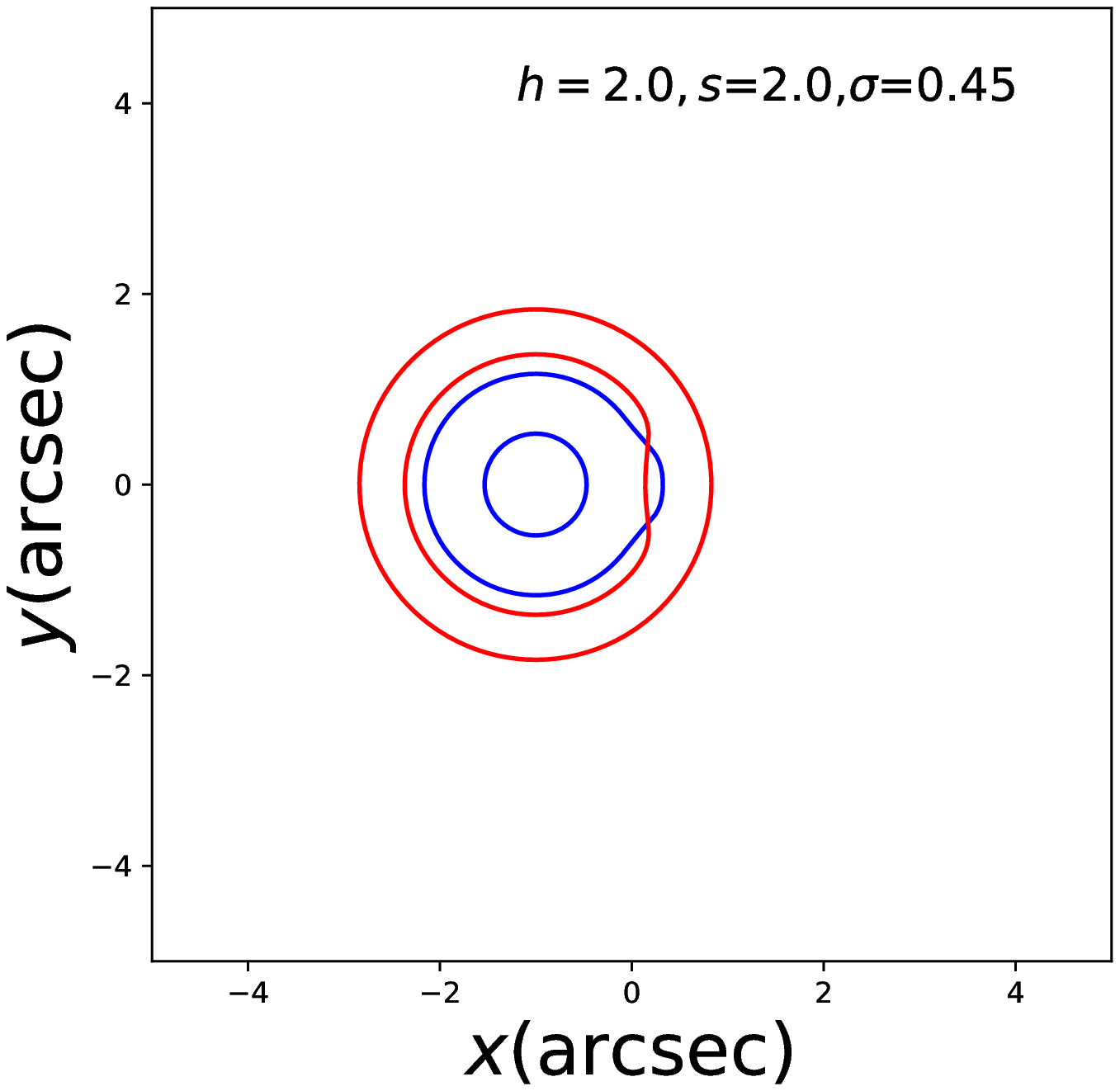}
  \includegraphics[width=4.7cm]{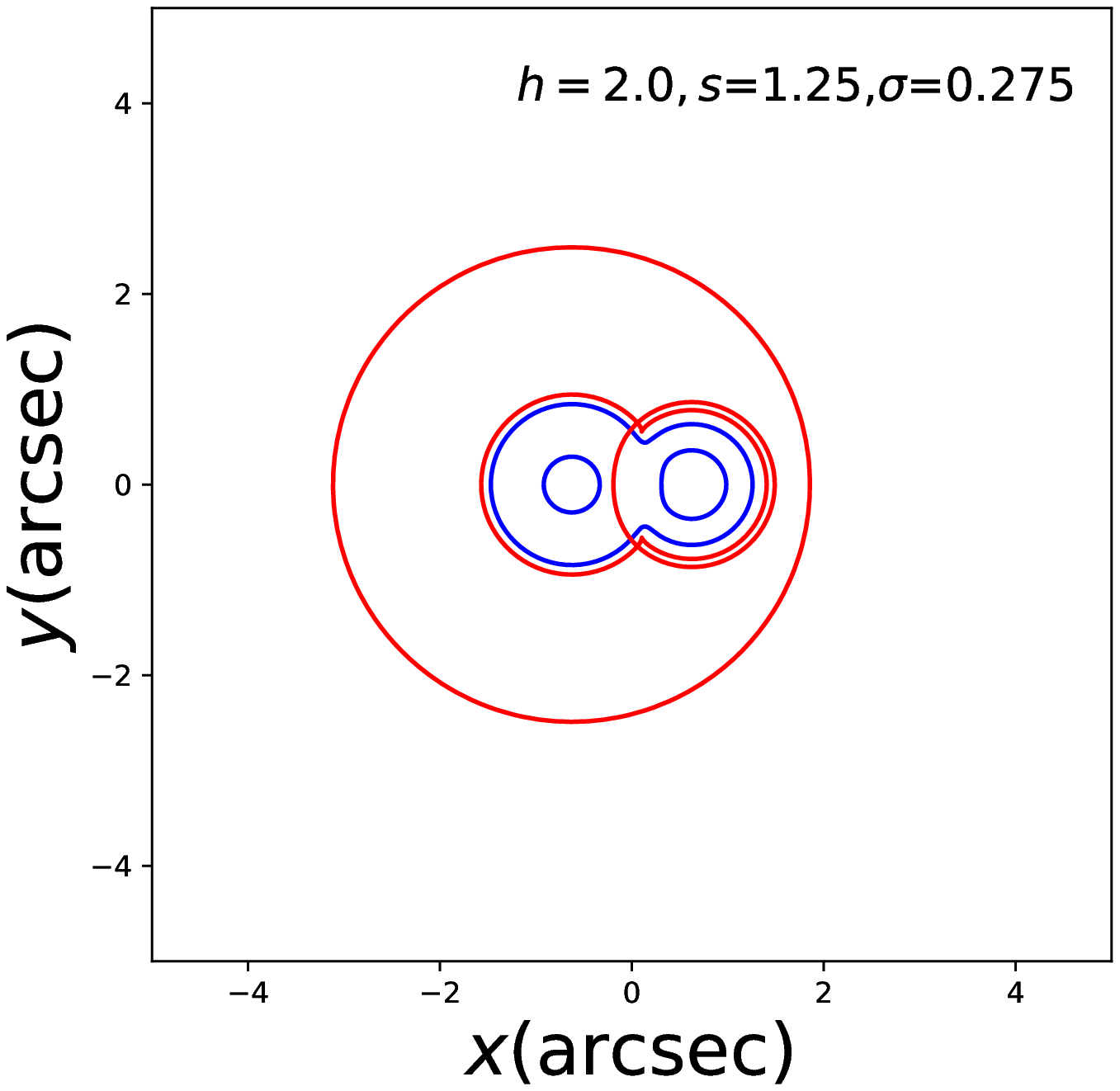}}
  \centerline{\includegraphics[width=4.7cm]{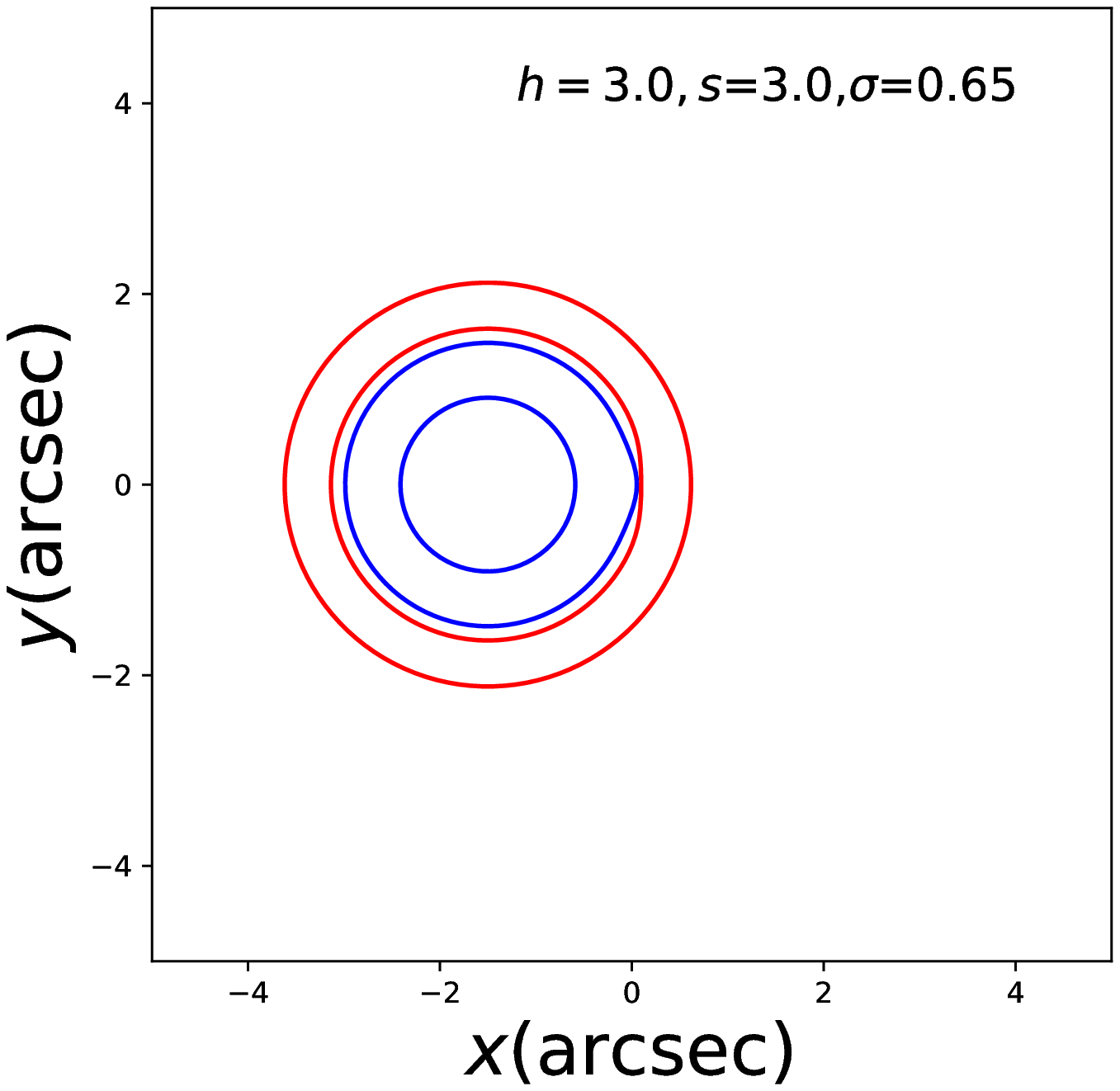}
  \includegraphics[width=4.7cm]{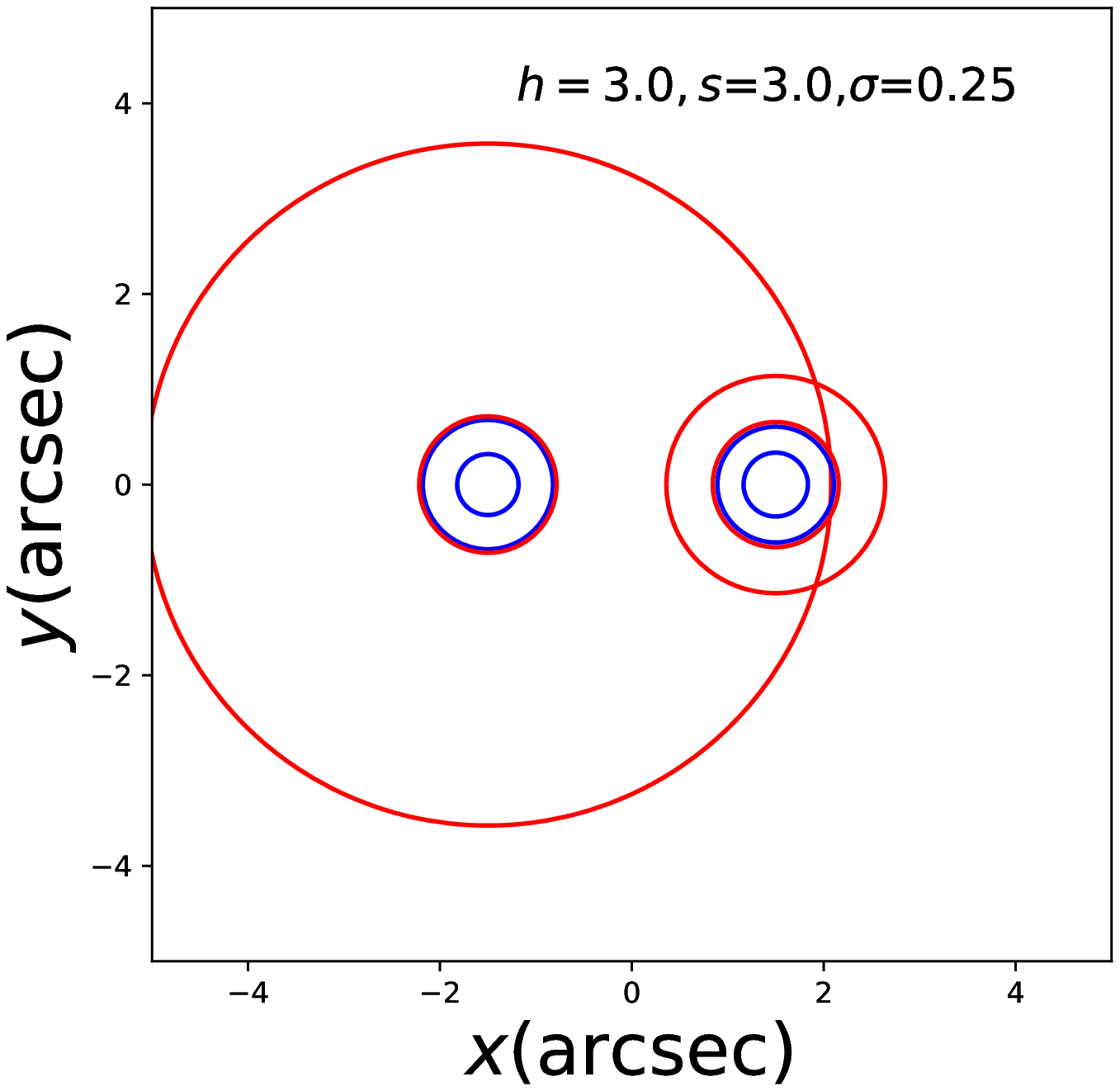}
  \includegraphics[width=4.7cm]{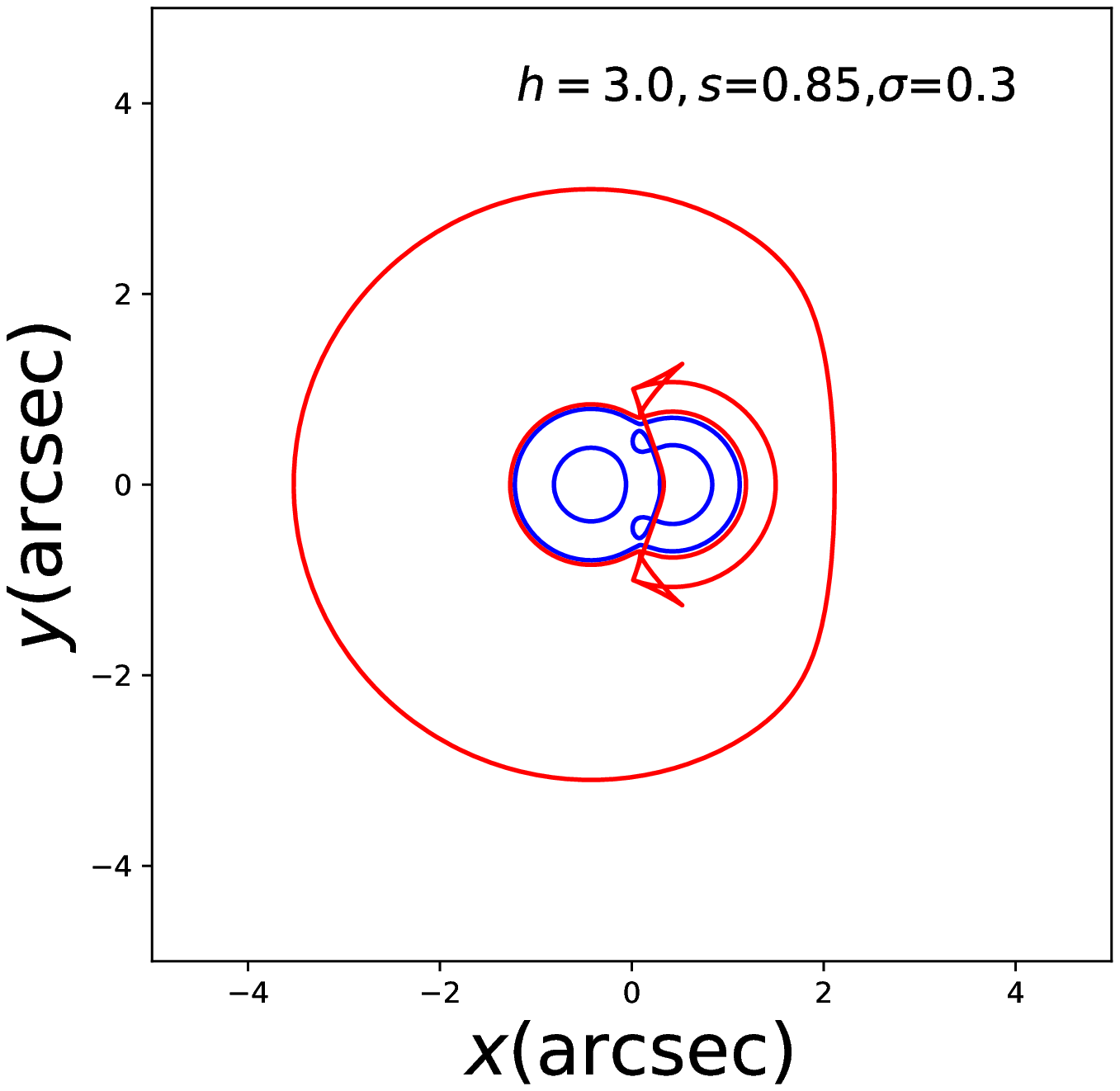}
  \includegraphics[width=4.7cm]{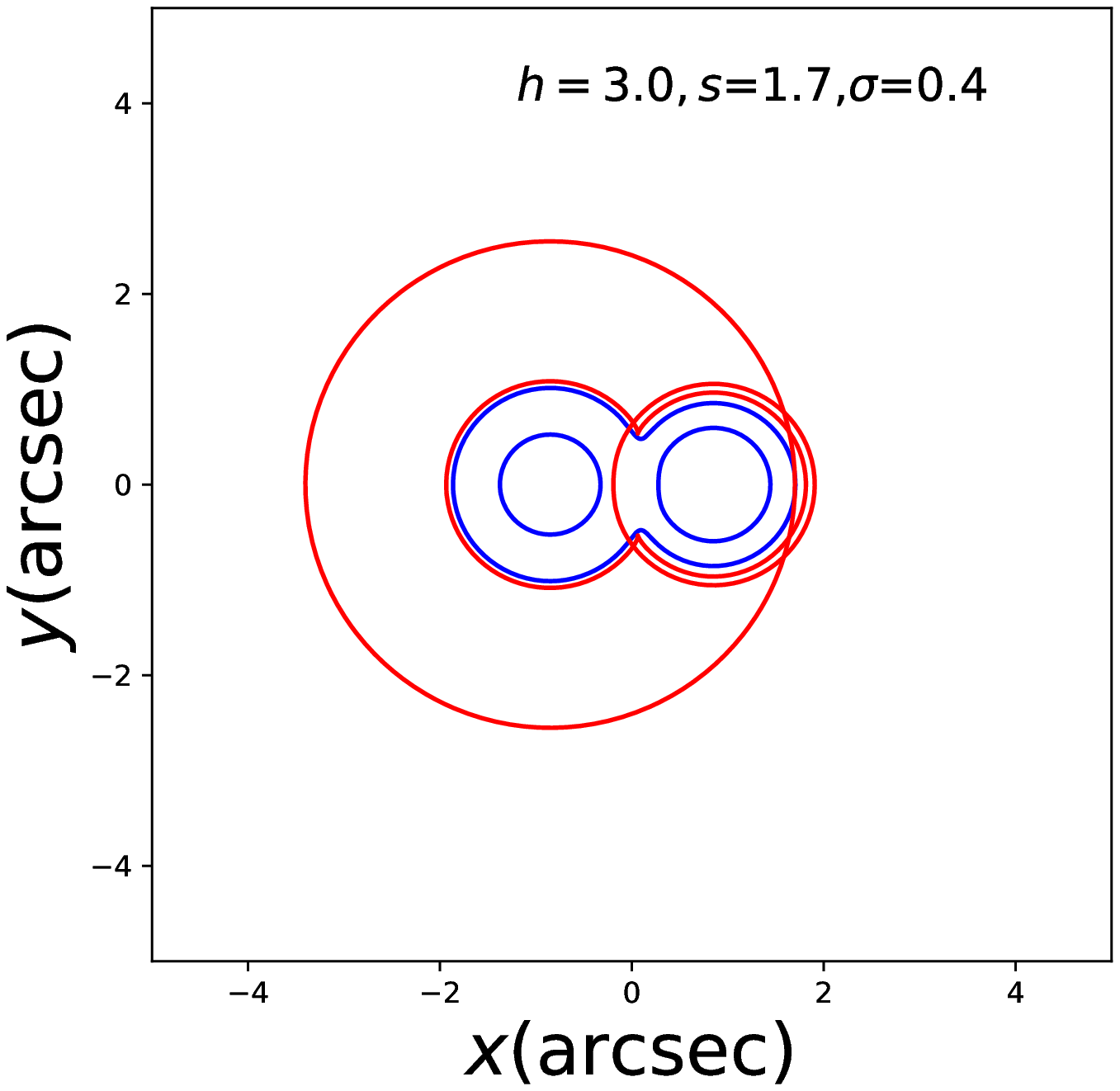}}
  \caption{Same as Fig.\,\ref{fig:expte11} but the dual-components have different $\theta_{0}(1.0,0.5)$. }
  \label{fig:expte12}
\end{figure*}

\begin{figure*}
  \centerline{\includegraphics[width=4.7cm]{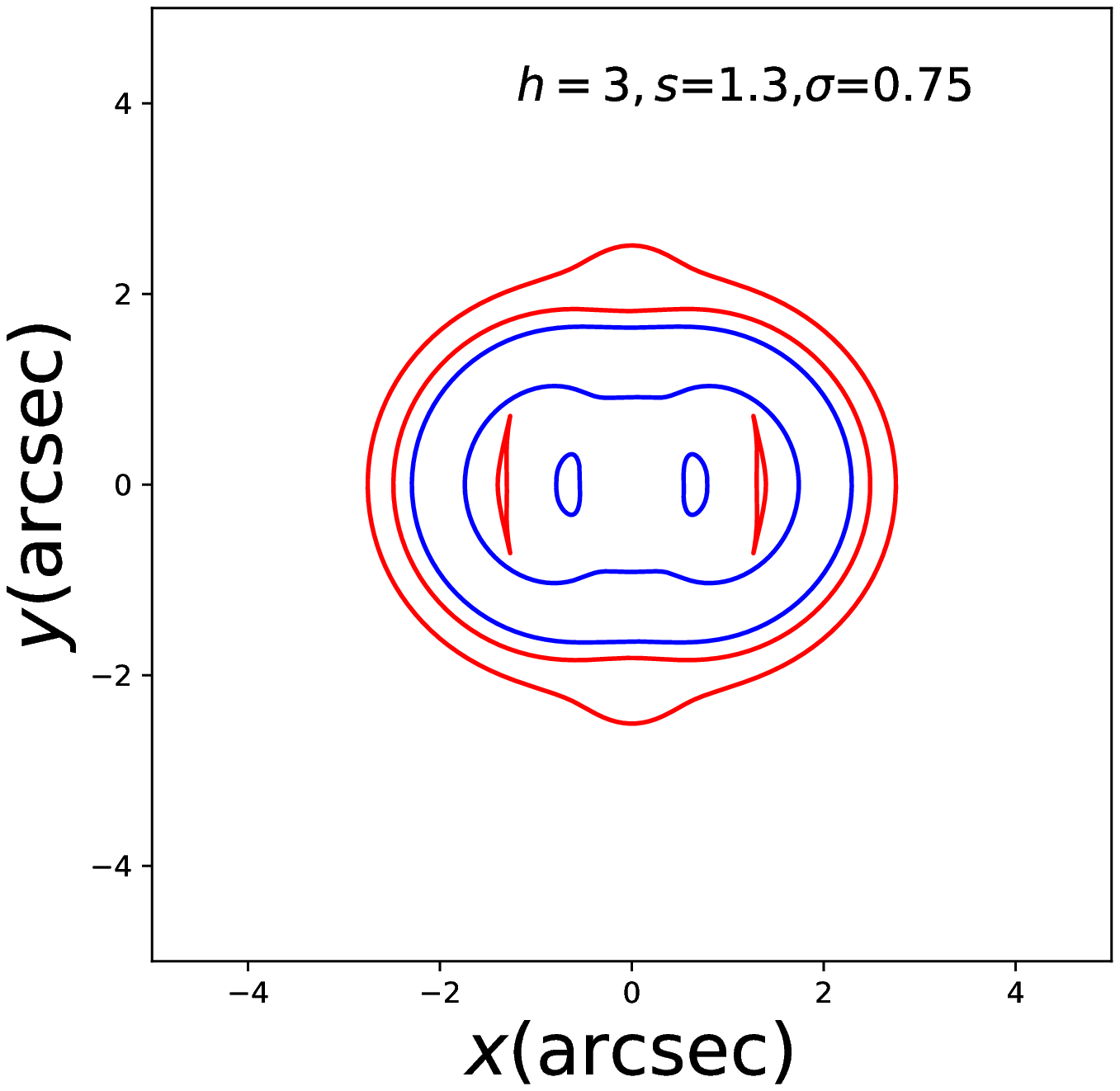}
  \includegraphics[width=4.7cm]{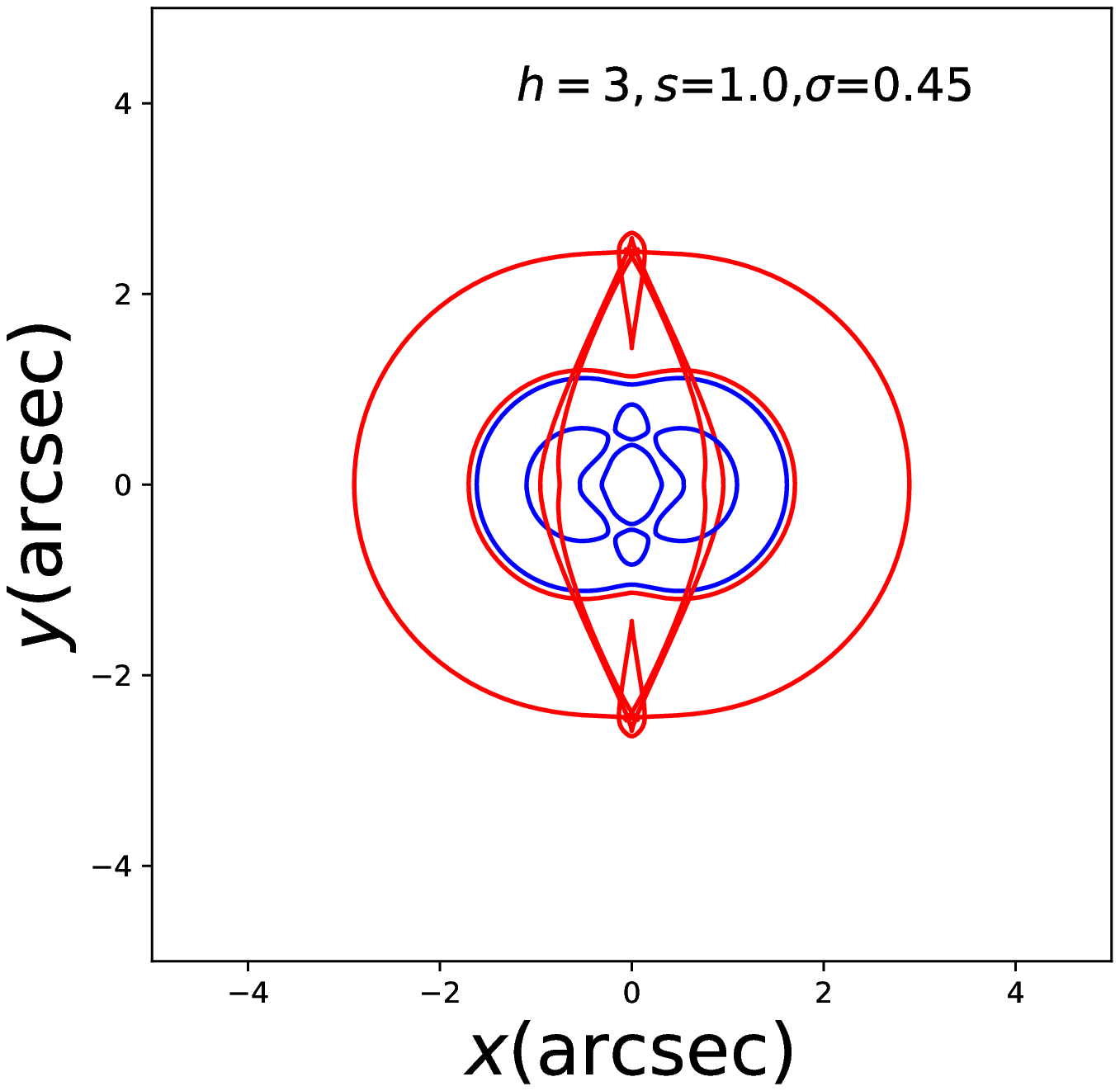}
  \includegraphics[width=4.7cm]{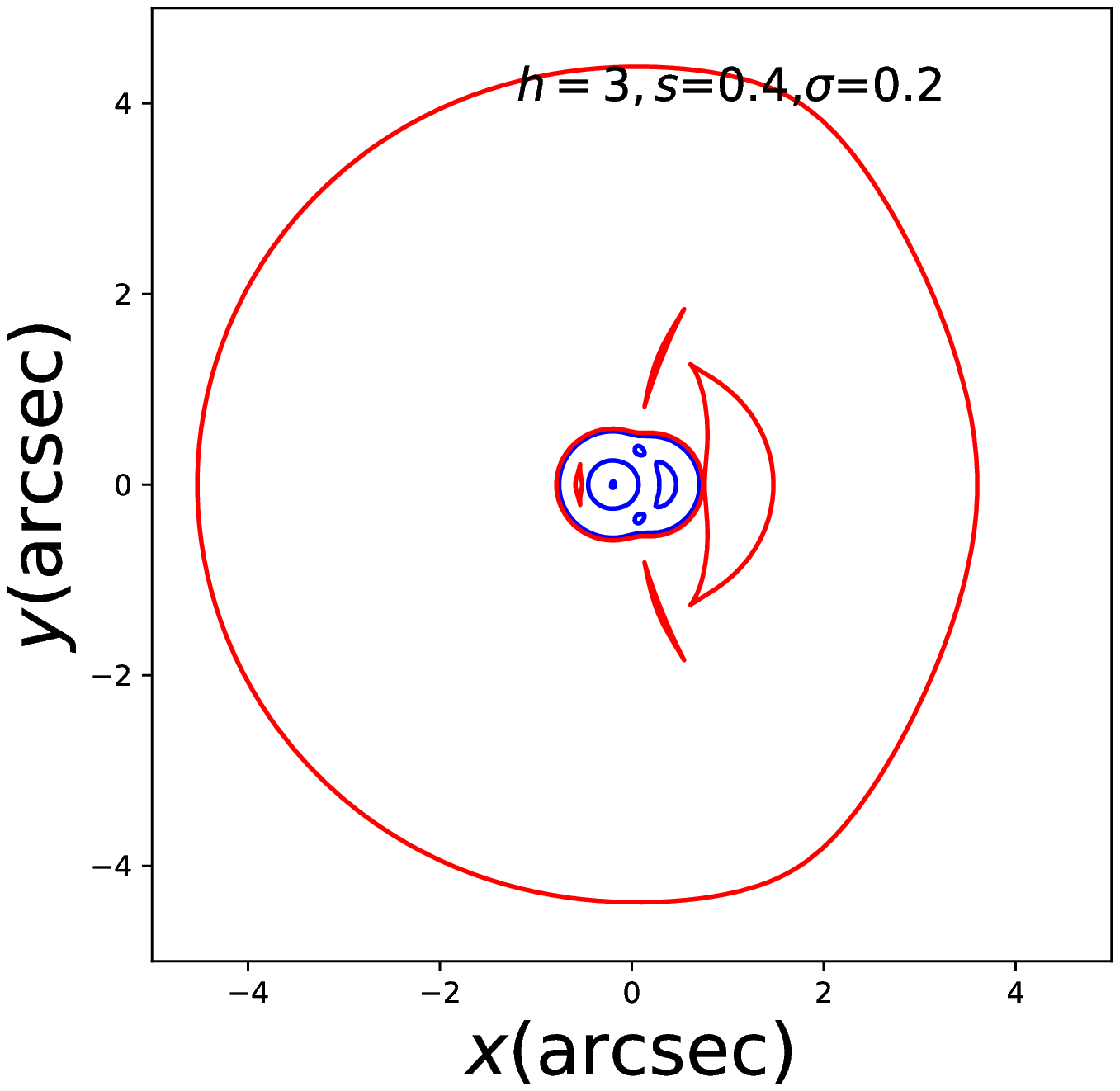}
  \includegraphics[width=4.7cm]{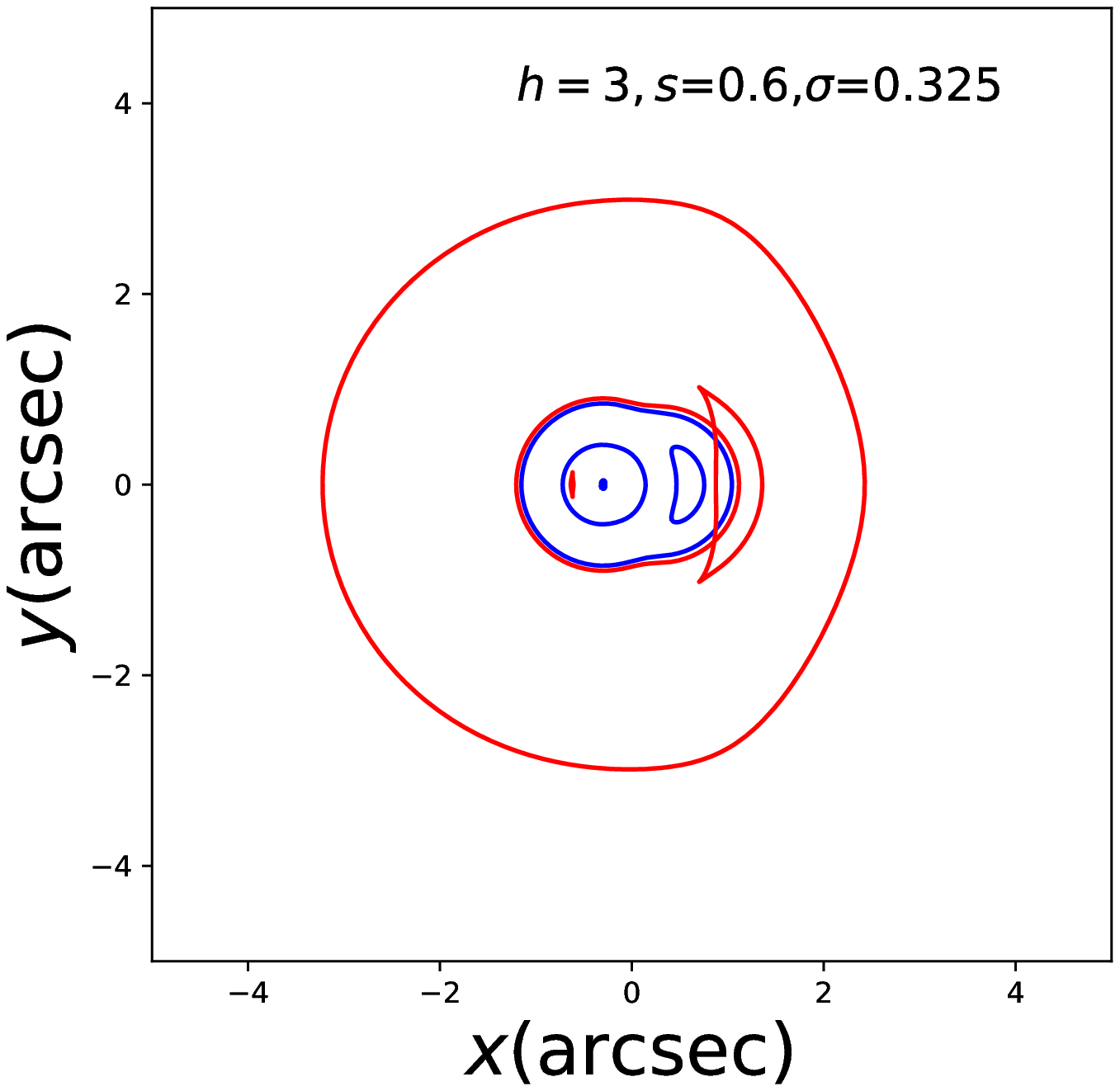}}
  \caption{Some more critical and caustics of the $h=3$
      exponential lens: in the two panels on the left, the dual
    components have the same $\theta_0=1.0$, and in the two right
    panels, the dual components have different $\theta_0 (1.0,0.5)$. }
  \label{fig:expte3}
\end{figure*}

Let us investigate some specific examples of the exponential lens. First, we plot the equal strength case, $\theta_{01}=\theta_{02}=1$ and label the plots with particular $s$ and $\sigma$ values in Fig.\,\ref{fig:expte11}. In Fig.\,\ref{fig:expte12}, we plot unequal strength components with $\theta_{01}=1$ and $\theta_{02}=0.5$. These combinations show a more interesting diversity of critical and caustic shapes than the equal strength lens, which are symmetric about the $y$-axis. The asymmetry produces a much larger array of caustics that are generally asymmetric about the $y$-axis. In Fig.\,\ref{fig:expte3} we show critical and caustic morphologies for the $h=3$ exponential dual component lenses that have four or more caustics. The left two panels have $\theta_{01}=\theta_{02}=1$ and the right two panels asymmetric $\theta_{01}=1$ and $\theta_{02}=0.5$.

\subsection{Dual-Component Softened Power-Law Lens}
\begin{figure*}
  \centerline{\includegraphics[width=4.7cm]{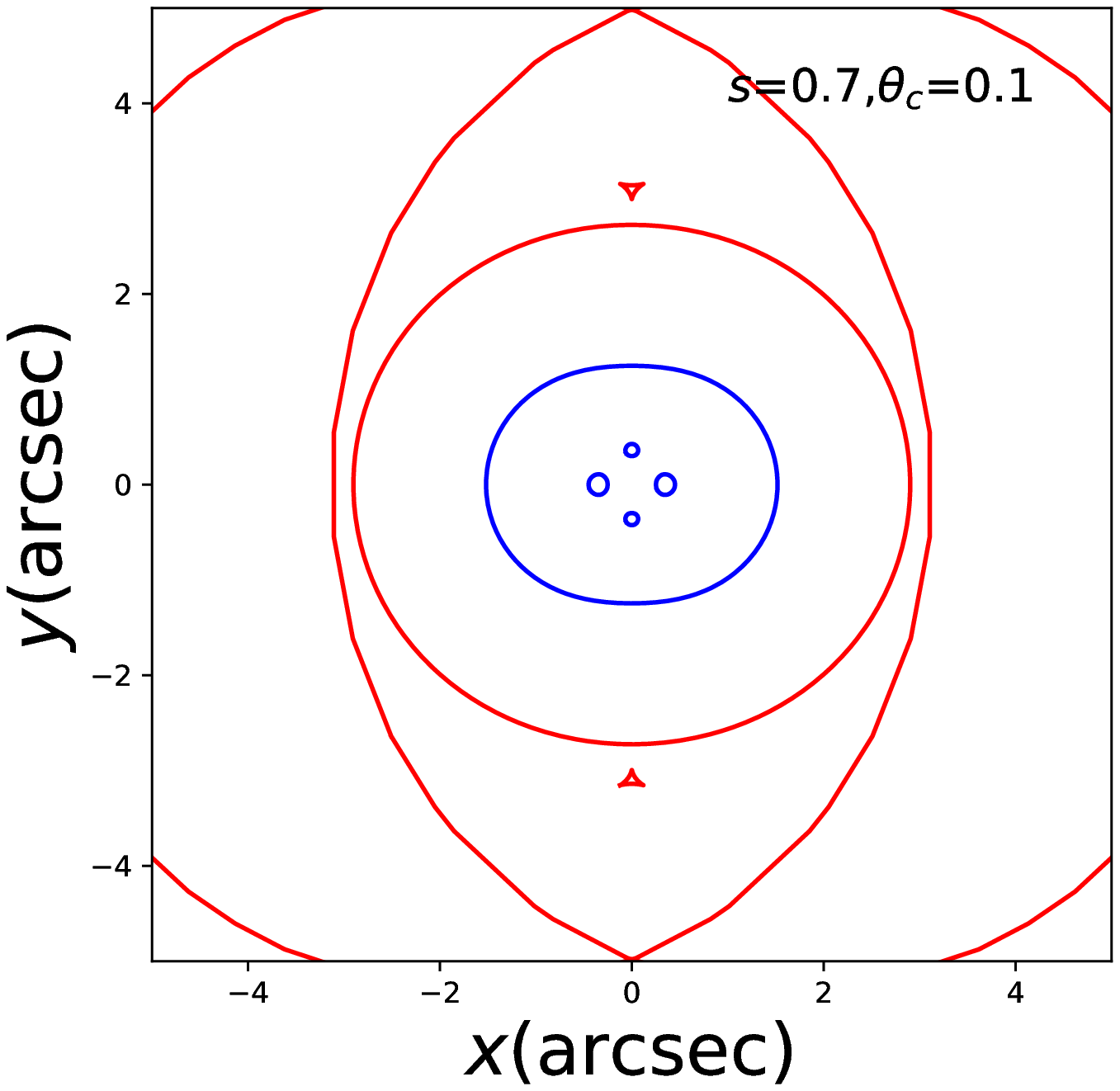}
  \includegraphics[width=4.7cm]{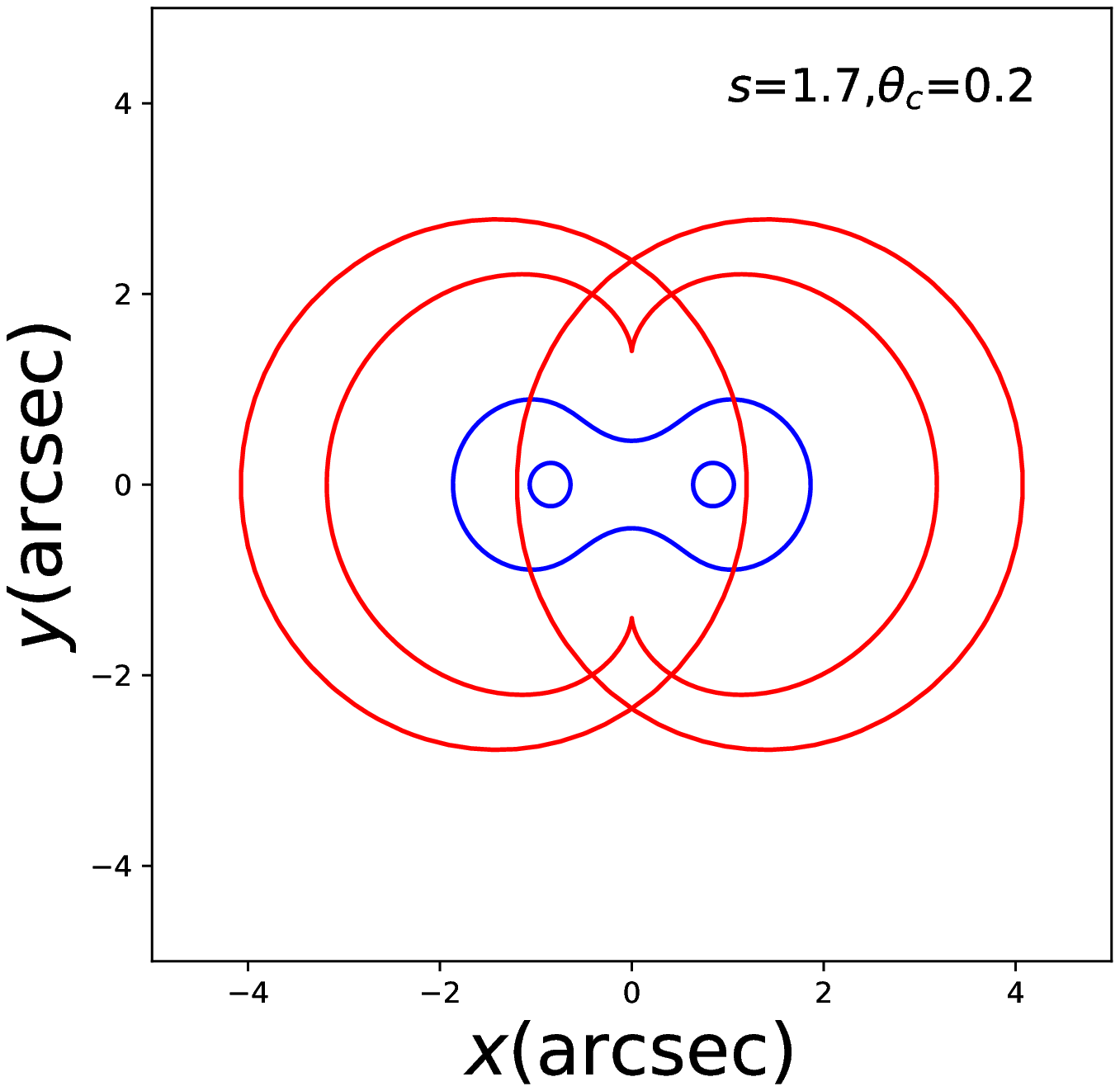}
  \includegraphics[width=4.7cm]{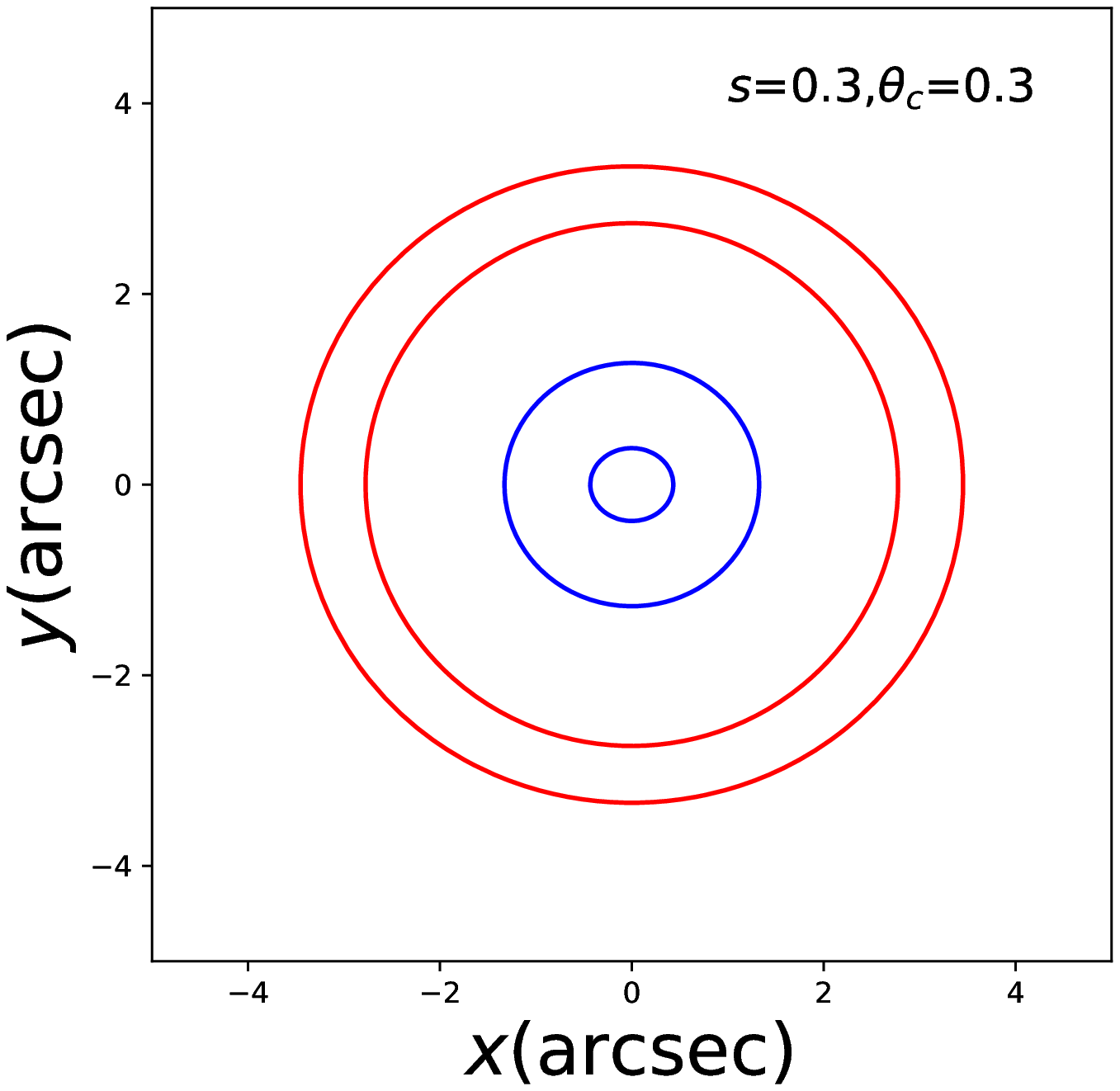}
  \includegraphics[width=4.7cm]{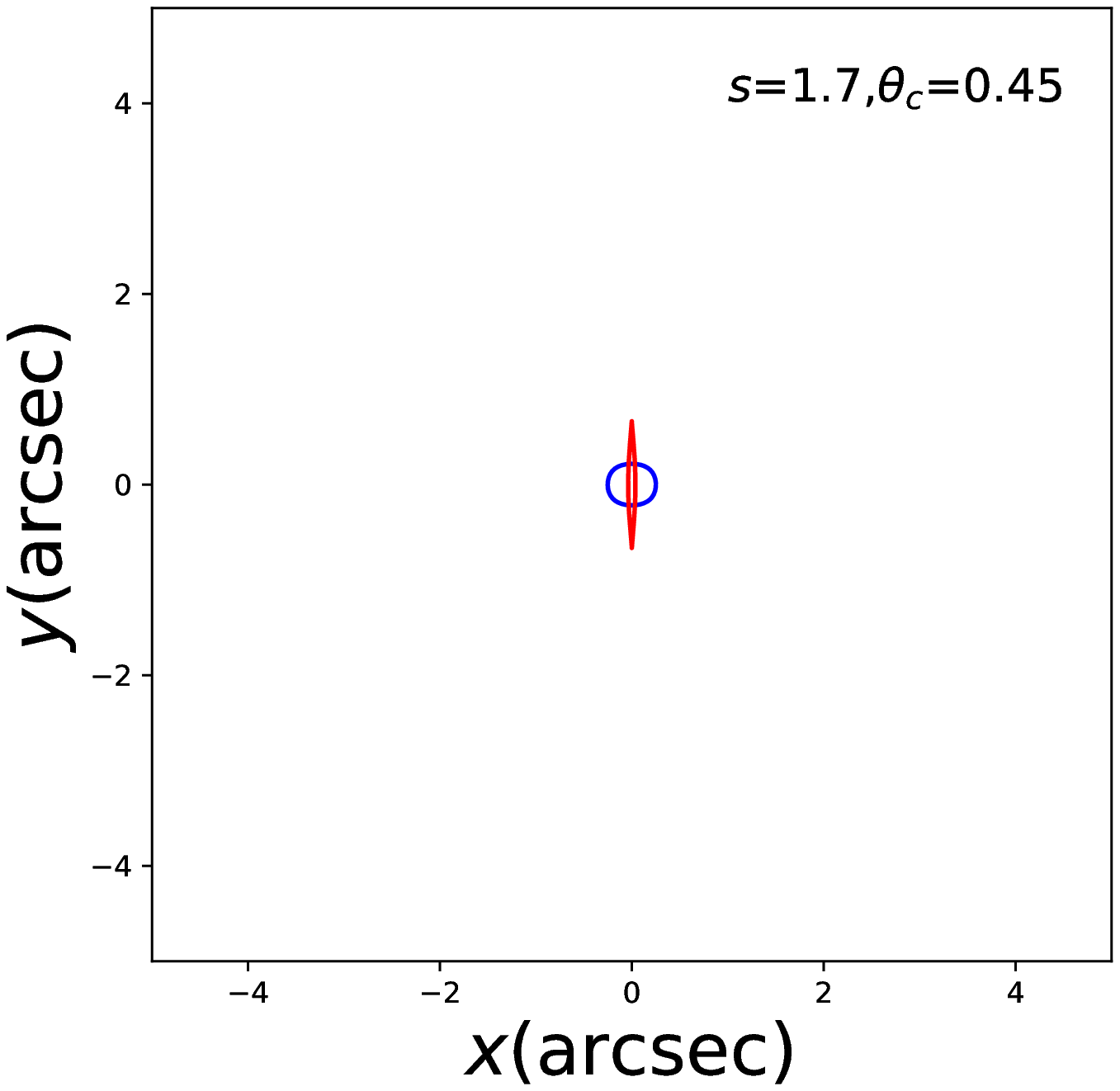}}
  \centerline{\includegraphics[width=4.7cm]{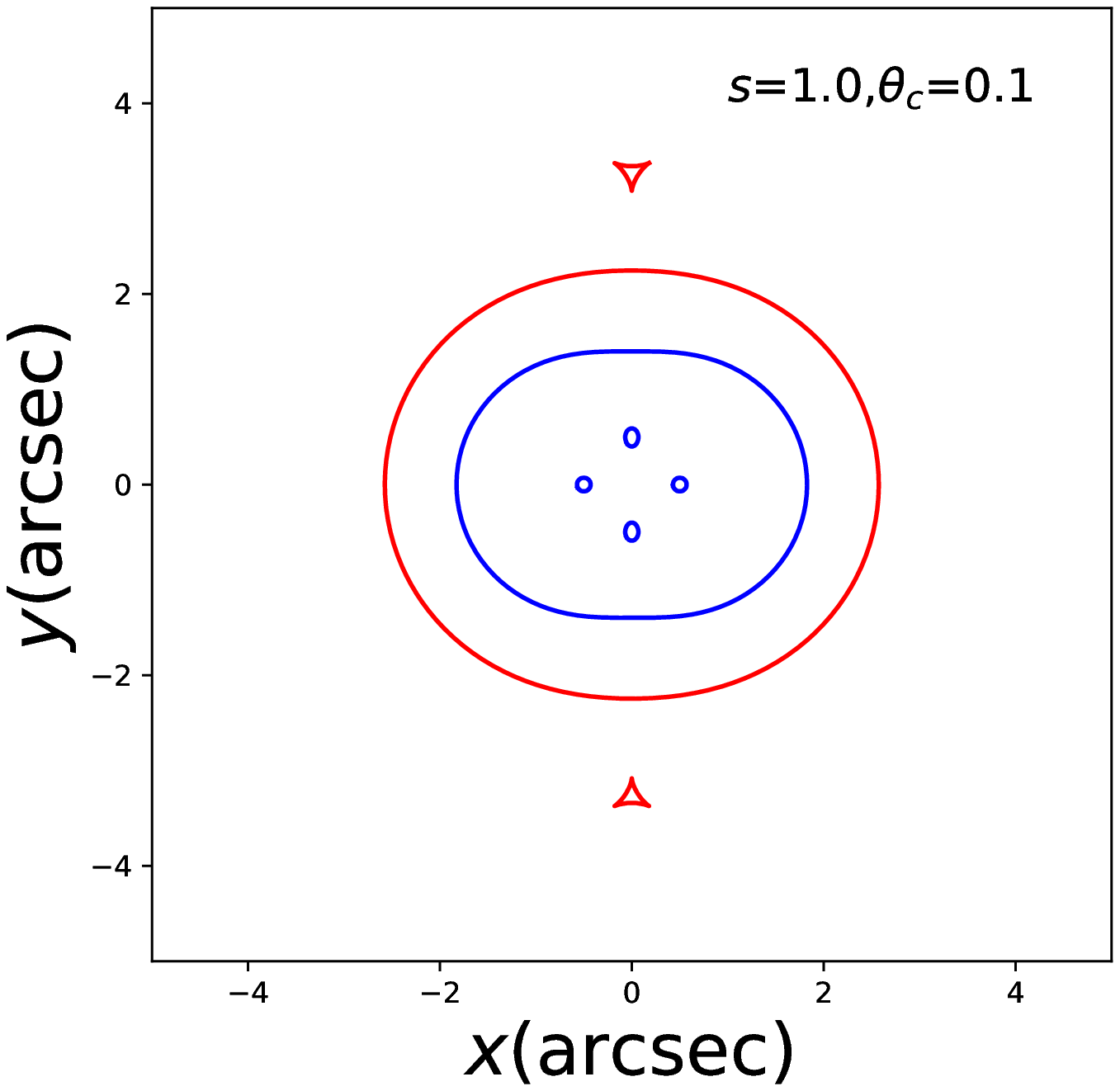}
  \includegraphics[width=4.7cm]{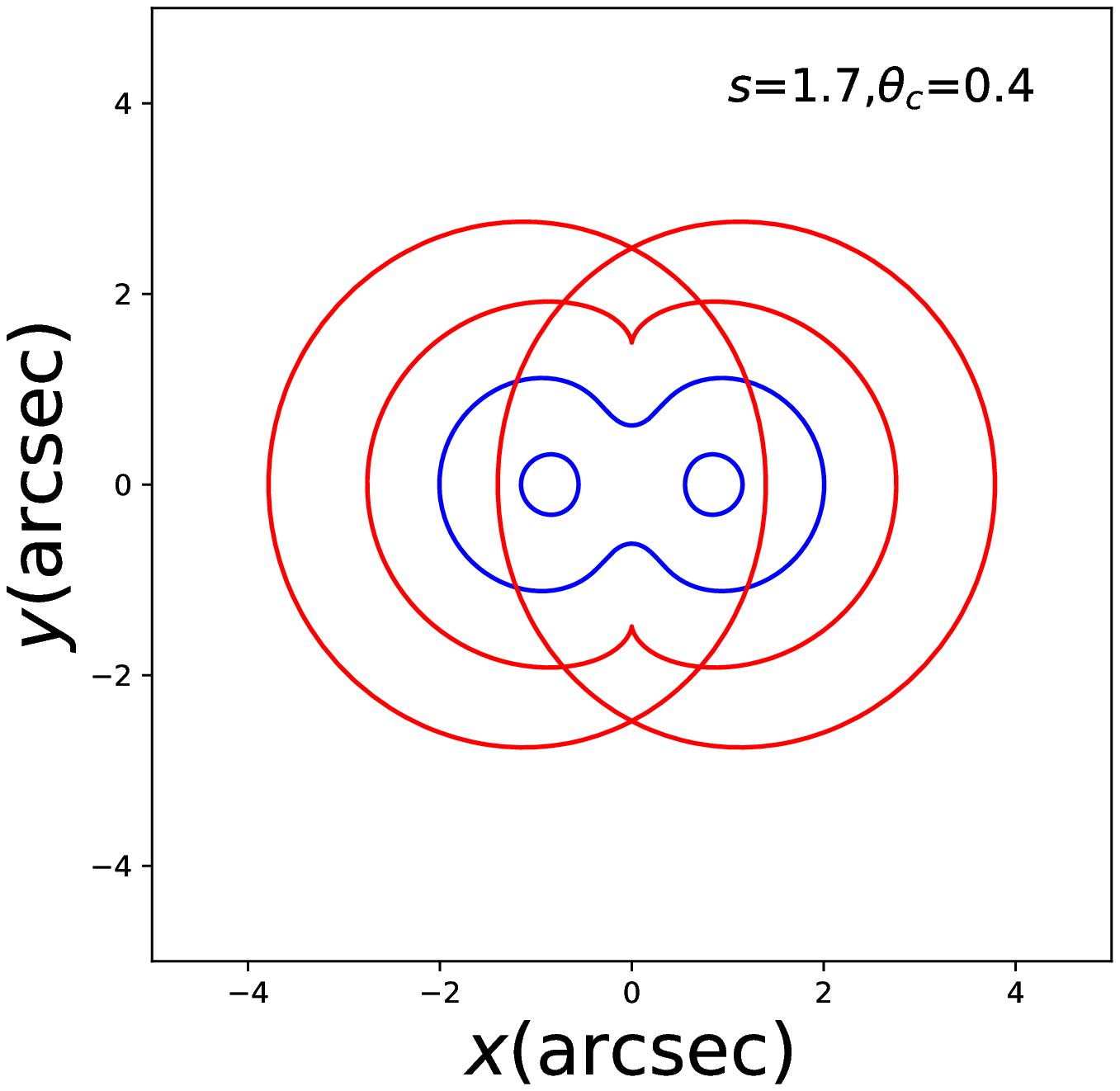}
  \includegraphics[width=4.7cm]{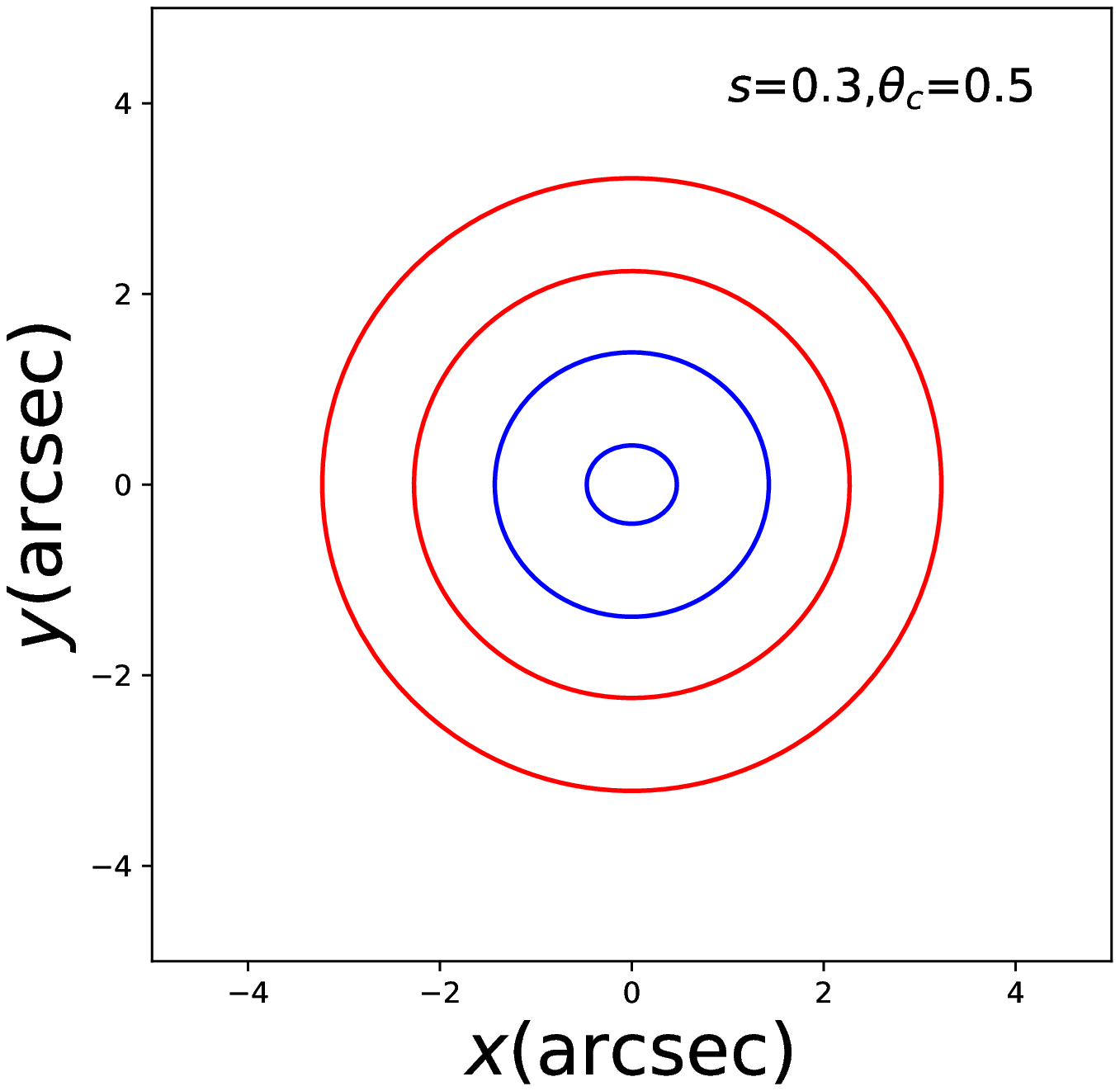}
  \includegraphics[width=4.7cm]{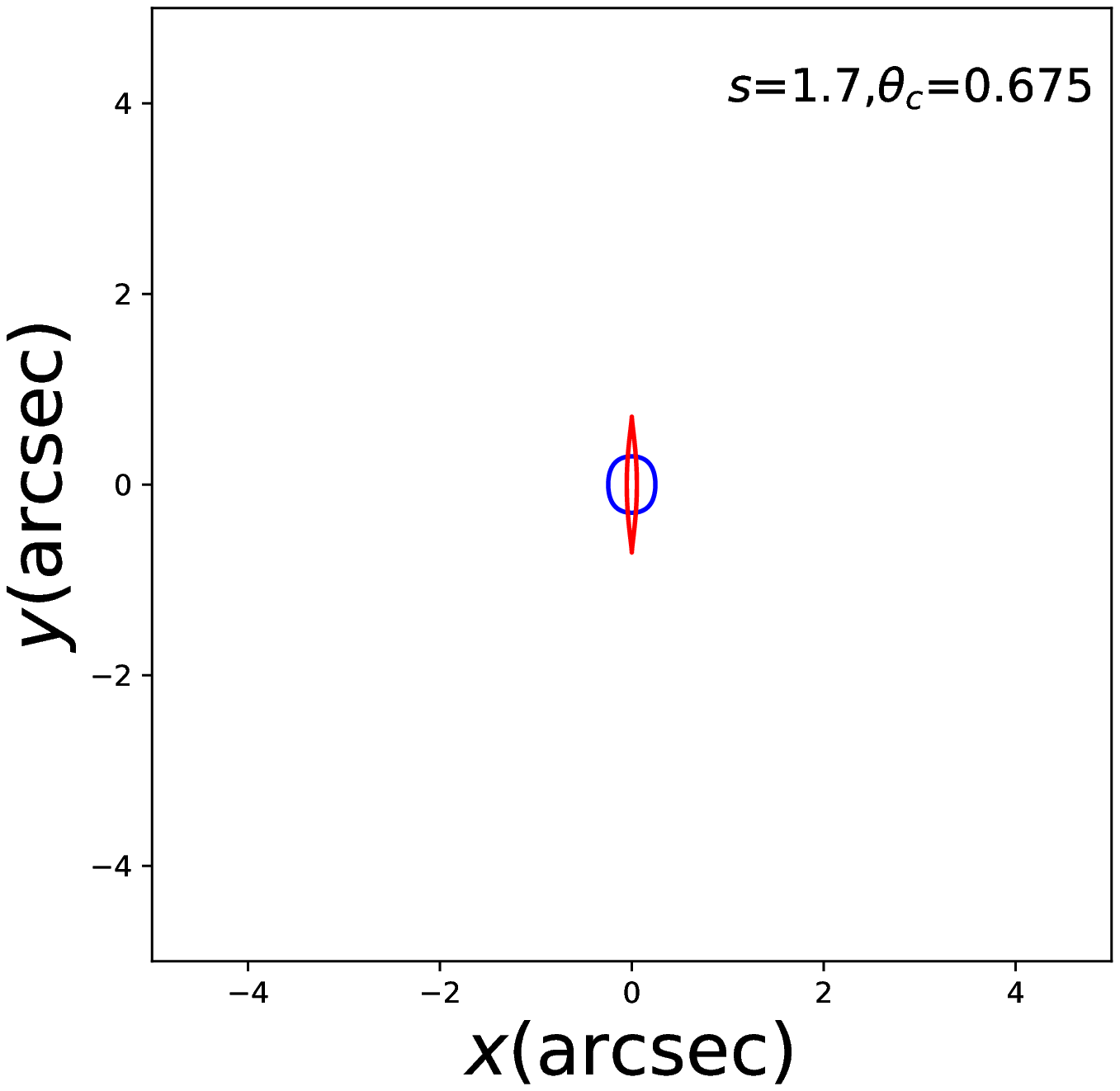}}
  \centerline{\includegraphics[width=4.7cm]{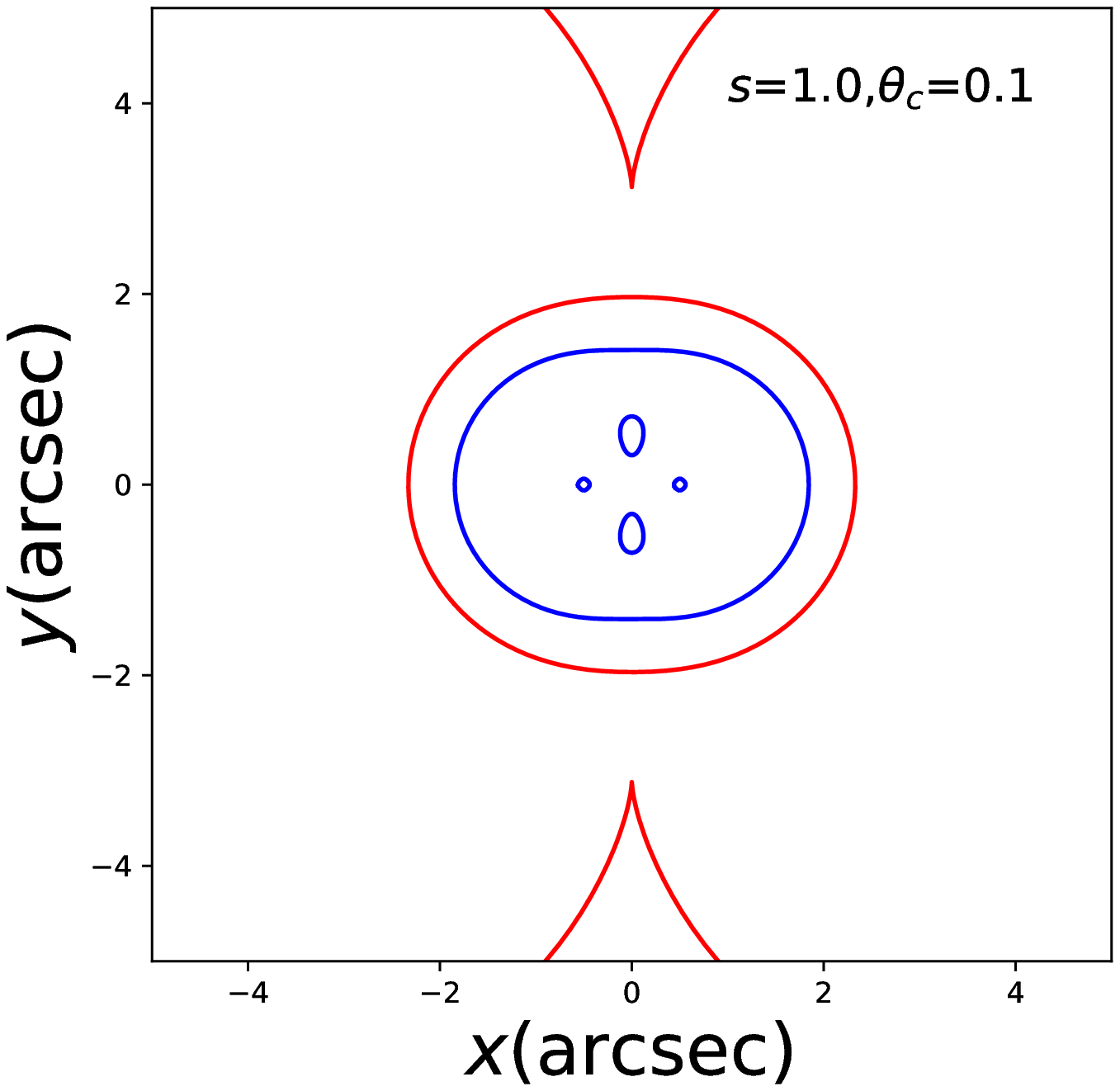}
  \includegraphics[width=4.7cm]{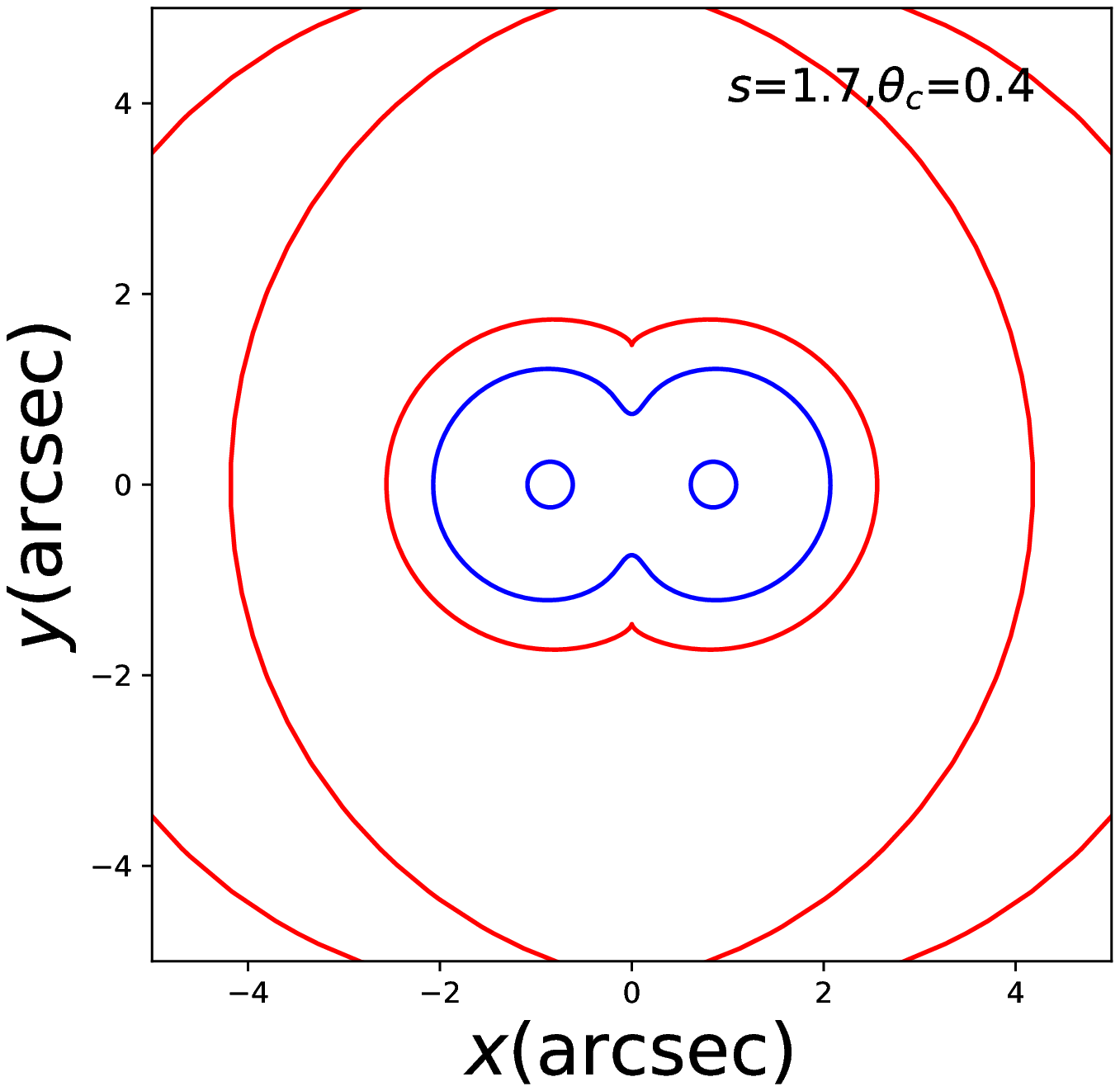}
  \includegraphics[width=4.7cm]{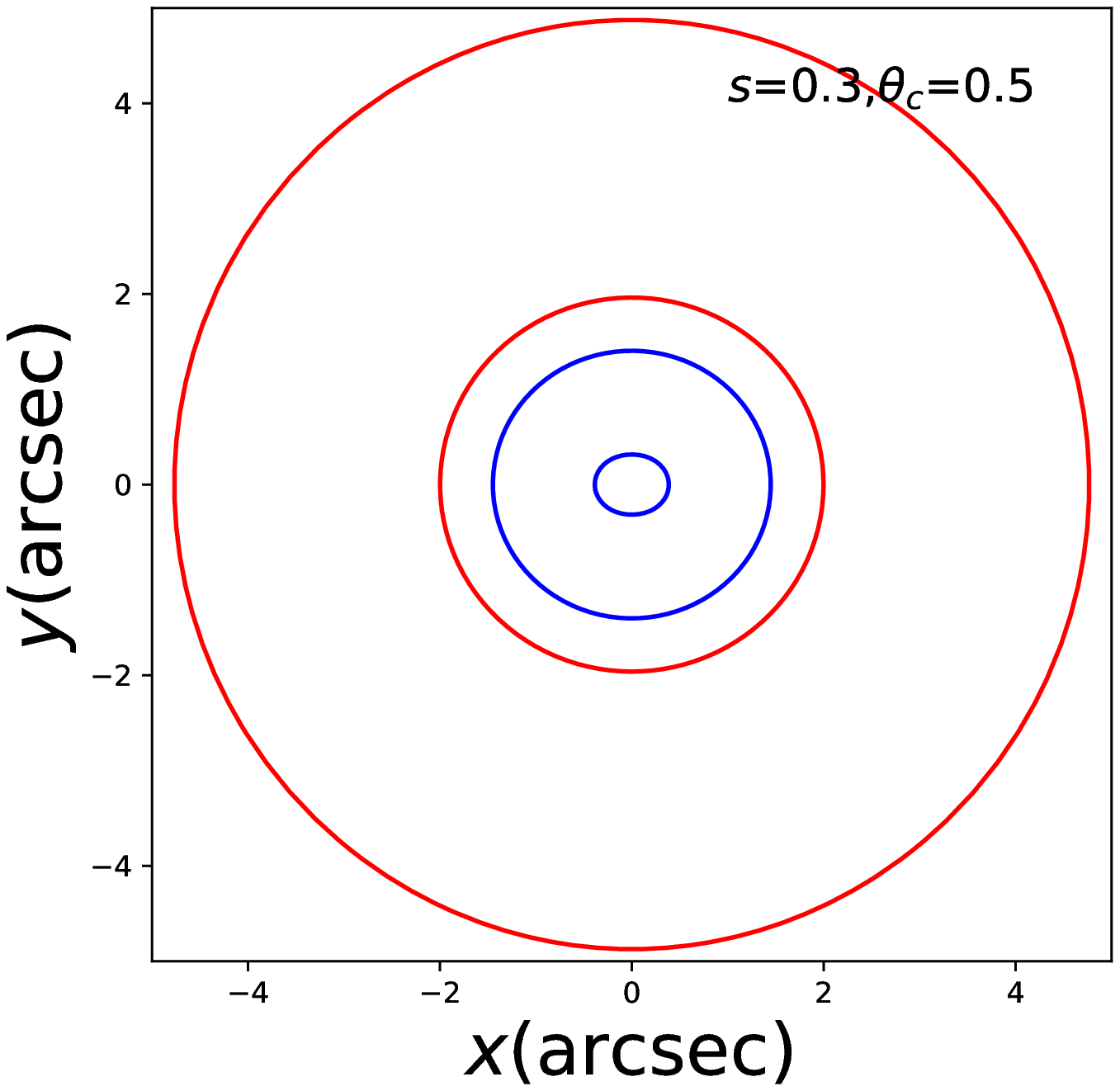}
  \includegraphics[width=4.7cm]{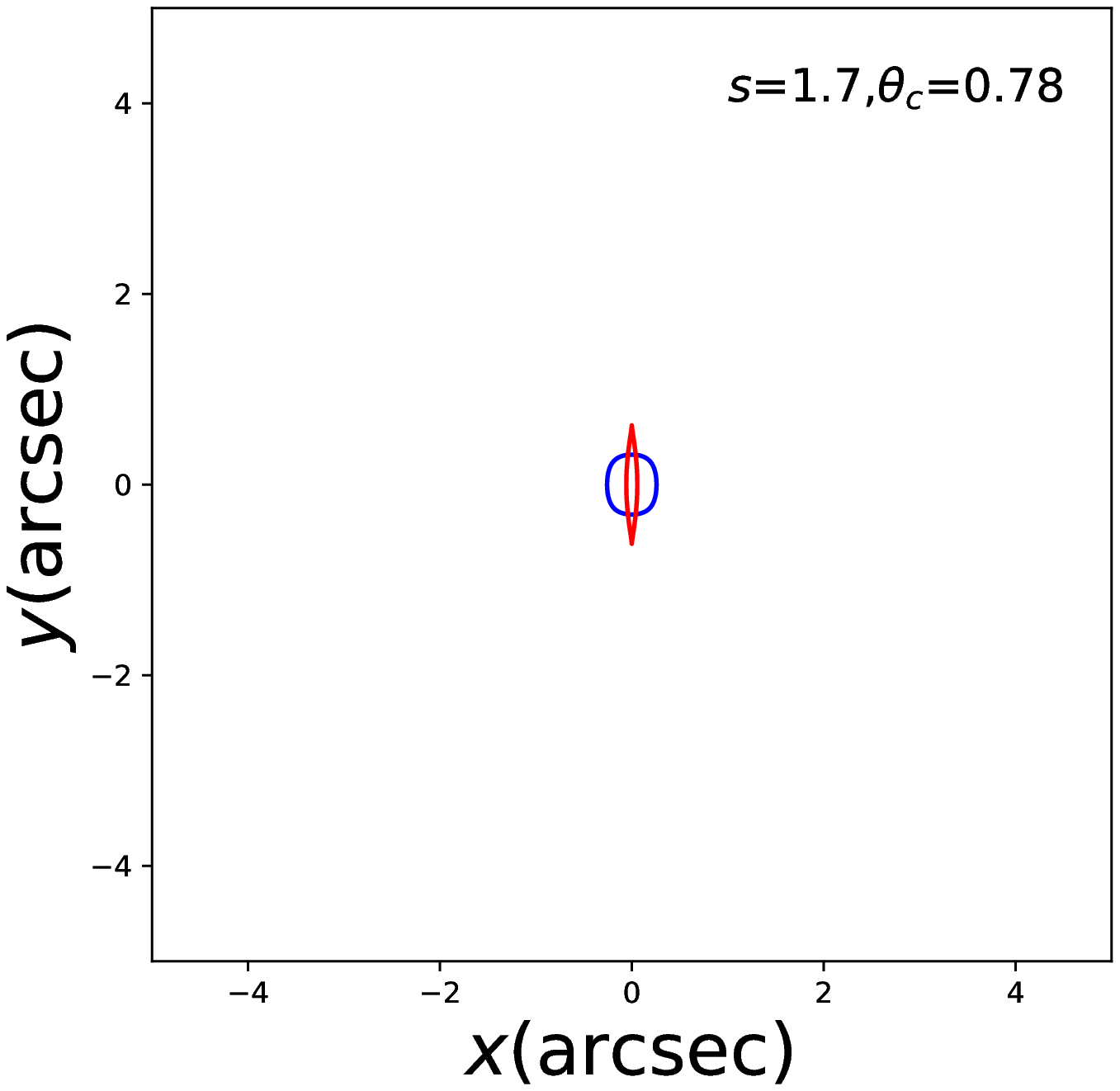}}
  \caption{The critical curve (blue) and caustics (red) of the dual
SPL lens with equal strength components $\theta_{01}=\theta_{02}$. The top row has power-index $h=1$, the middle row has $h=2$ and the bottom row has $h=3$ respectively.}
  \label{fig:powte11}
\end{figure*}

\begin{figure*}
  \centerline{\includegraphics[width=3.2cm]{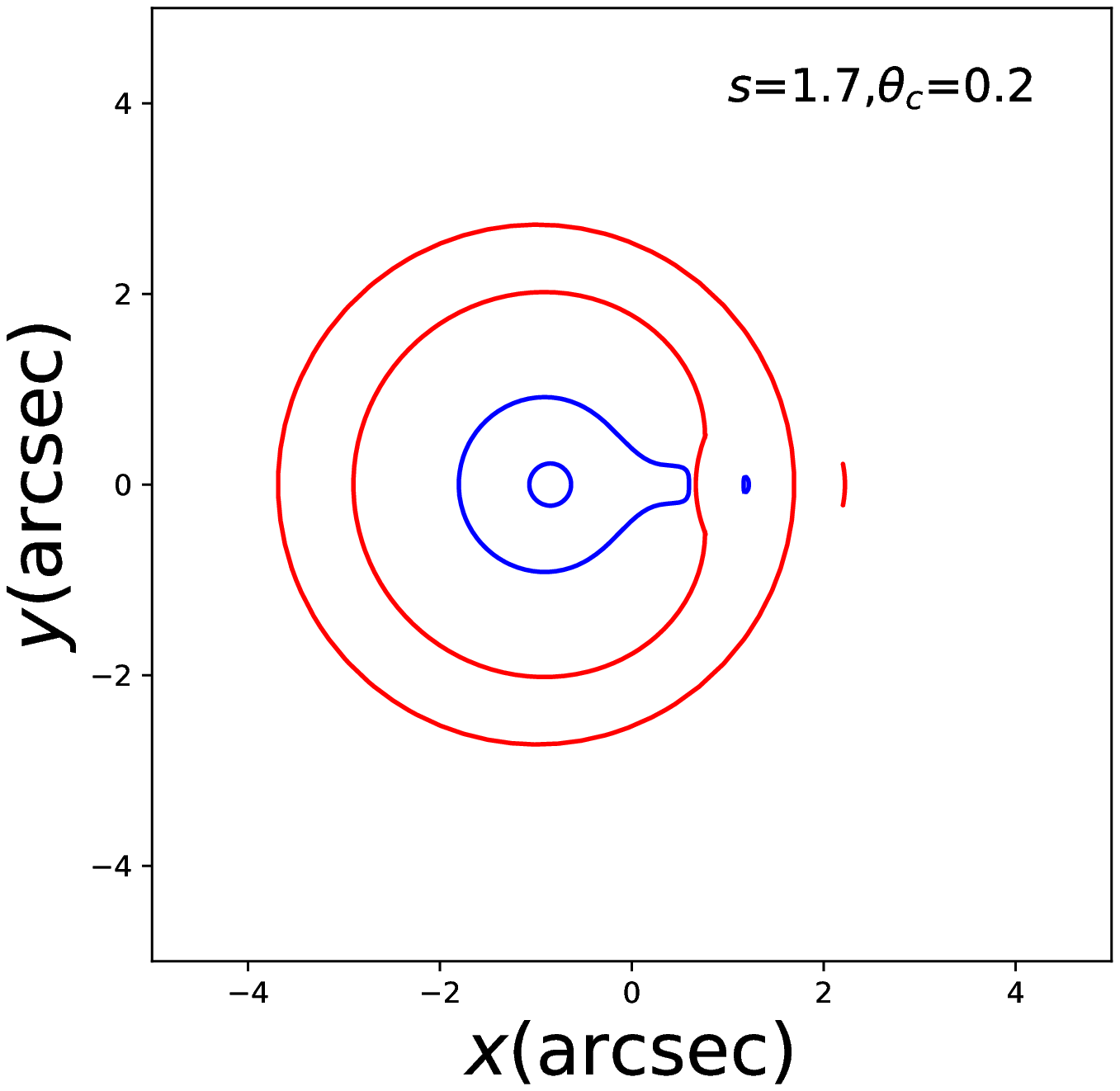}
  \includegraphics[width=3.2cm]{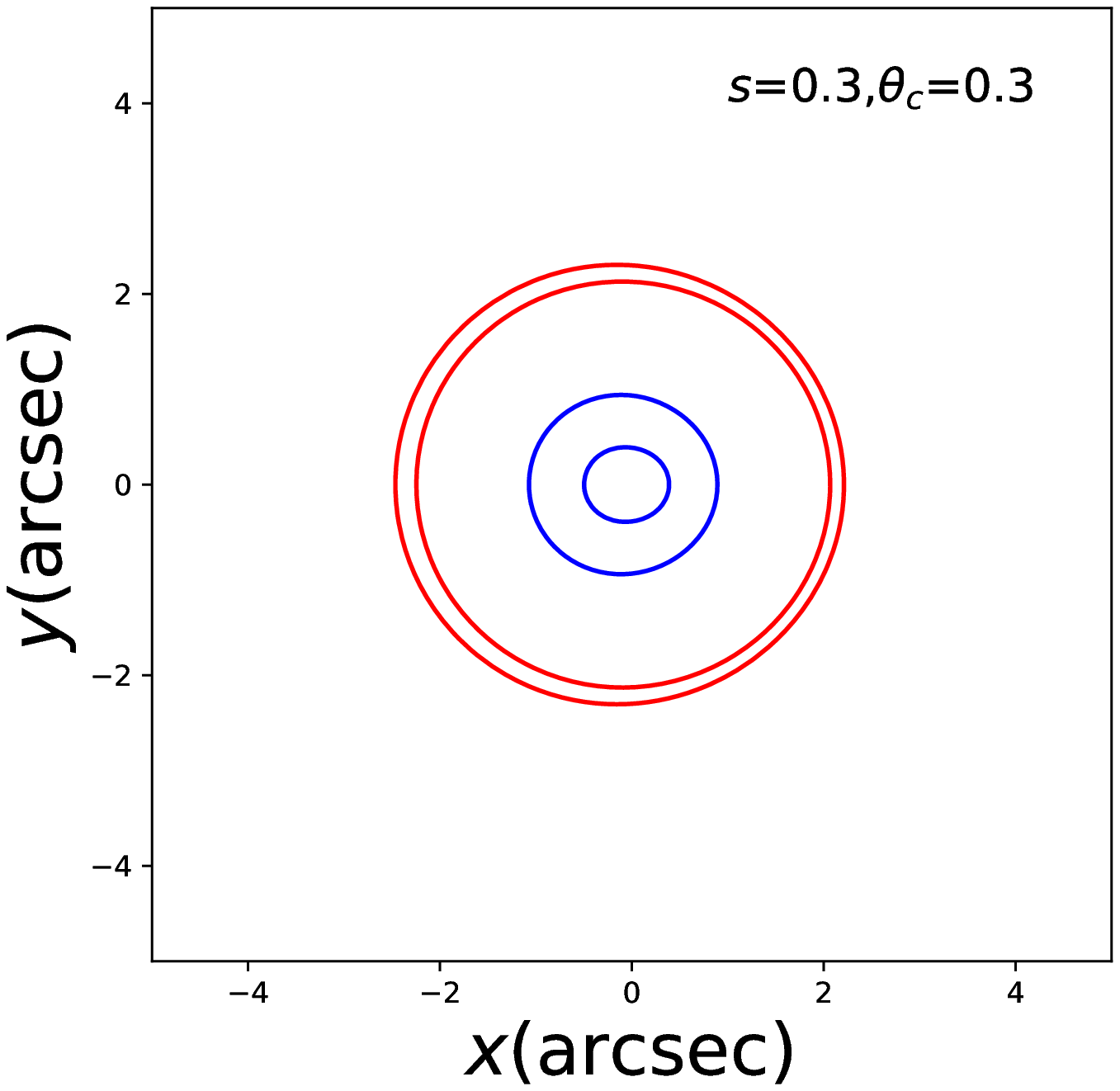}
  \includegraphics[width=3.2cm]{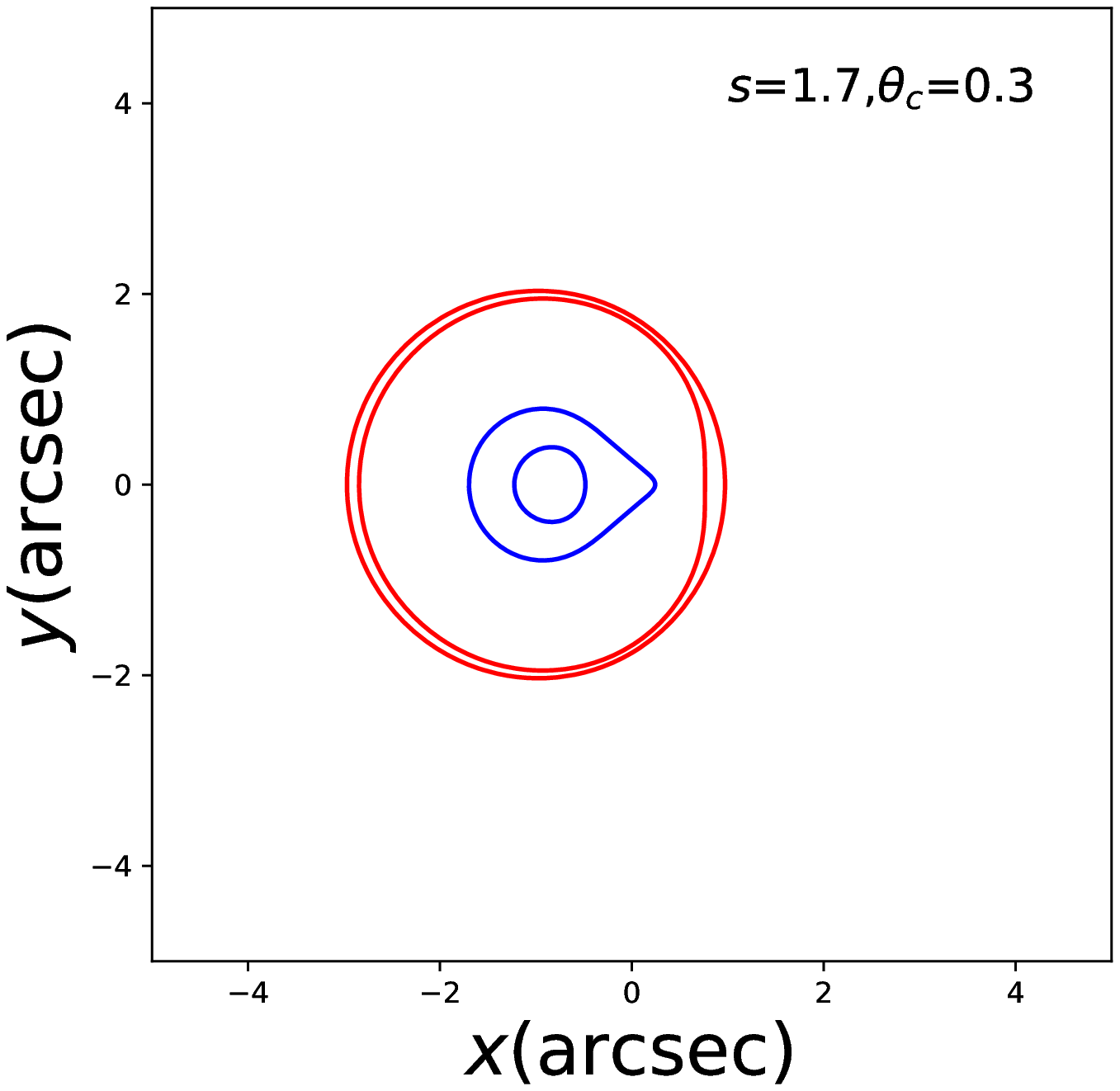}
  \includegraphics[width=3.2cm]{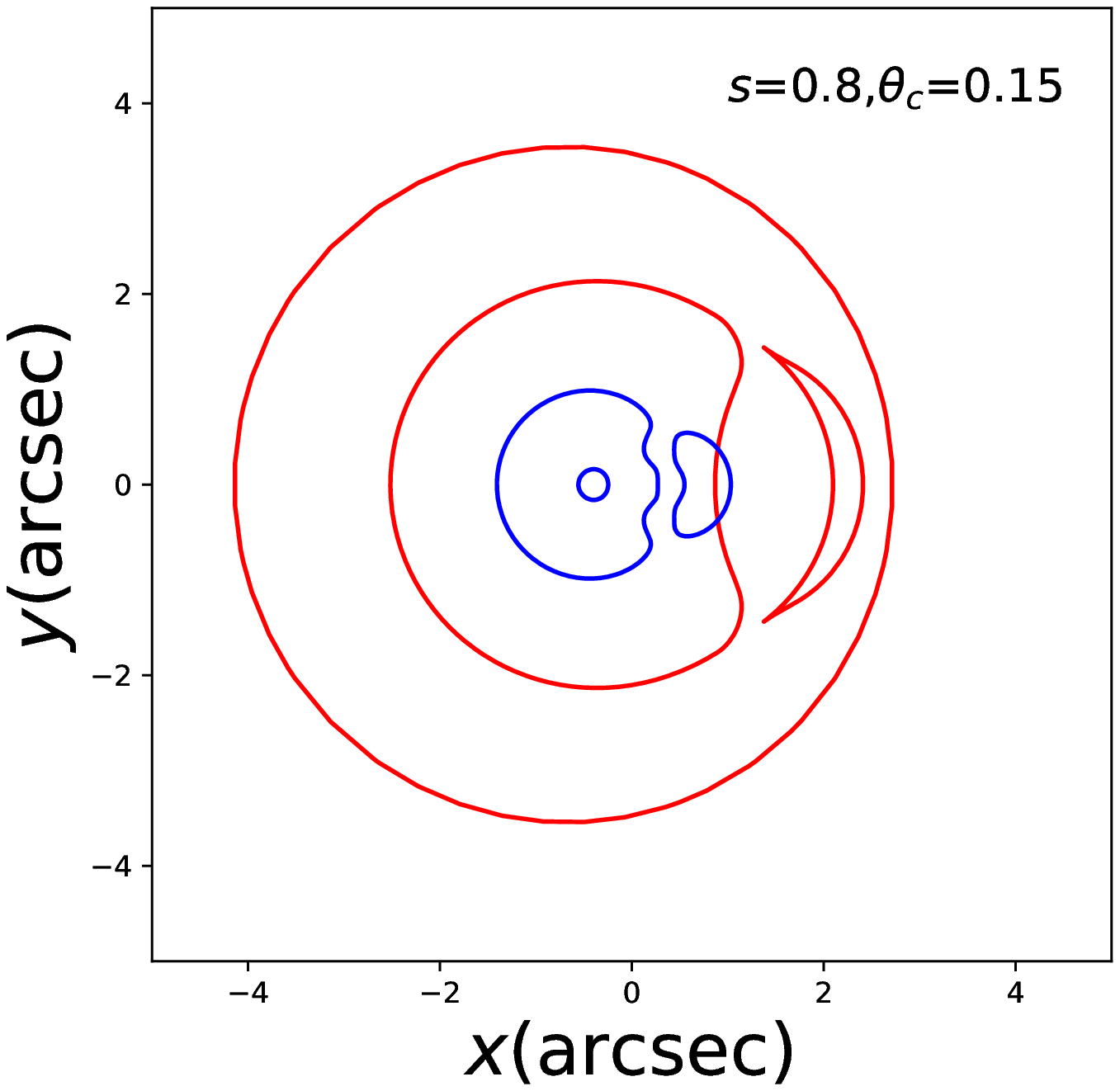}
  \includegraphics[width=3.2cm]{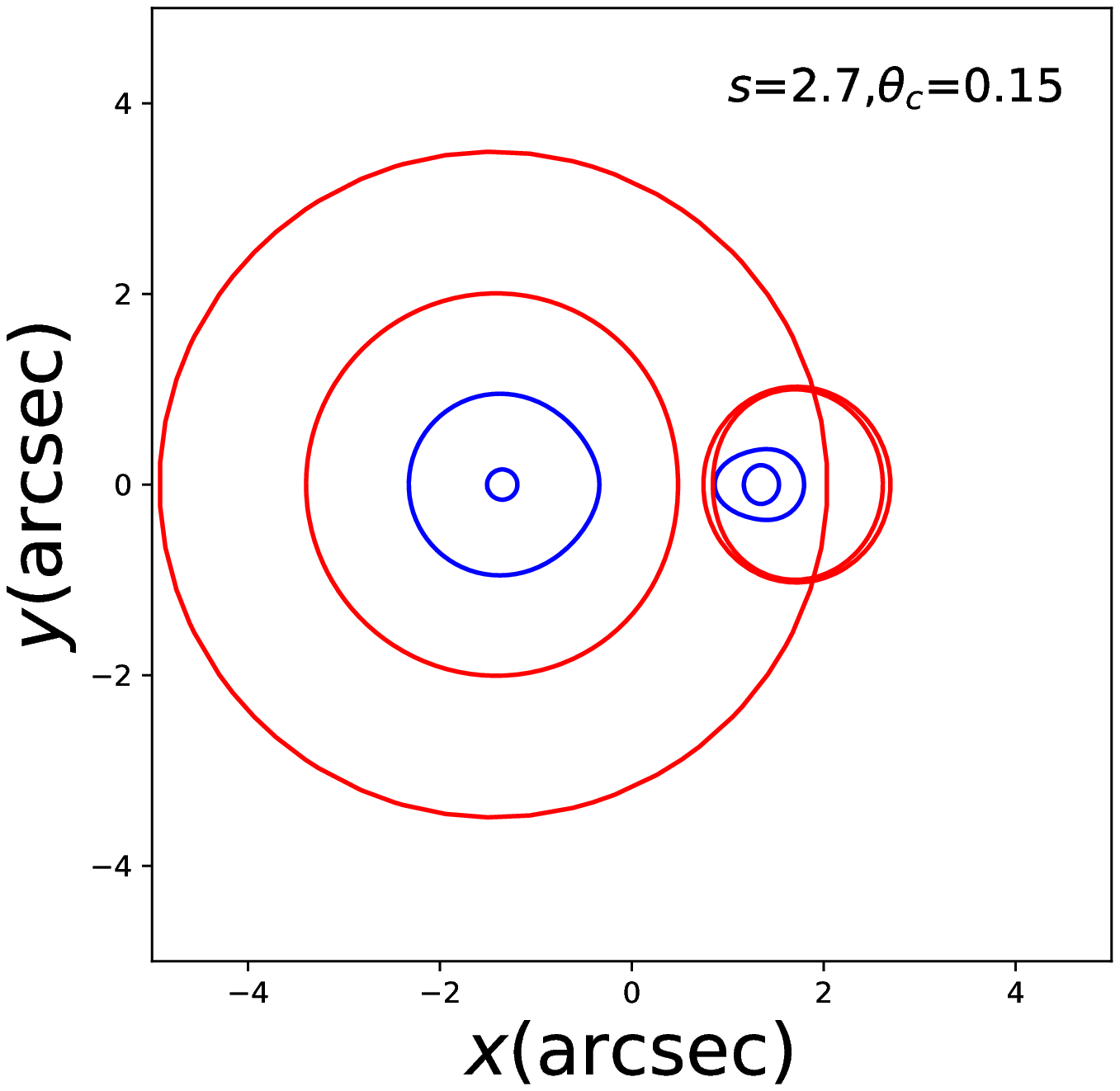}
  \includegraphics[width=3.2cm]{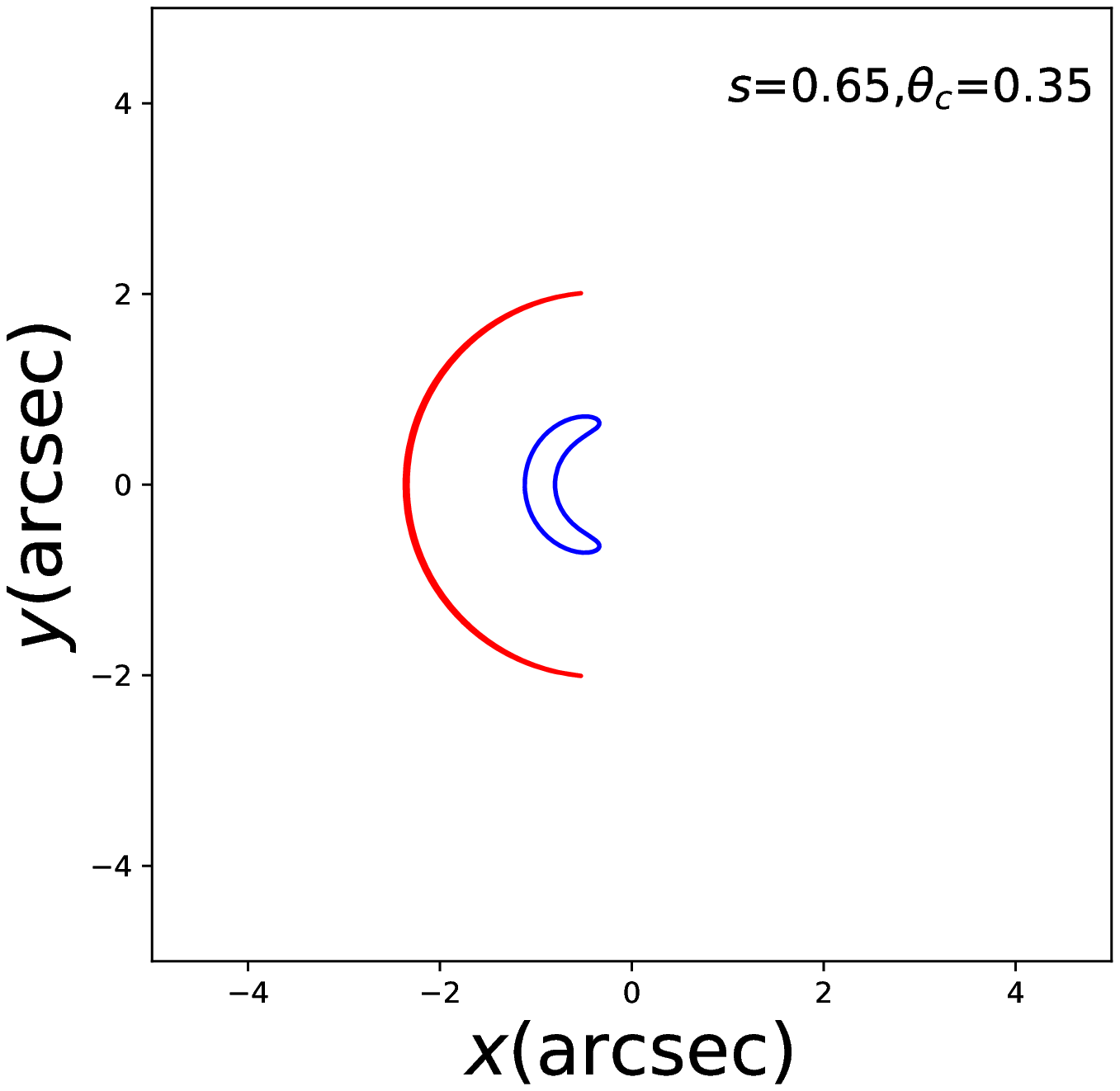}}
  \centerline{\includegraphics[width=3.2cm]{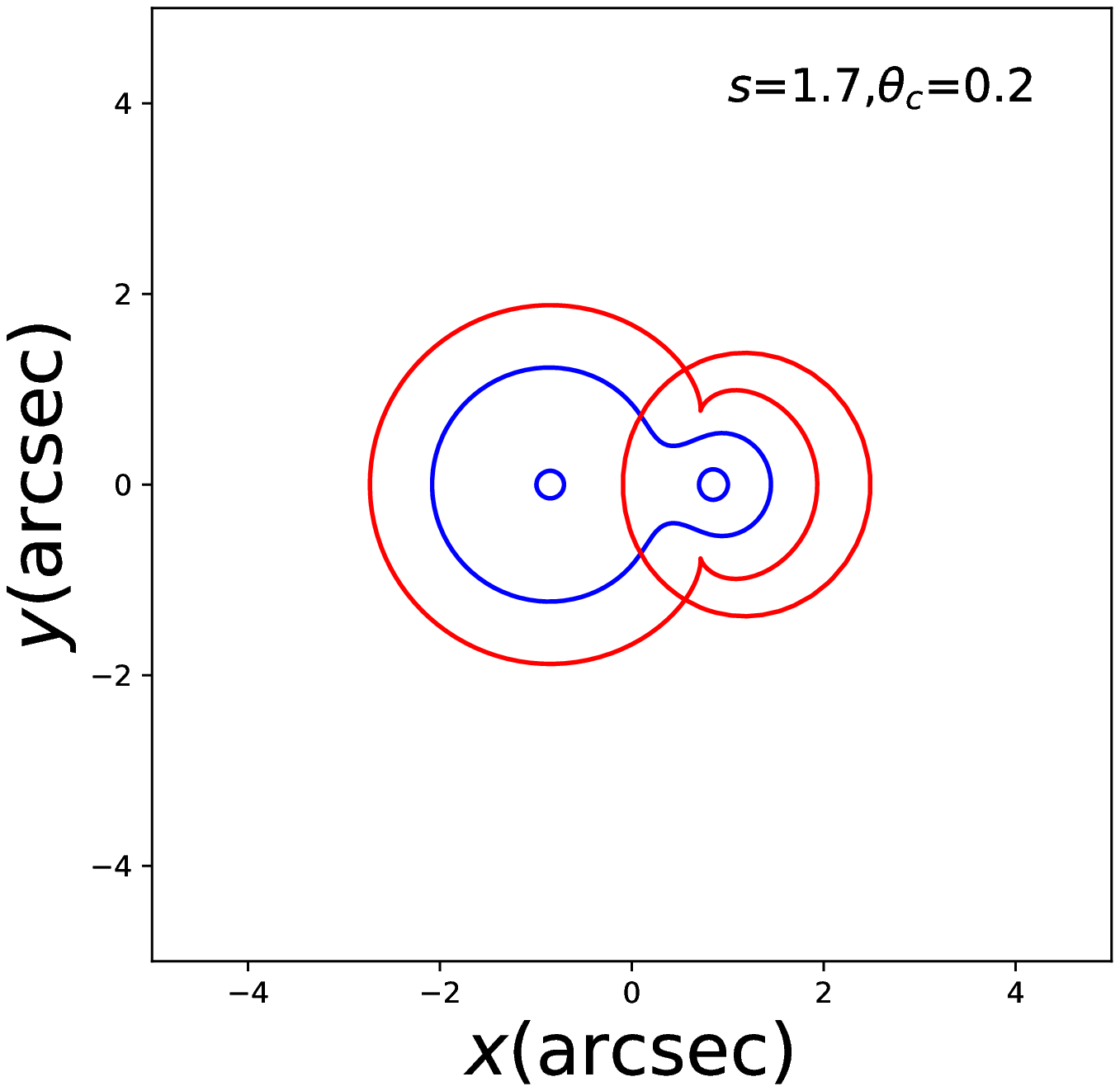}
  \includegraphics[width=3.2cm]{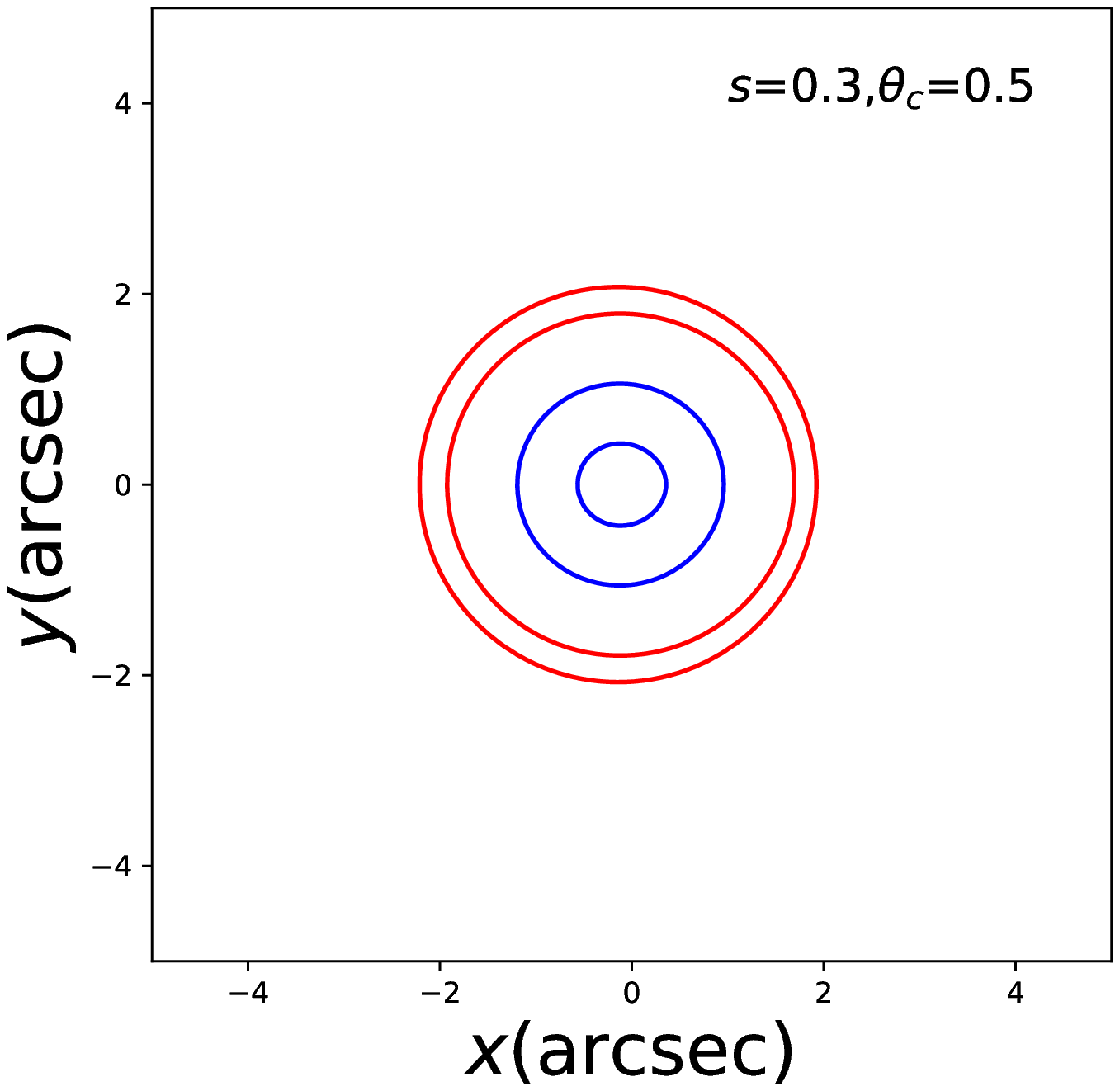}
  \includegraphics[width=3.2cm]{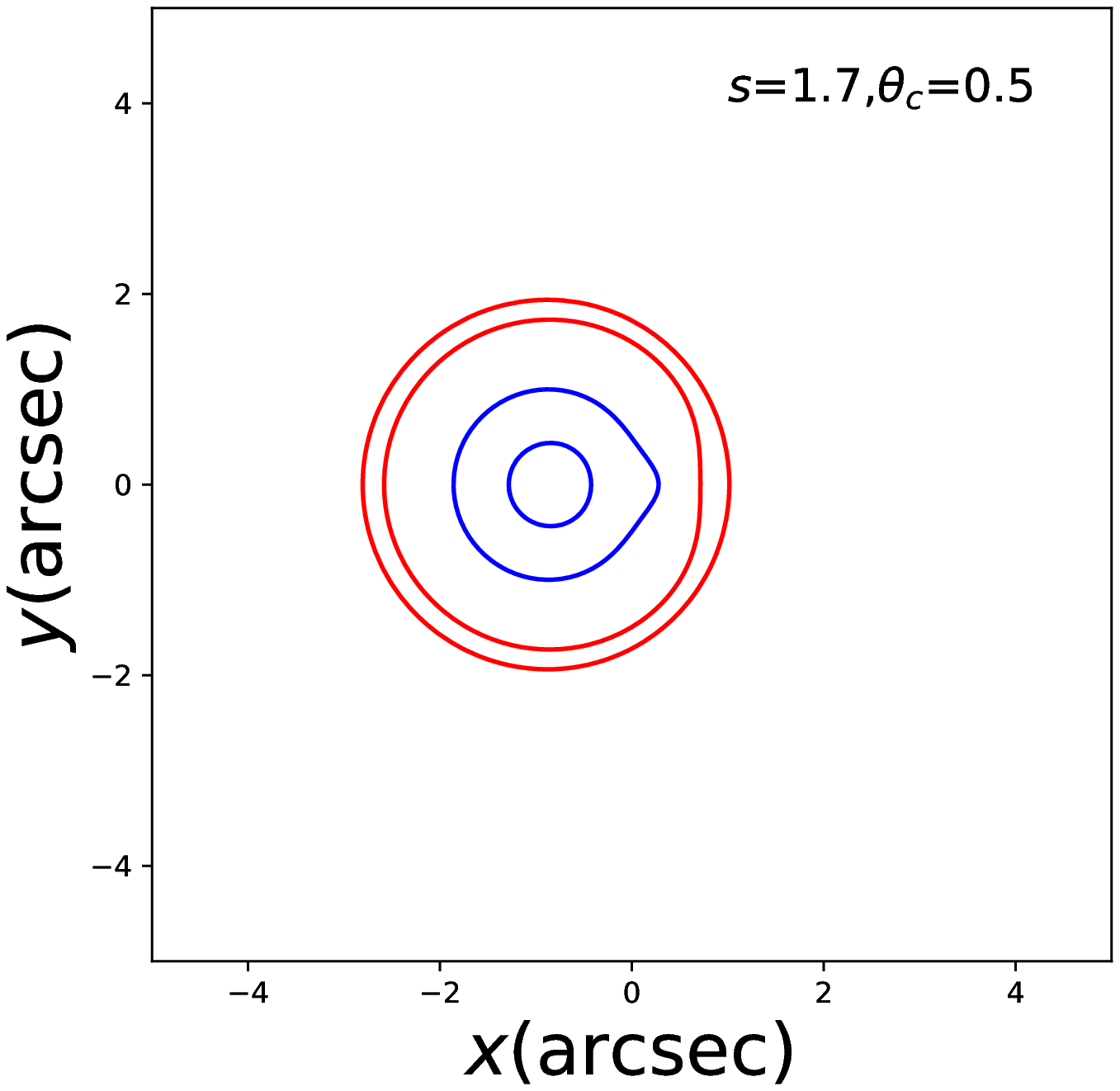}
  \includegraphics[width=3.2cm]{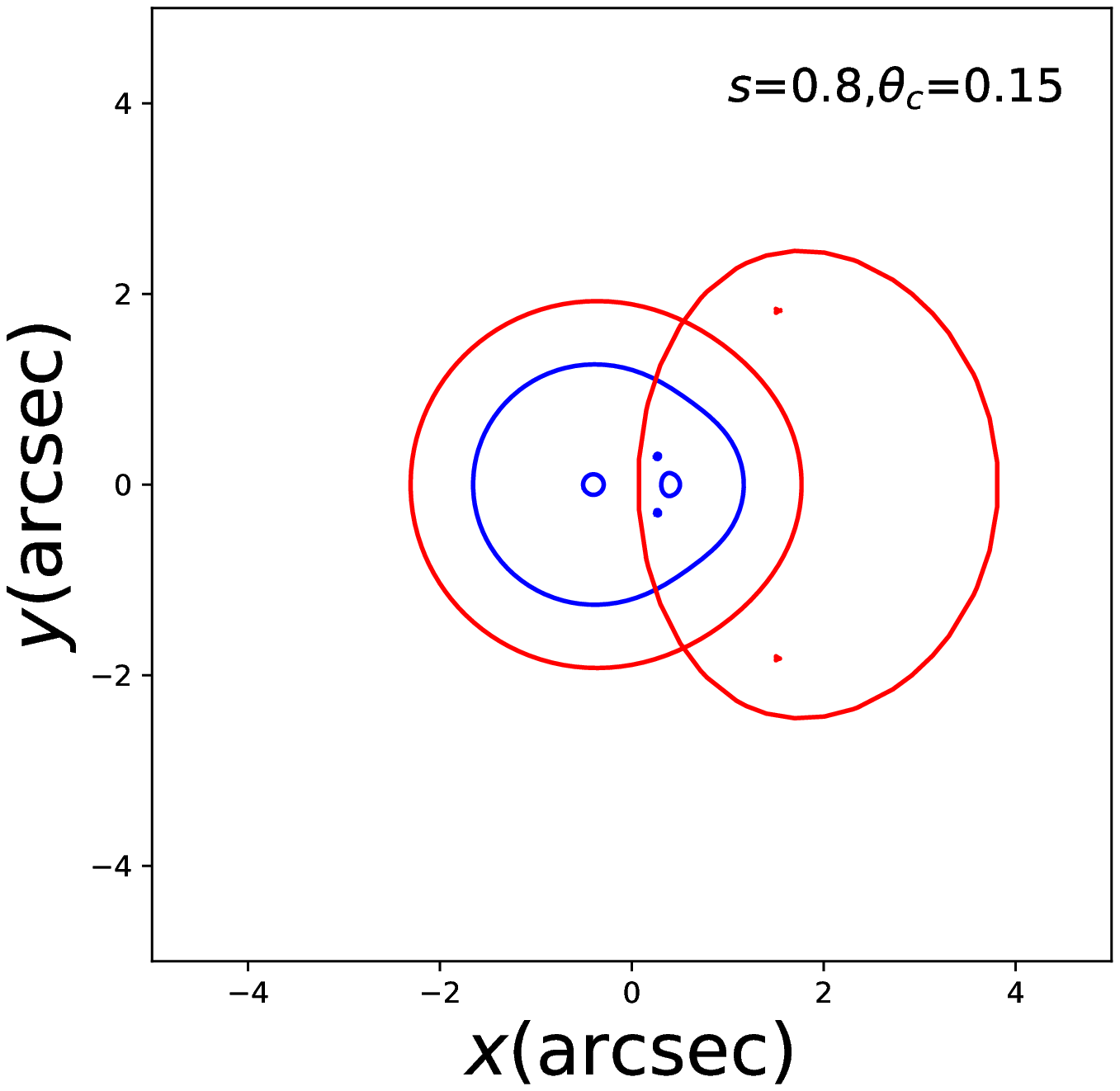}
  \includegraphics[width=3.2cm]{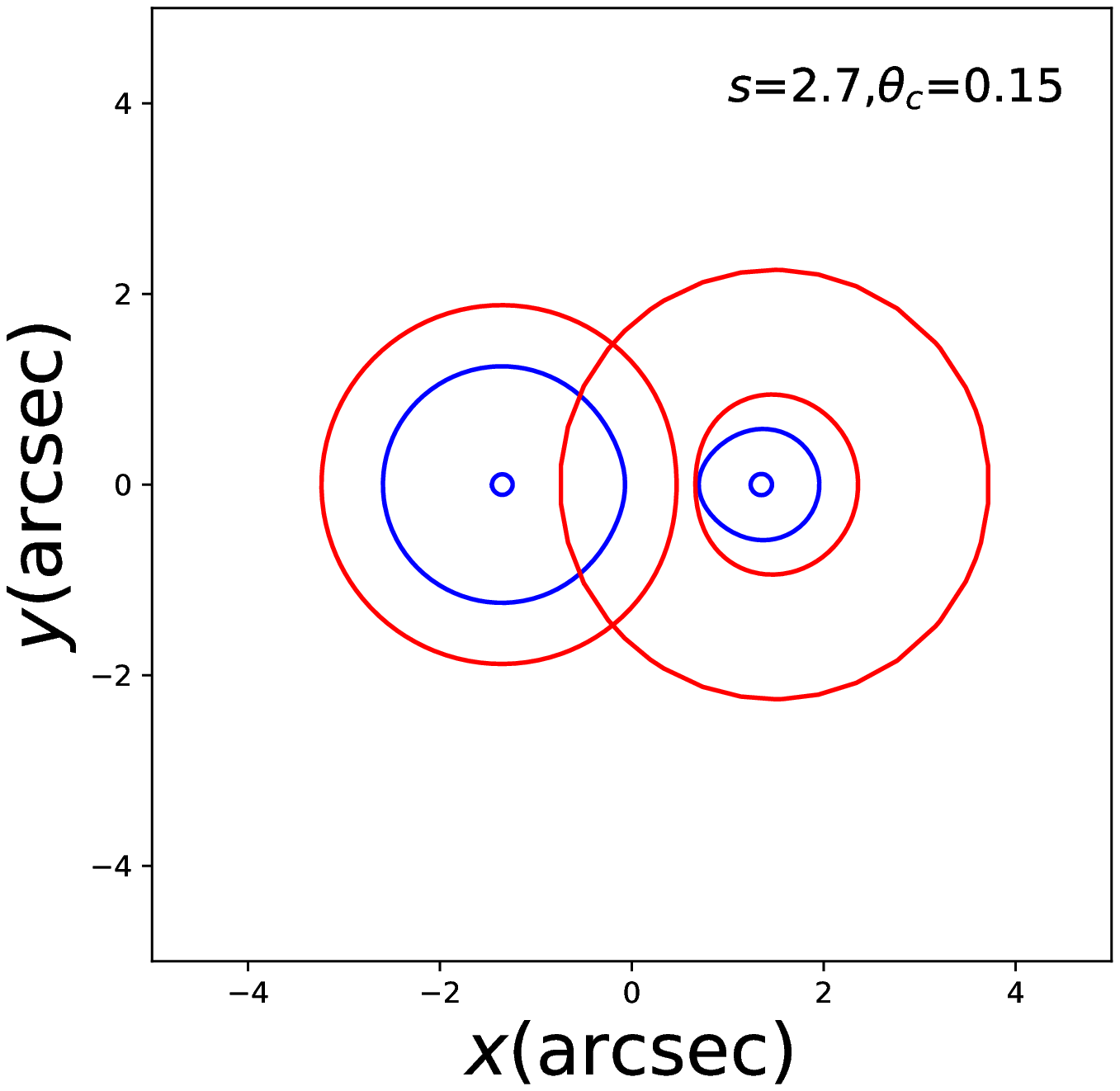}
  \includegraphics[width=3.2cm]{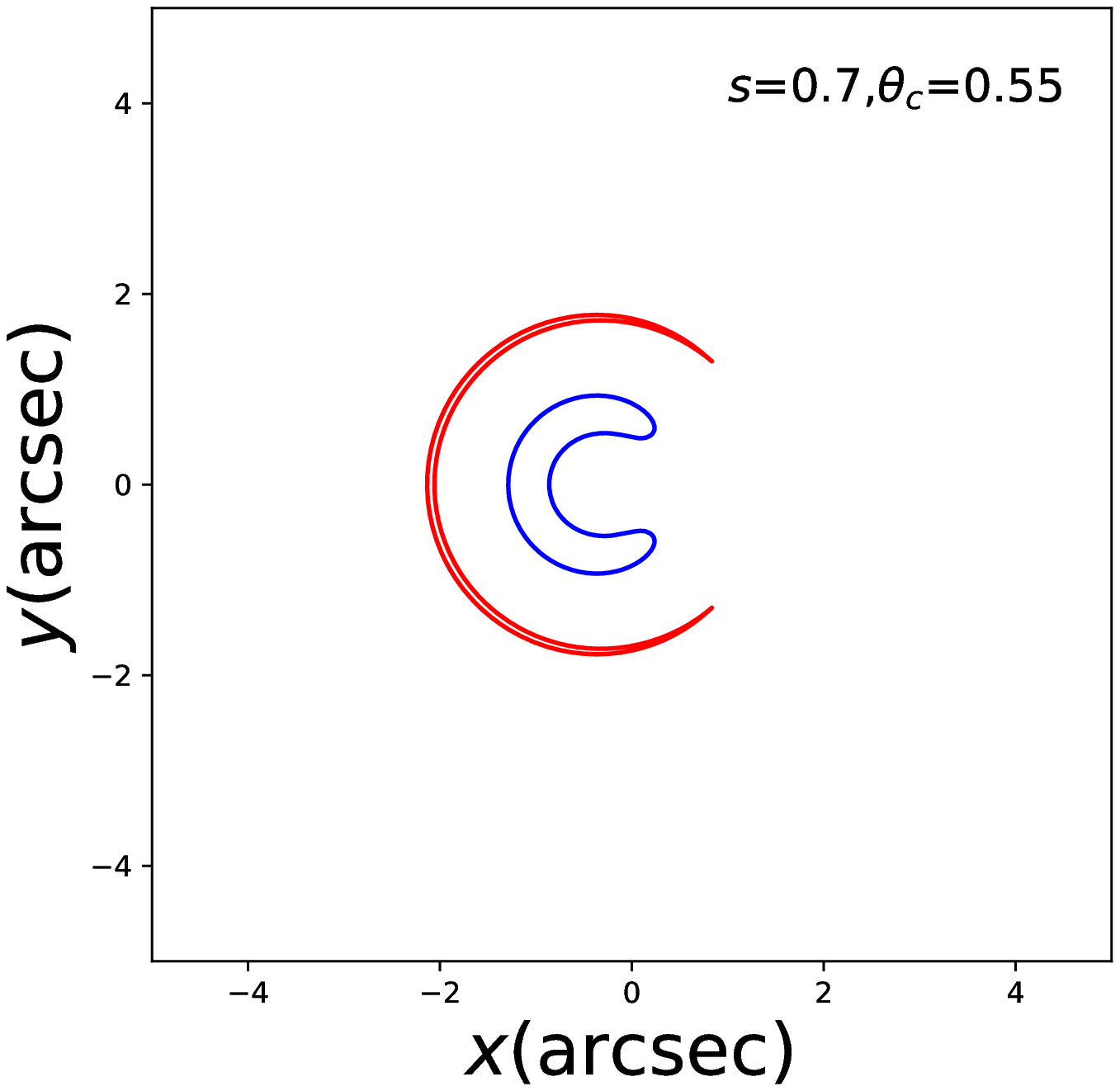}}
  \centerline{\includegraphics[width=3.2cm]{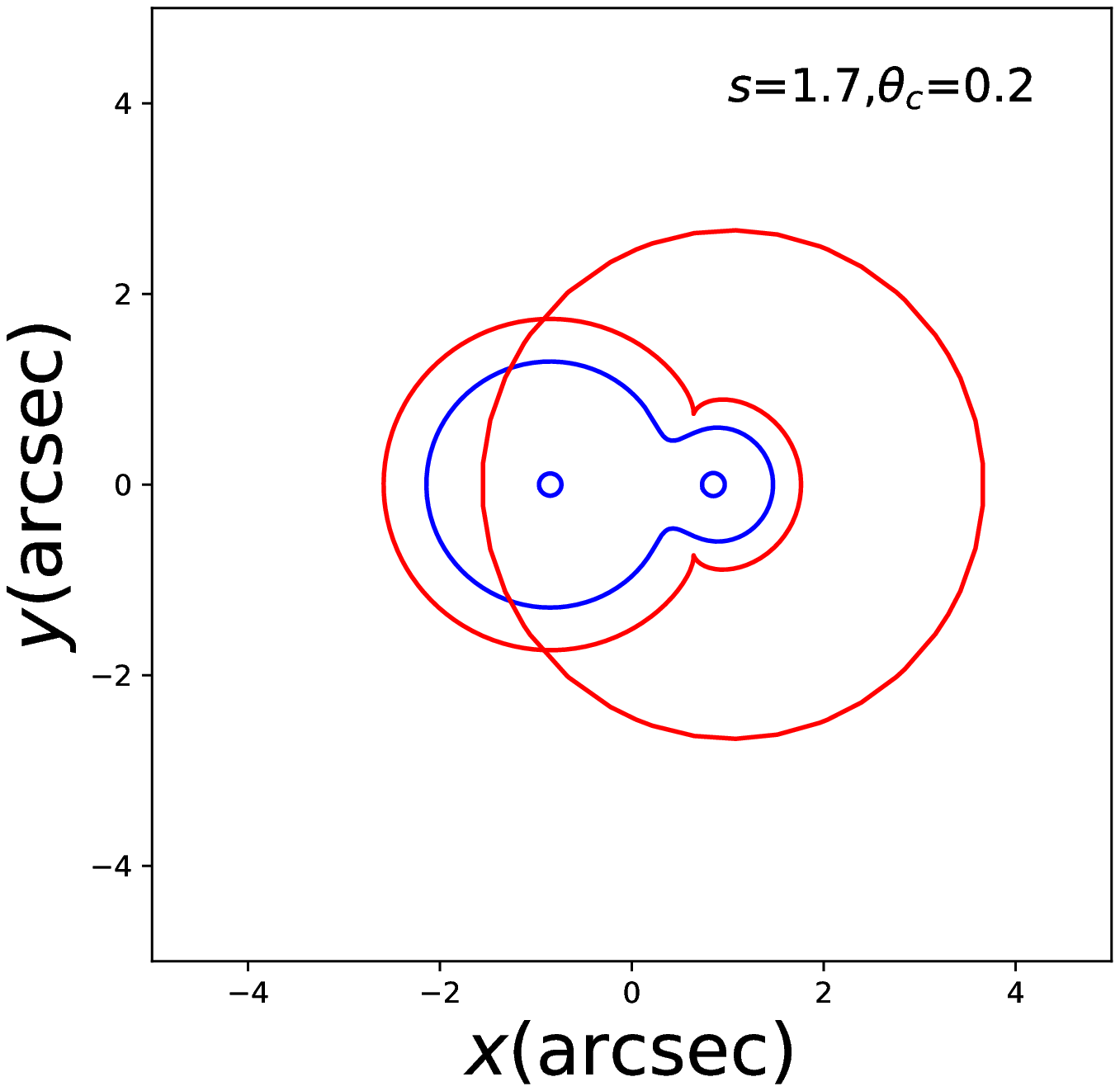}
  \includegraphics[width=3.2cm]{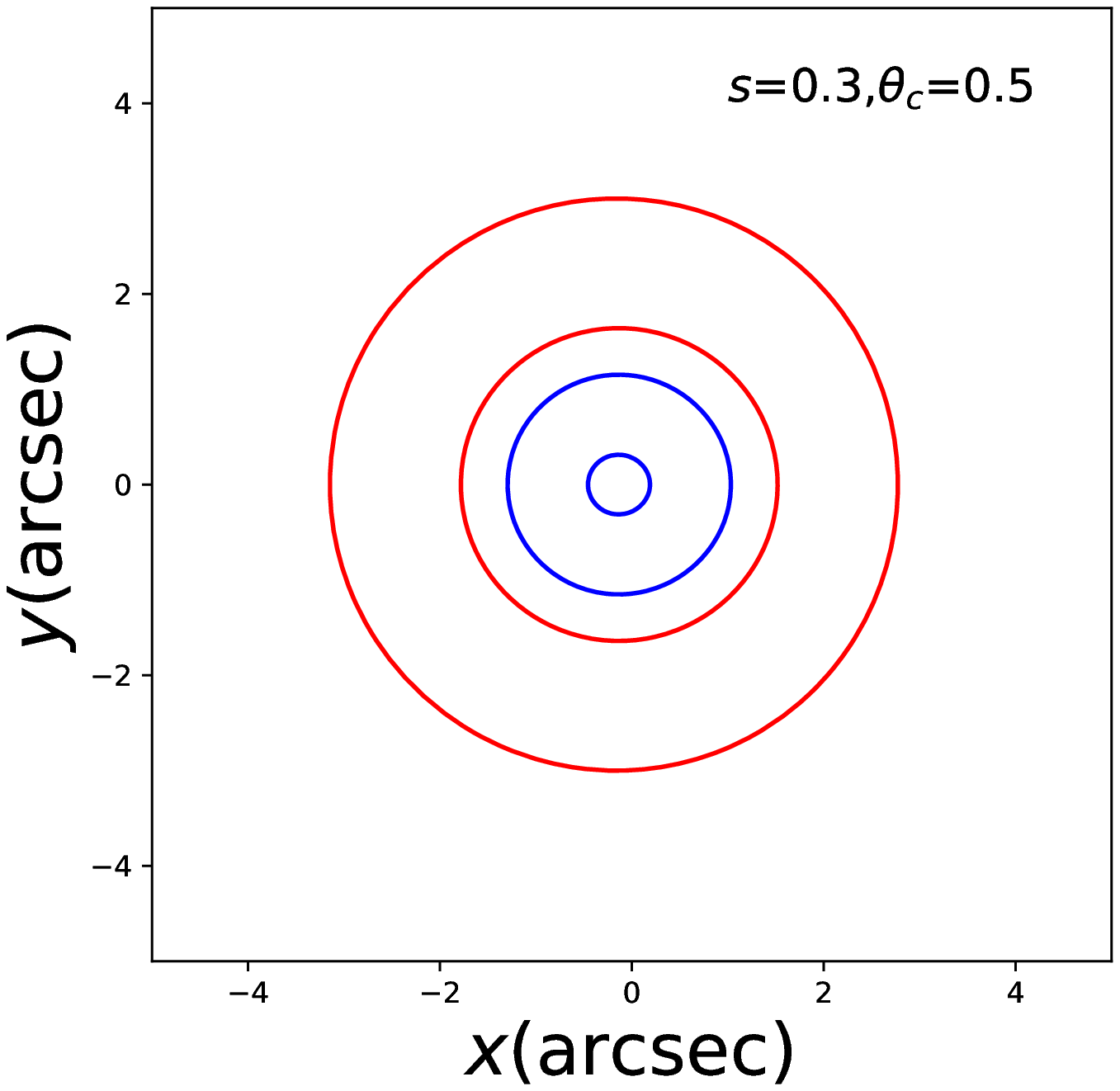}
  \includegraphics[width=3.2cm]{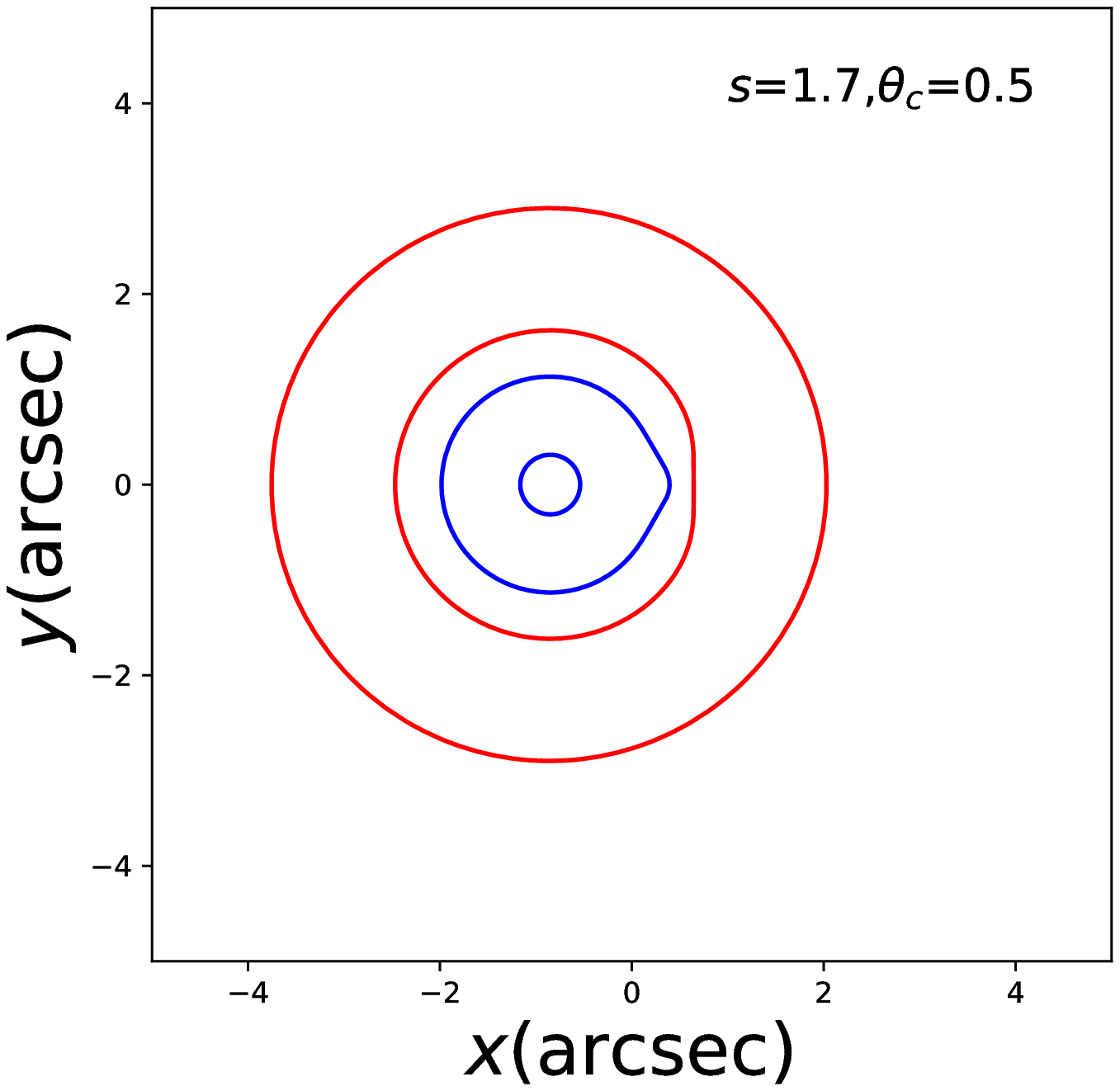}
  \includegraphics[width=3.2cm]{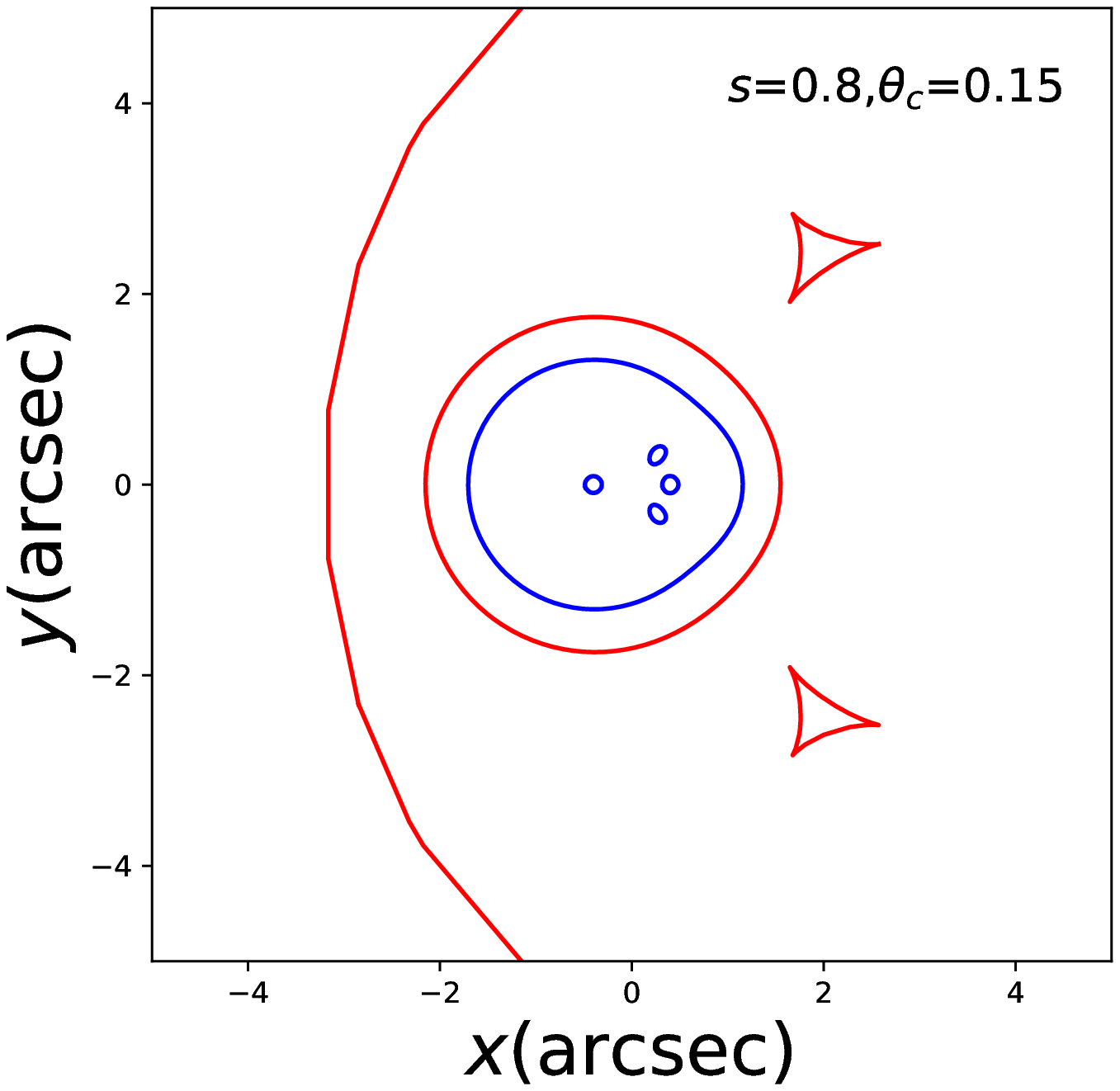}
  \includegraphics[width=3.2cm]{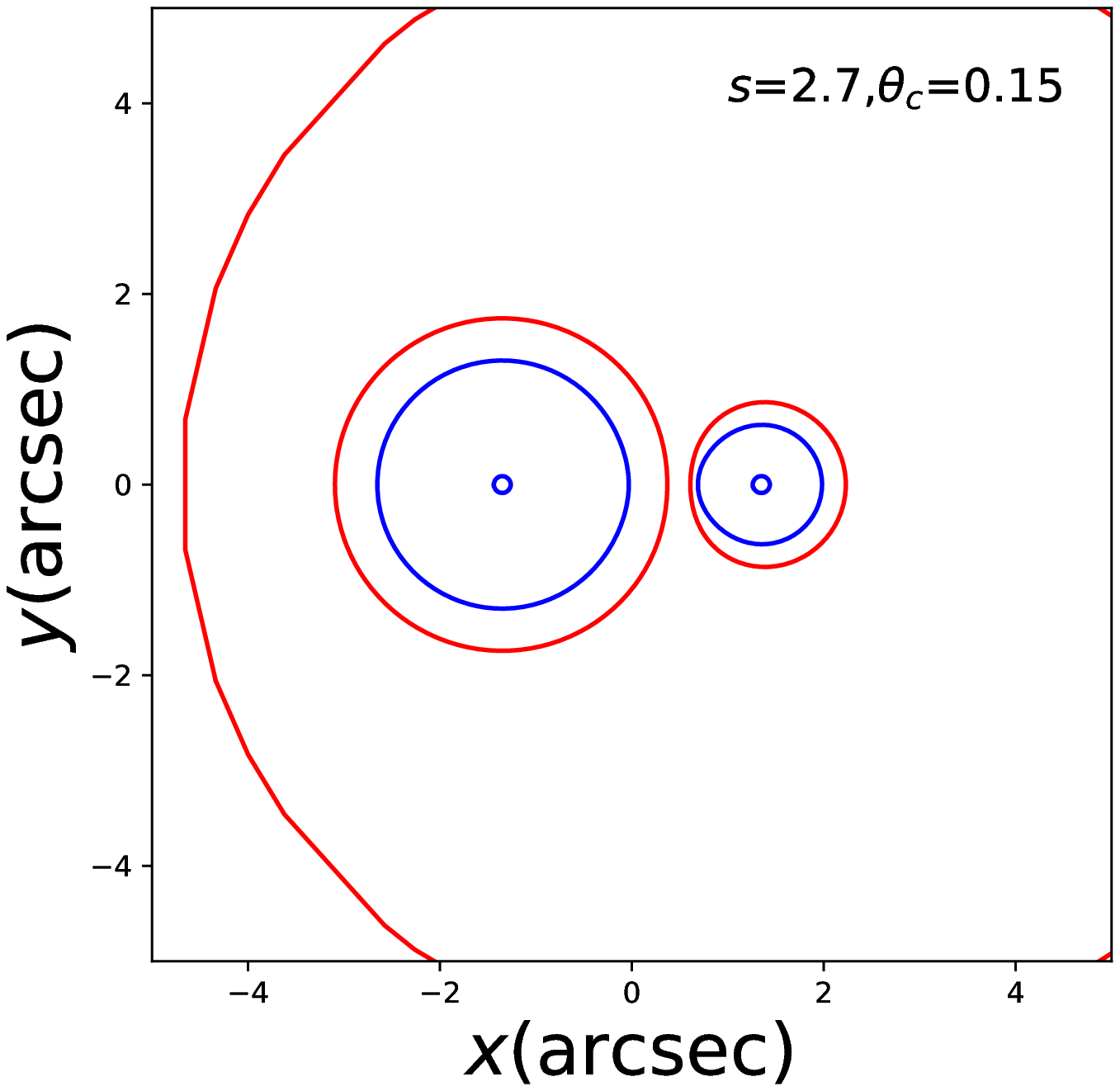}
  \includegraphics[width=3.2cm]{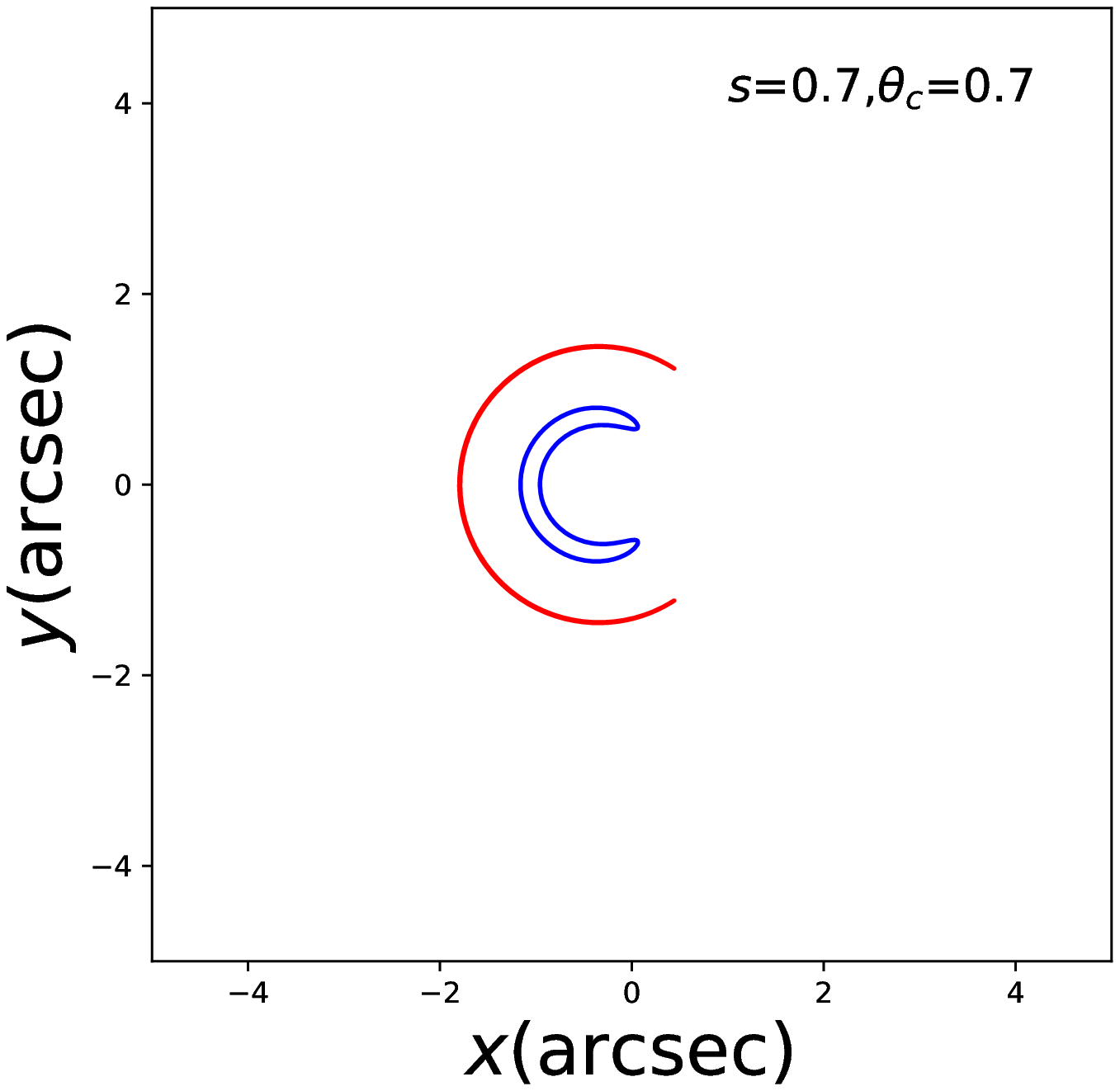}}
  \caption{A collection of criticals (blue) and caustics (red) for SPL lenses with $\theta_{01}=1$ and $\theta_{02}=0.5$. Examples with power-index $h=1$ is on the top row, $h=2$ on the second row, and $h=3$ on the third row. The value of $s$ and $\theta_\text{c}$ for each plot is labelled.}
  \label{fig:powte13}
\end{figure*}

\begin{figure*}
  \centerline{\includegraphics[width=4.0cm]{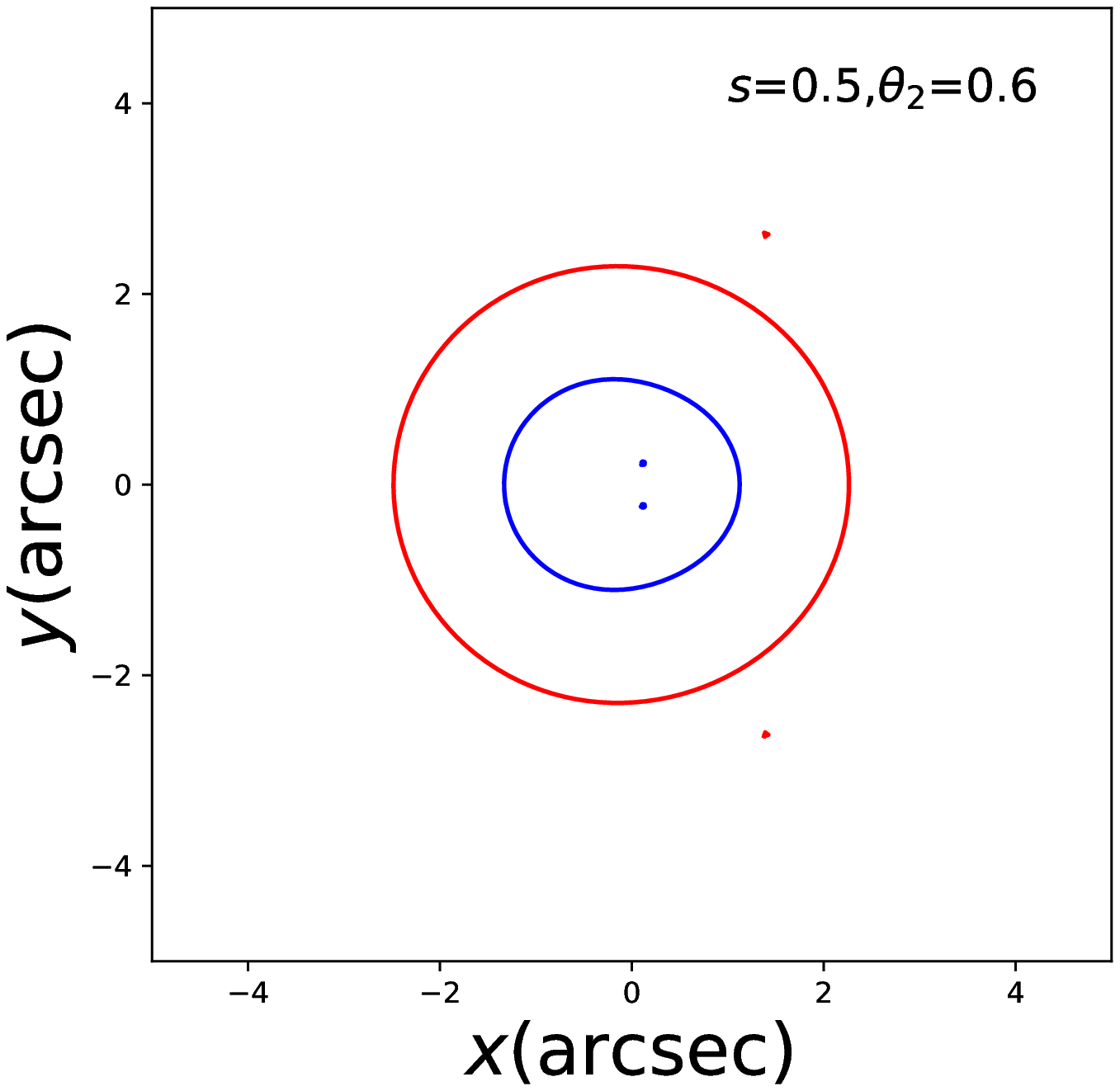}
  \includegraphics[width=4.0cm]{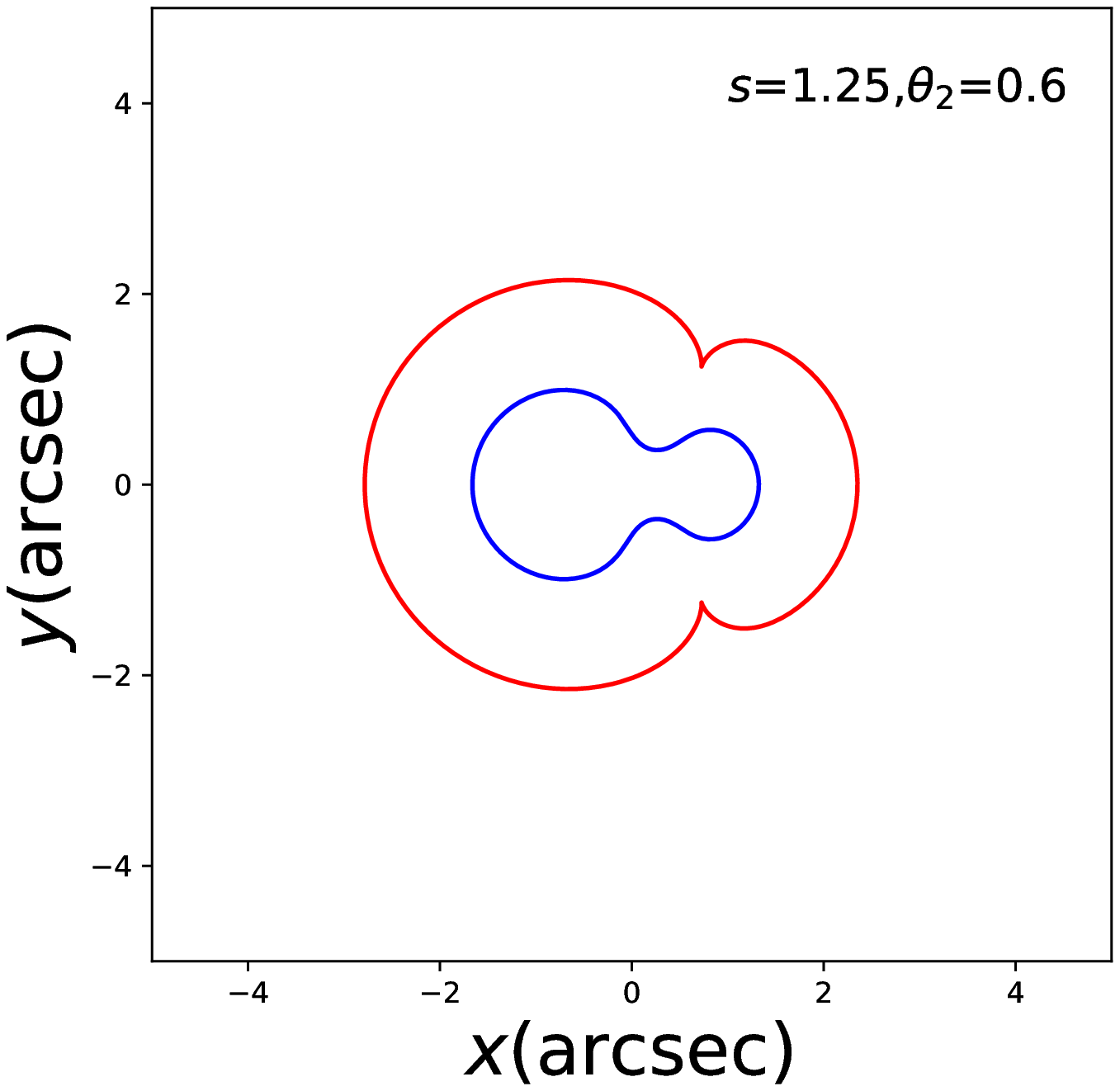}
  \includegraphics[width=4.0cm]{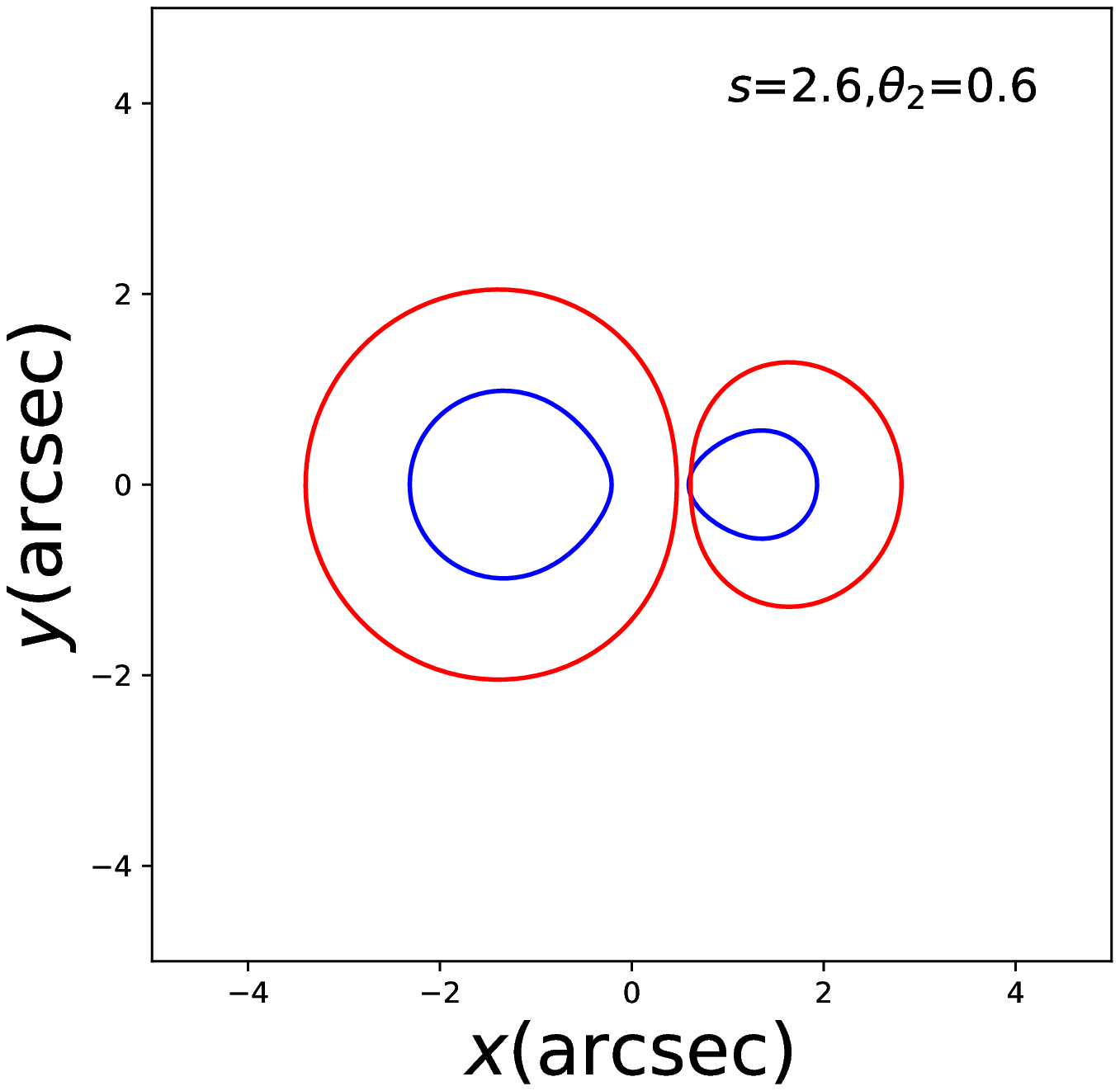}
  \includegraphics[width=4.0cm]{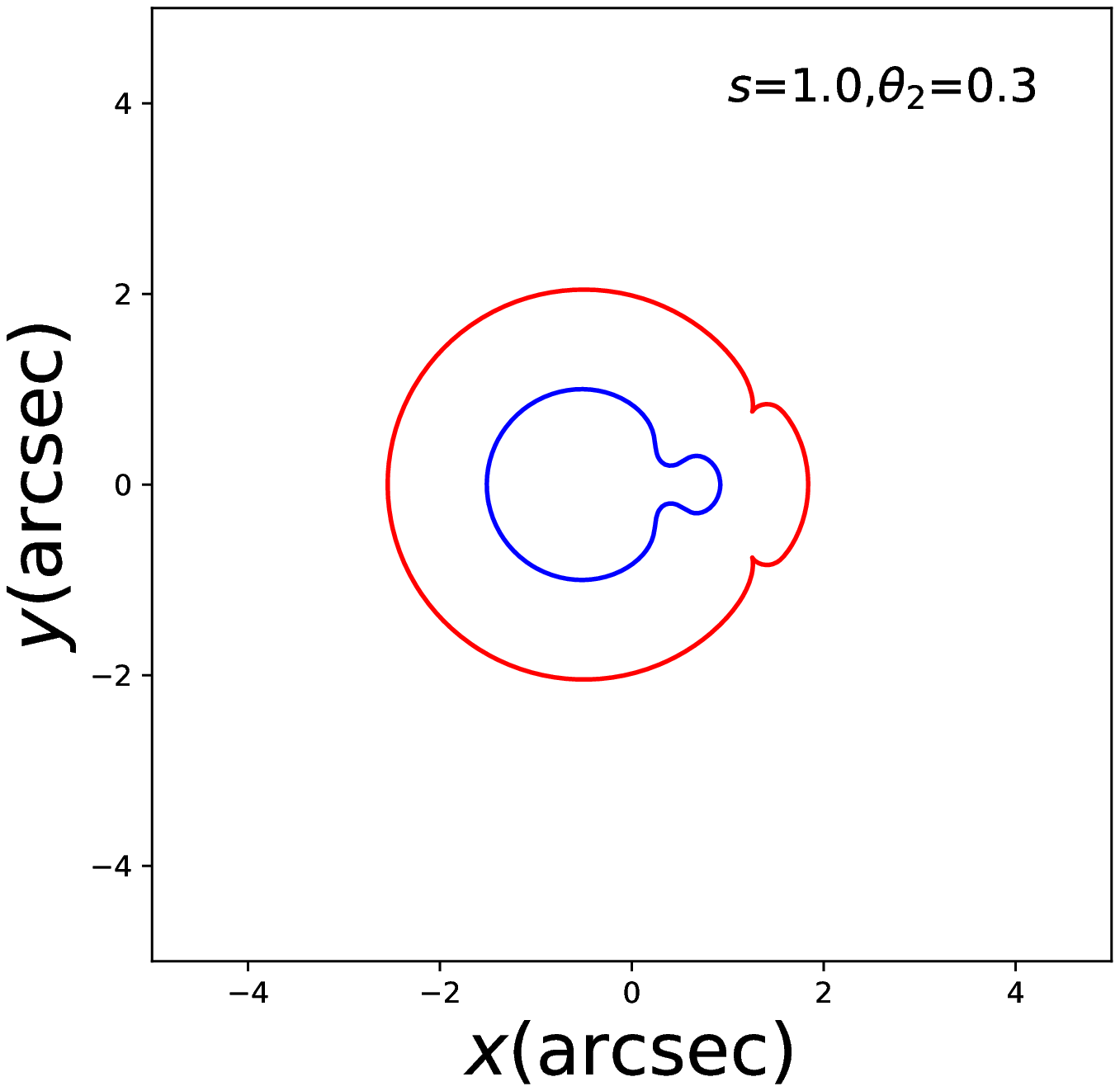}}
  \centerline{\includegraphics[width=4.0cm]{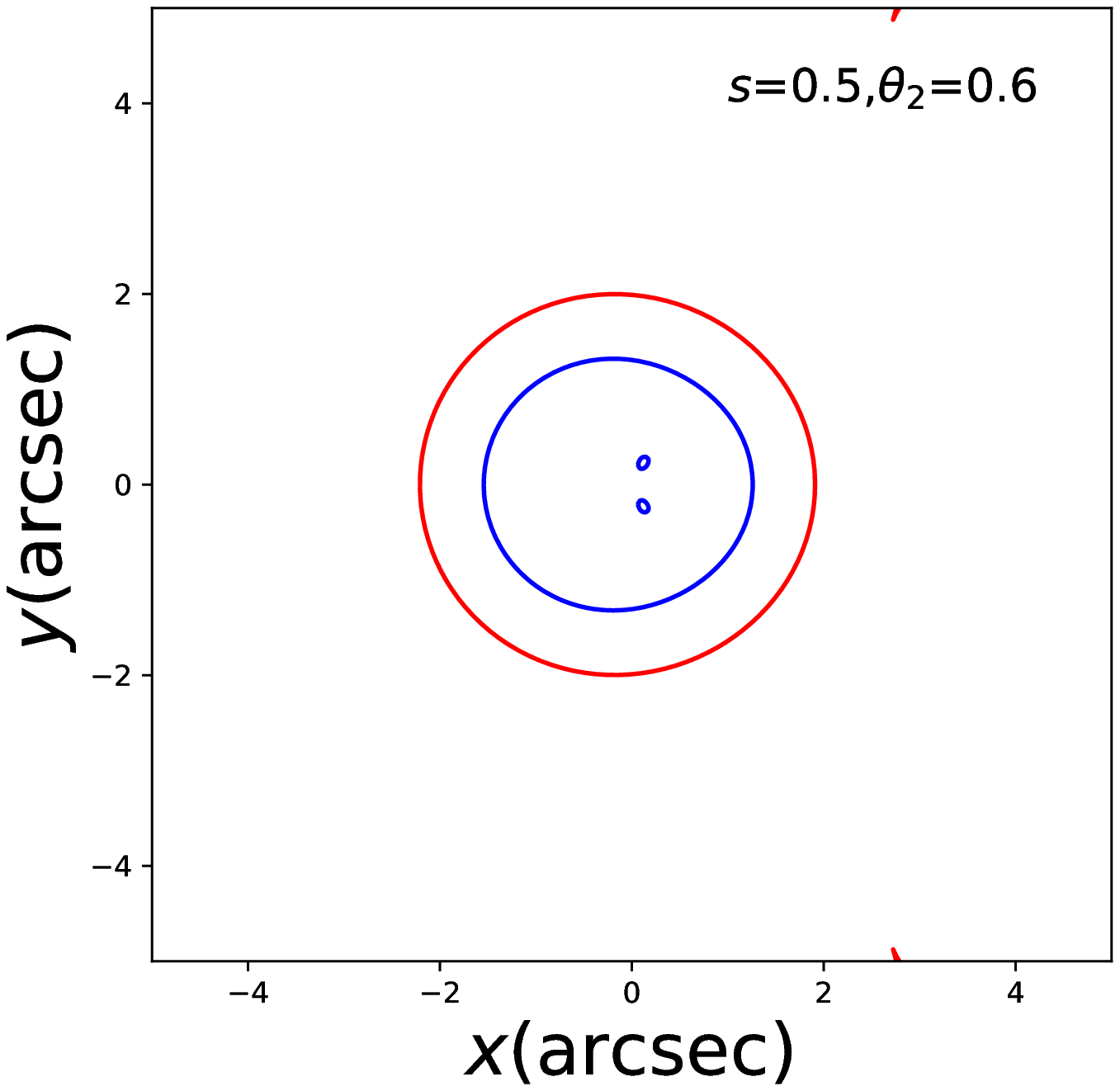}
  \includegraphics[width=4.0cm]{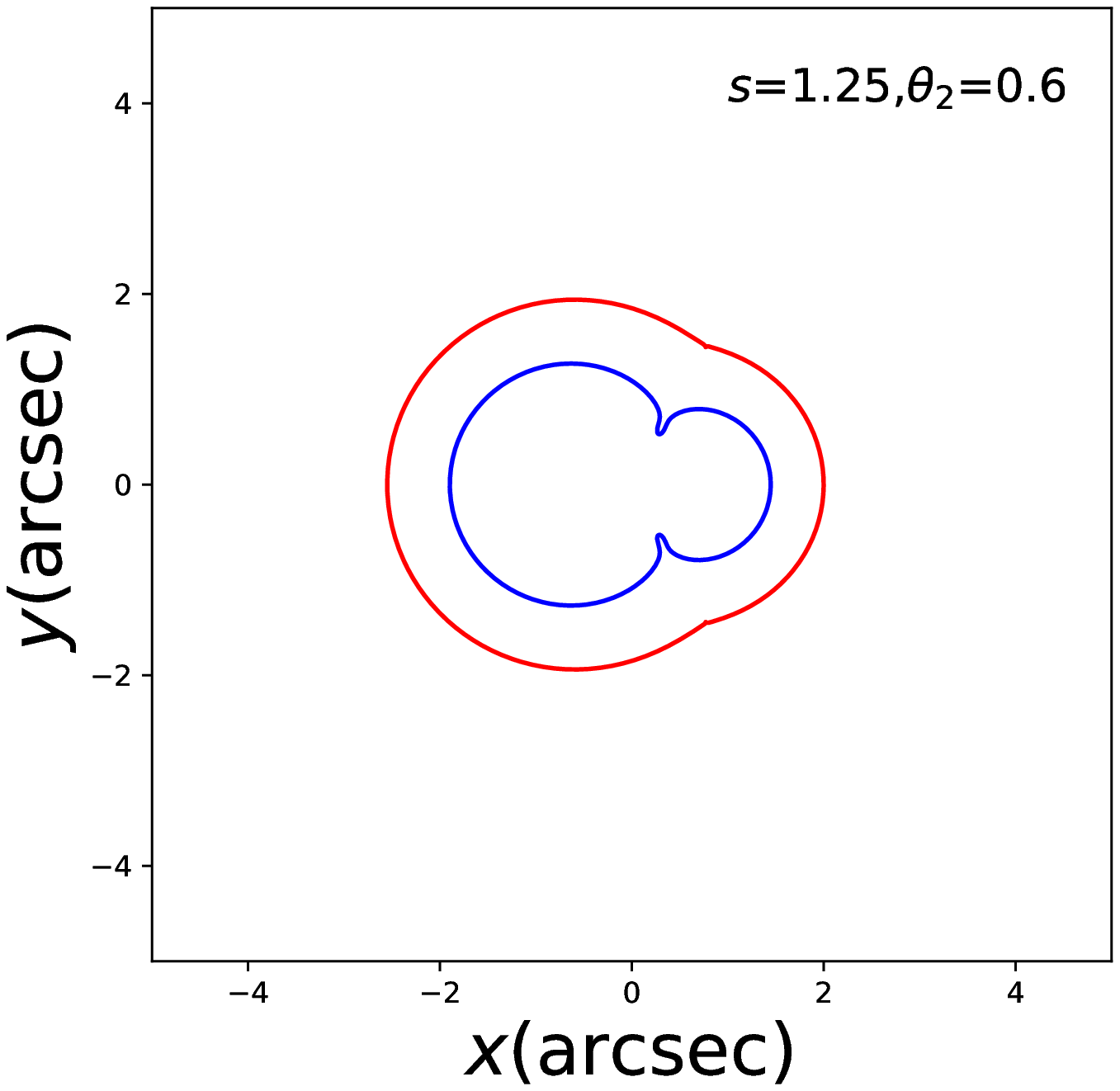}
  \includegraphics[width=4.0cm]{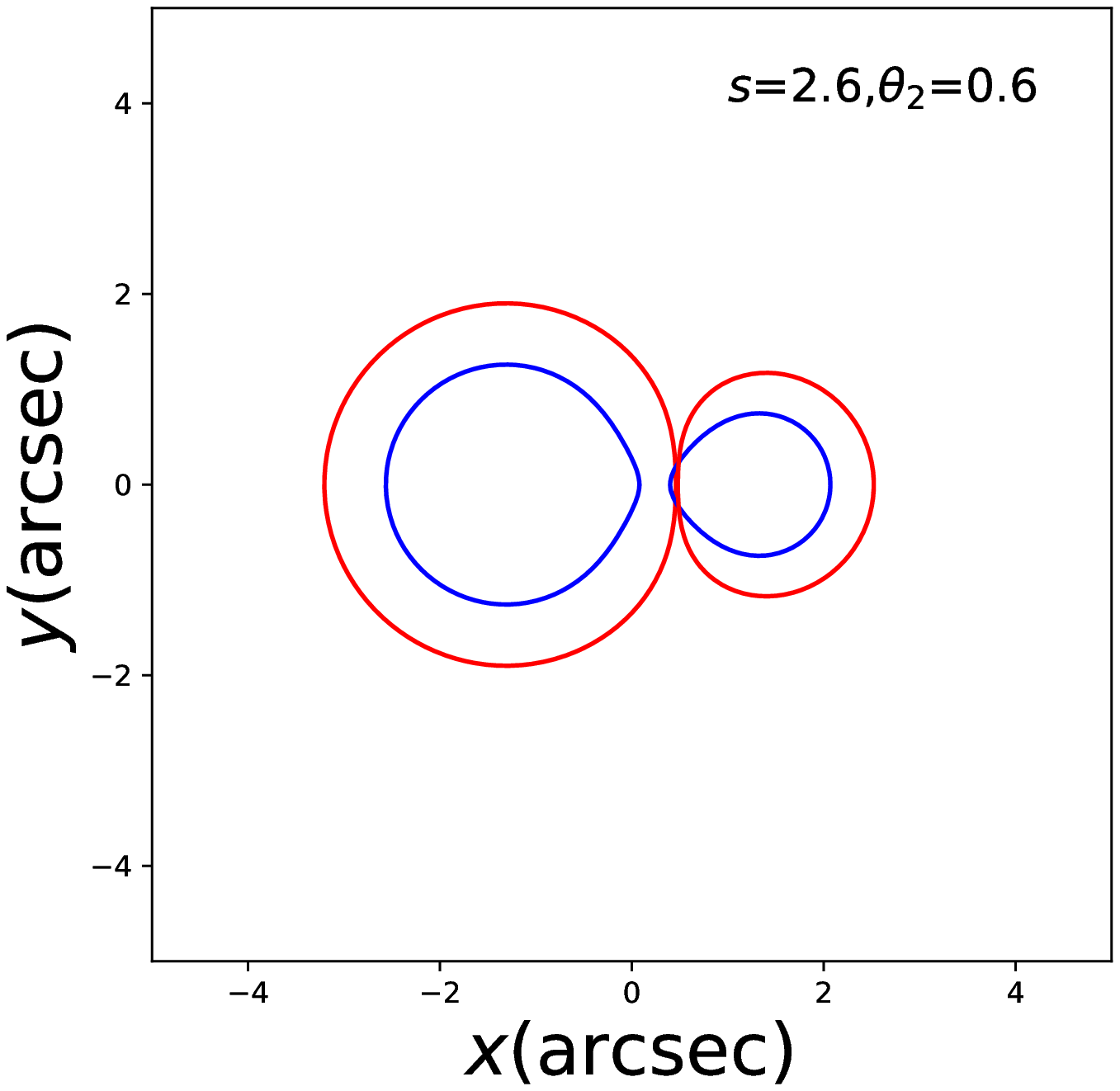}
  \includegraphics[width=4.0cm]{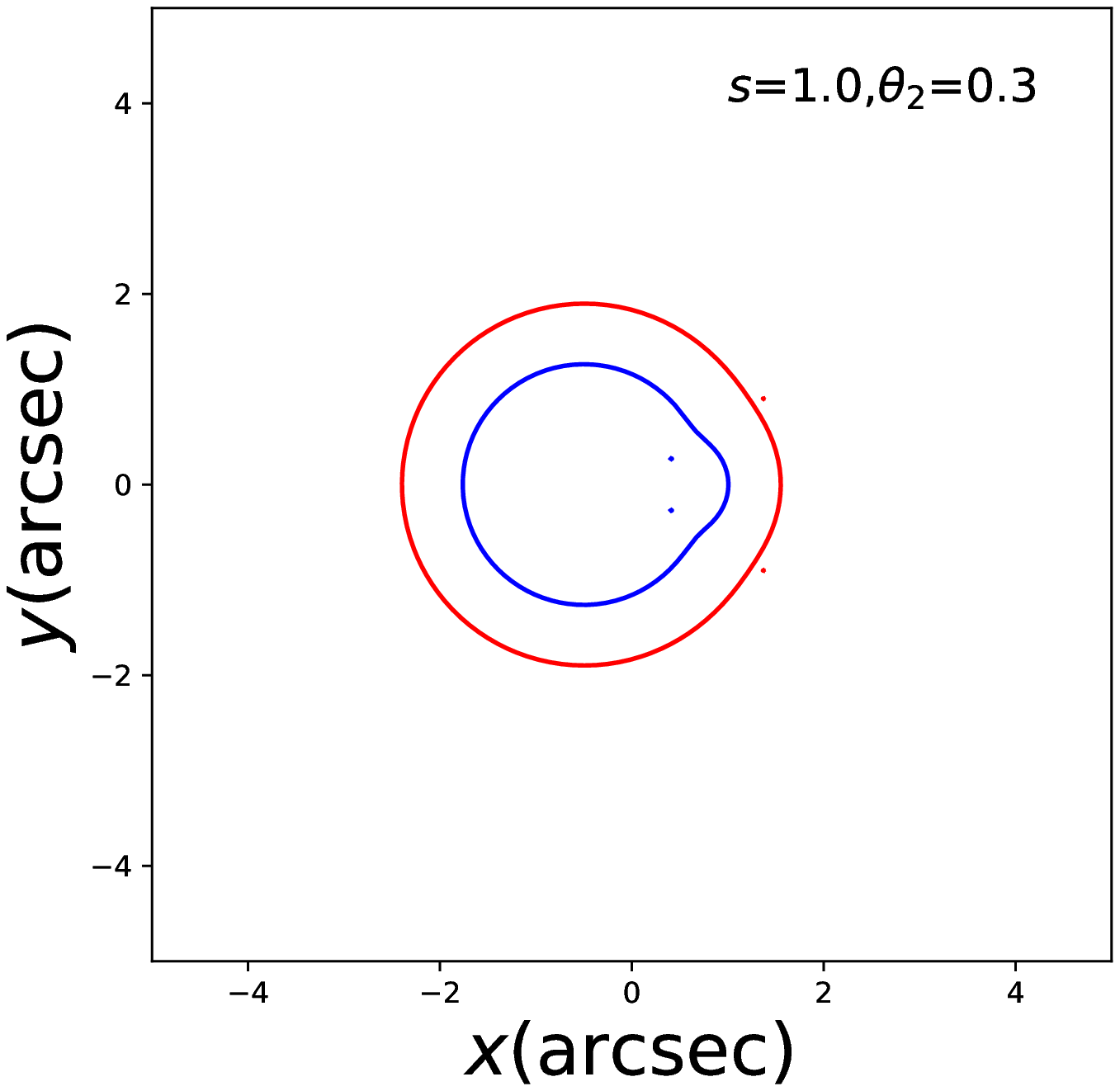}}
  \centerline{\includegraphics[width=4.0cm]{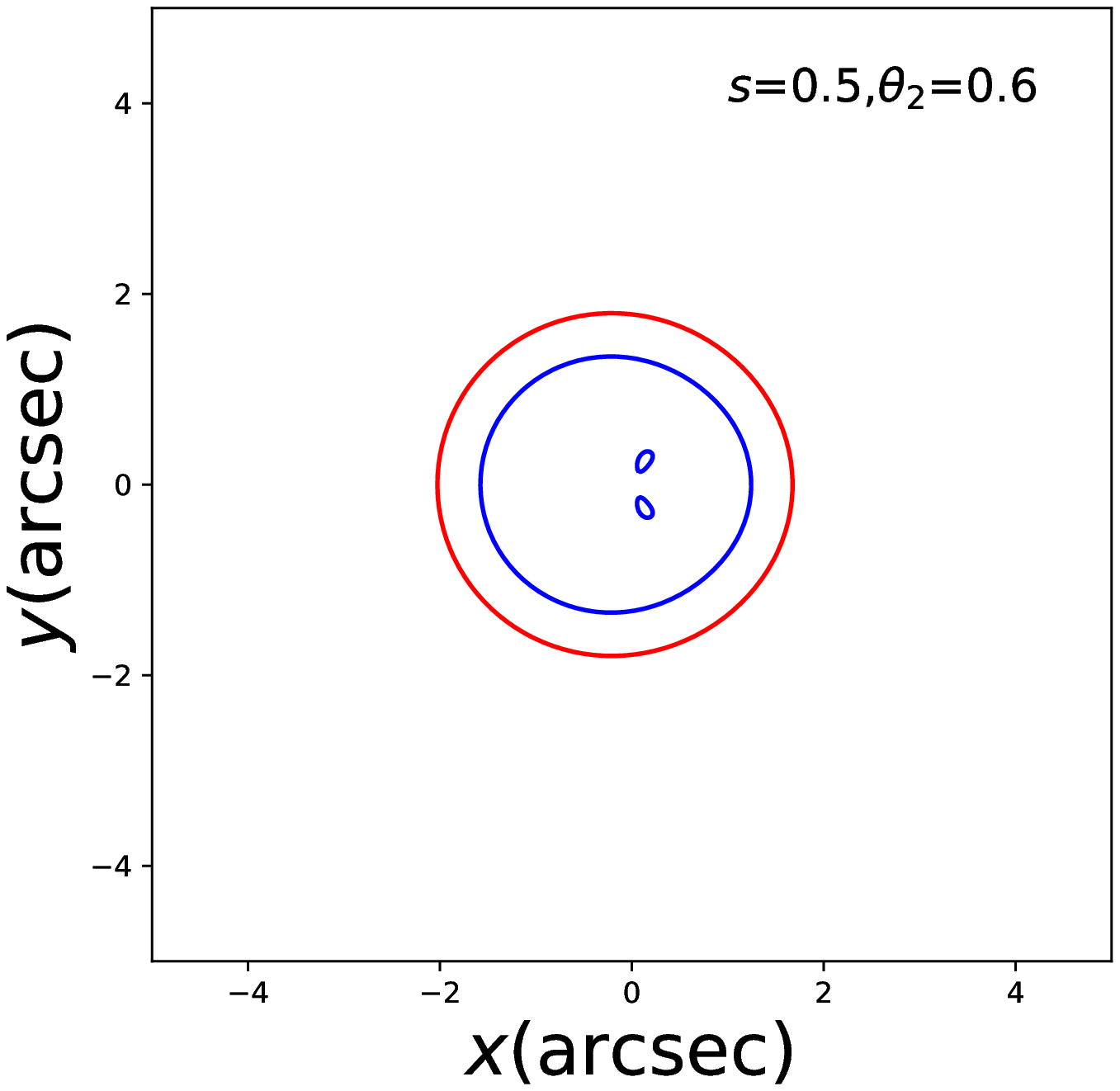}
  \includegraphics[width=4.0cm]{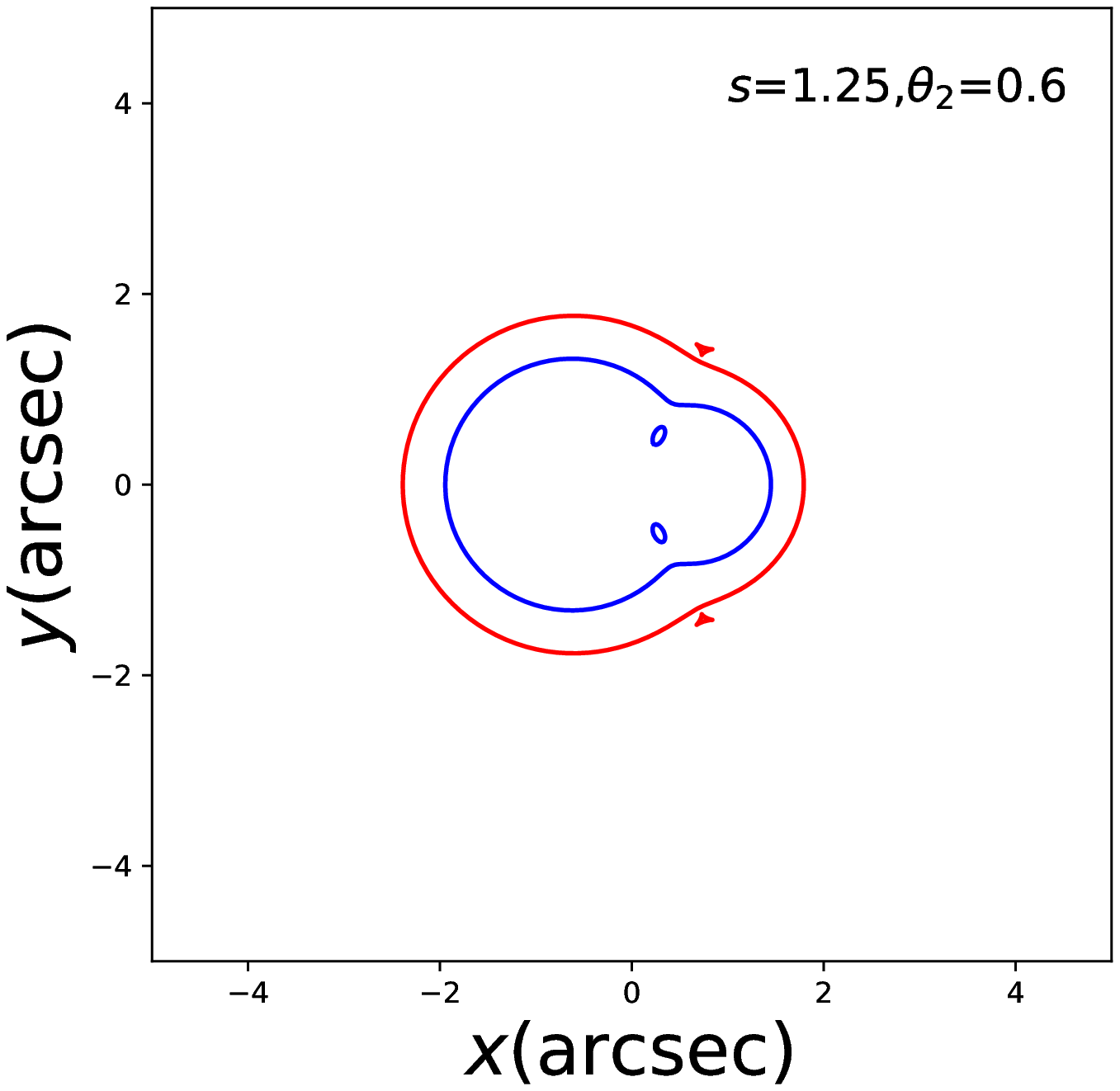}
  \includegraphics[width=4.0cm]{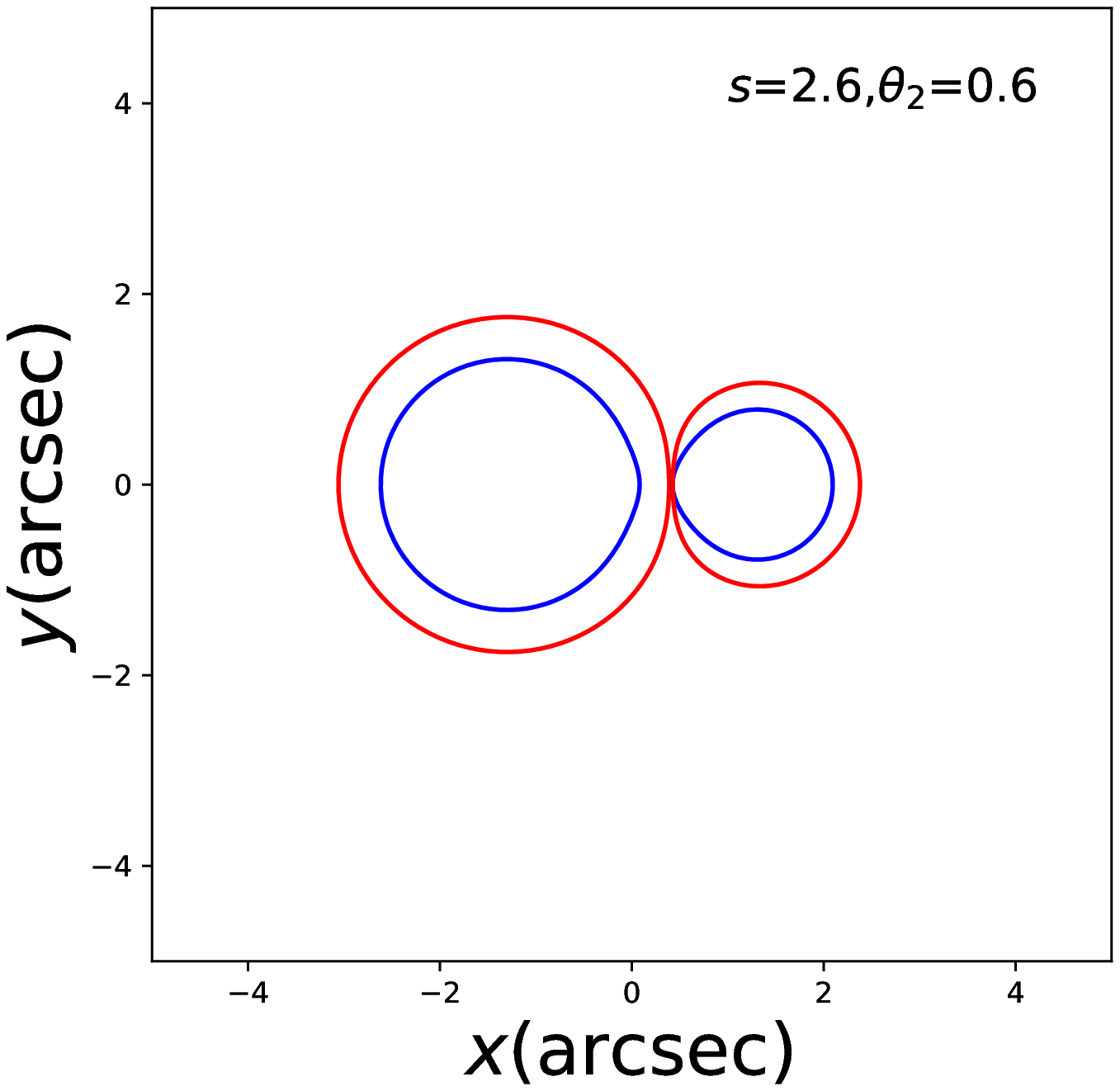}
  \includegraphics[width=4.0cm]{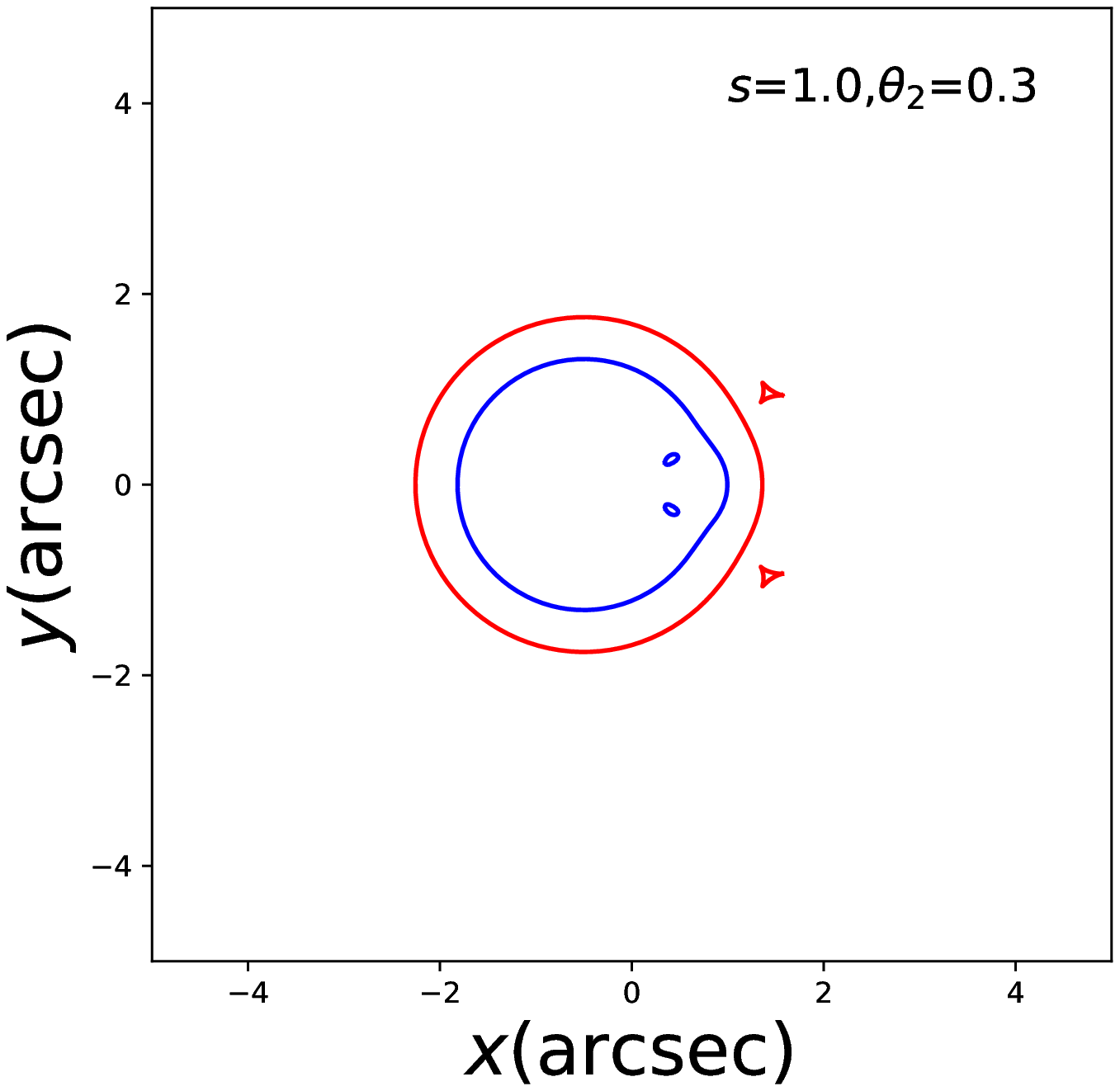}}
  \caption{A collection of critical (blue) and caustic curves (red) for the dual component singular power-law lens with $\theta_{01}=1$. The secondary lens has $\theta_{02}$ which is labelled on each plot. The top row has power-index $h=1$, the middle row has $h=2$ and the bottom row has $h=3$.}
  \label{fig:powte12}
\end{figure*}

In Fig.\,\ref{fig:powte11} we demonstrate examples of criticals and caustics of the dual component SPL lens with equal strength components $\theta_{01}=\theta_{02}=1$. We label the lens separation $s$ and core size $\theta_\text{c}$ on each panel of the figures. In Fig.\,\ref{fig:powte13}, we demonstrate some examples of unequal components with $\theta_{01}=1$ and $\theta_{02}=0.5$. In Fig.\,\ref{fig:powte12} we show some examples of unequal strength components $\theta_{01}=1$ for the singular lens. In this case we vary the lens separation and the secondary component strength $\theta_{02}$.

\section*{Acknowledgments}
We thank our anonymous referee for providing an insightful and constructive report that also provided a new direction to explore. Their suggestions improved the overall flow of the paper. We thank Artem Tuntsov and Oleg Yu. Tsupko for many valuable comments and suggestions on our draft. A.R. would like to acknowledge Samar
Safi-Harb for support during the development of this work. X.E. is
supported by NSFC Grant No. 11473032 and No. 11873006.

\bibliographystyle{mn2e}
\bibliography{ms_rogers_er_v3}
\end{document}